\documentclass[prd,
reprint,
superscriptaddress,
preprintnumbers,
nofootinbib,
 amsmath,amssymb,
 aps,
]{revtex4-2}

\usepackage[english]{babel}
\usepackage{xcolor}
\usepackage[linktocpage, colorlinks, citecolor=blue, linkcolor=blue, urlcolor=blue]{hyperref} 
\usepackage{enumerate}  
\usepackage{float}
\usepackage{calc}
\usepackage{dutchcal}
\usepackage{revsymb}
\usepackage{comment}
\usepackage{upgreek}
\usepackage{IEEEtrantools}
\usepackage{Commands}
\usepackage{soul} 
\usepackage[normalem]{ulem}
\usepackage{amsthm}

\usepackage[acronym]{glossaries}
\usepackage{siunitx}
\usepackage{multirow}

\DeclareSymbolFont{cmbrightop}{OT1}{cmbr}{m}{n}
\DeclareMathSymbol{\sfPsi}{\mathalpha}{cmbrightop}{9}

\usepackage{amsfonts, amssymb, amsthm, mathtools, mathrsfs}
\usepackage{physics}

\usepackage{graphicx}
\usepackage{dcolumn}
\usepackage{bm}
\usepackage[dvipsnames]{xcolor}

\newcommand{\orcid}[1]{\href{https://orcid.org/#1}{\includegraphics[width=8pt]{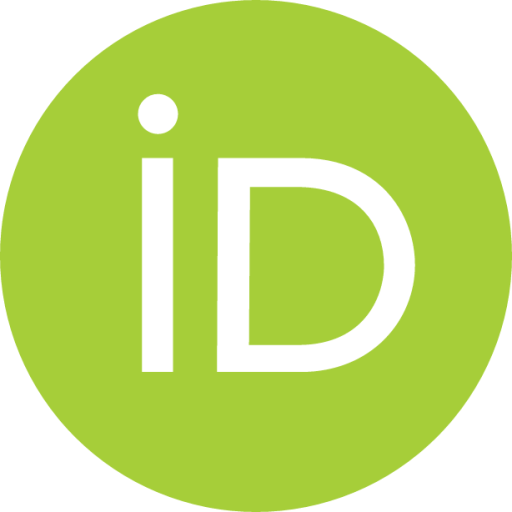}}}

\newcommand{\SNRmem}{\textrm{SNR}_{\textrm{mem}}}

\newcommand{\myhyperref}[1]{\hyperref[#1]{\ref{#1}}}

\begin{document}


\preprint{CERN-TH-2026-042}

\title{Toward claiming a detection of gravitational memory}

\author{Jann Zosso \orcid{0000-0002-2671-7531}}
\email{jann.zosso@nbi.ku.dk}
\affiliation{Center of Gravity, Niels Bohr Institute, Blegdamsvej 17, 2100 Copenhagen, Denmark}
\author{Lorena \surname{Maga\~na Zertuche} \orcid{0000-0003-1888-9904}}
\affiliation{Center of Gravity, Niels Bohr Institute, Blegdamsvej 17, 2100 Copenhagen, Denmark}
\author{Silvia Gasparotto~\orcid{0000-0001-7586-1786}}
\affiliation{CERN, Theoretical Physics Department, Esplanade des Particules 1, Geneva 1211, Switzerland}
\affiliation{Institut de F\'isica d’Altes Energies (IFAE), The Barcelona Institute of Science and Technology, Campus UAB, 08193 Bellaterra (Barcelona), Spain}
\author{Adrien Cogez \orcid{0009-0007-1316-9648}}
\affiliation{IRFU, CEA, Université Paris-Saclay, 91191, Gif-sur-Yvette, France}
\affiliation{Centre national d’études spatiales (CNES), Paris, France}
\author{Henri Inchauspé \orcid{0000-0002-4664-6451}}
\affiliation{Institute for Theoretical Physics, KU Leuven,
Celestijnenlaan 200D, B-3001 Leuven, Belgium}
\affiliation{Leuven Gravity Institute, KU Leuven,
Celestijnenlaan 200D box 2415, 3001 Leuven, Belgium}
\author{Milo Jacobs}
\affiliation{Institute for Theoretical Physics, KU Leuven,
Celestijnenlaan 200D, B-3001 Leuven, Belgium}
\affiliation{Leuven Gravity Institute, KU Leuven,
Celestijnenlaan 200D box 2415, 3001 Leuven, Belgium}

\date{\today}

\begin{abstract}
Gravitational memory is a zero-frequency effect associated with a permanent change in the asymptotic spacetime metric induced by radiation. While its universal manifestation is a net change in the proper distances between freely falling test masses, gravitational wave detectors are intrinsically insensitive to the final offset and can only probe the associated transition. A central challenge for any claim of detection, therefore, lies in defining a physically meaningful and operationally robust model of the corresponding time-dependent signal that is uniquely attributable to gravitational memory and clearly distinguishable from purely oscillatory radiation. We show that while the Bondi–van der Burg–Metzner–Sachs balance laws rigorously establish the total memory offset, a robust definition of the observable memory rise requires an additional physical input: a separation of scales between high-frequency gravitational waves and the lower-frequency buildup of memory. We formulate this separation using the Isaacson description of gravitational wave energy momentum.
Motivated by this observation, we develop a theoretical framework for defining and modeling the time-dependent memory rise, building on a self-contained review of the theory of gravitational memory and focusing in particular on compact binary coalescences. Specializing to space-based detectors, we analyze the response of LISA to gravitational radiation including a memory contribution, with emphasis on mergers of supermassive black hole binaries, which offer the most promising prospects for a first single-event detection. The framework developed here provides the theoretical foundation for statistically well-defined hypothesis testing between memory-free and memory-full radiation and enables quantitative assessments of detection prospects. These results establish a principled pathway toward a future observational claim of gravitational memory.
\end{abstract}

\maketitle

\tableofcontents

\section{Introduction}

\emph{Gravitational memory}~\cite{Zeldovich:1974gvh,Turner:1977gvh,Braginsky:1985vlg,Braginsky:1987gvh,Christodoulou:1991cr,Ludvigsen:1989cr,Blanchet:1992br,Thorne:1992sdb,PhysRevD.44.R2945,Favata:2008yd,Favata:2009ii,Favata:2010zu,Barnich:2009se,Bieri:2013ada,Pasterski:2015tva,Nichols:2017rqr,Garfinkle:2022dnm,Zosso:2024xgy}
is defined as a permanent offset in the radiation-carrying metric component of an asymptotic spacetime, which manifests itself as a permanent change in proper distances between freely falling test masses.
More precisely, it represents a \emph{net effect} associated with the zero-frequency limit of the gauge field supporting the radiation.
As a formal zero-frequency mode, memory is closely connected to soft theorems, first identified by Weinberg~\cite{Weinberg_PhysRev.140.B516}, which universally govern the infrared structure of scattering processes.
This universal property can be understood through its connection to Bondi–van der Burg–Metzner–Sachs (BMS) spacetime symmetries defined in an idealized asymptotic limit~\cite{Bondi:1962px,Sachs:1962wk,FrauendienerJ,Ashtekar:2014zsa,Compere:2019gft,DAmbrosio:2022clk,Goncharov:2023woe,Mitman:2024uss,DeLuca:2024cjl,Bhattacharjee:2019jaf}.
From that point of view, the memory effect is naturally interpreted as a transition between inequivalent vacua characterized by a nonzero vacuum expectation value.
This threefold connection between memory, soft factors, and asymptotic symmetries is known as the infrared triangle~\cite{Strominger:2014pwa,Strominger:2017zoo}.
Although subleading infrared effects are sometimes also referred to as ``memory''\footnote{More specifically, the subleading gravitational effect is known as \emph{spin memory}, while the leading low-frequency effect considered here is termed \emph{displacement memory}.},
we will reserve the term throughout this work for the leading, net zero-frequency effect.
In particular, we distinguish gravitational memory associated with gravitational radiation from analogous leading effects in other field theories, such as electromagnetic memory in electrodynamics~\cite{Bieri:2013hqa,He:2014cra,Campiglia:2015qka,Pasterski:2015zua,Kapec:2015ena,Zosso:2025orw}.

A direct measurement of gravitational memory would therefore provide an experimental window into foundational principles of radiation theory.
However, in the gravitational case, a permanent change in the vacuum configuration of the asymptotic spacetime metric is not directly detectable with current gravitational-wave observatories.
This limitation arises because gravitational-wave detectors are designed to measure time-dependent signals within a finite frequency band, dictated by instrumental and environmental noise sources.
As a consequence, they are fundamentally insensitive to the final, constant memory offset, in the sense that once a radiation burst has passed, it is not possible to verify a permanent displacement of the test masses.\footnote{This is equally true for ground-based and space-based detectors alike: in all cases, uncontrolled low-frequency noise prevents a direct observation of a permanent shift.}
Instead, detectors can only access the time-dependent part of a displacement memory signal, namely the transition between the initial and final constant values.

Crucially, there is no unique \emph{a priori} prescription for the time-dependent profile of this signal. Although the total memory offset is well defined, any smooth function interpolating between the initial and final values reproduces by definition the same net difference (see also Sec.~II~B of Ref.~\cite{Grant:2022bla}). A convincing claim of detection therefore requires more than a formal definition of the final offset: it calls for a physically motivated model of the time-dependent memory rise that can be distinguished from purely oscillatory radiation in a frequency-band-limited detector and whose detectability is directly associated with the amplitude of the total memory difference. In other words, a robust detection strategy should ensure that the statistical significance of the signal directly tracks the size of the physical memory offset. From this perspective, oscillatory components of the radiation are not suitable building blocks for a memory model, since they increase the signal-to-noise ratio without contributing to the defining zero-frequency memory offset.
As we will show in this work, current strategies for defining a memory model do not address this challenge in a fully satisfactory and self-contained manner.\footnote{While the BMS balance laws~\cite{Bondi:1962px,Sachs:1962wk,FrauendienerJ,Ashtekar:2014zsa,Compere:2019gft,DAmbrosio:2022clk,Goncharov:2023woe,Mitman:2024uss,DeLuca:2024cjl,Bhattacharjee:2019jaf,Newman:1968,Geroch:1981ut,Ashtekar:1981,Ashtekar:1982,Wald:1999wa,Flanagan:2015pxa,Strominger:2017zoo,Compere:2019gft,Ashtekar:2019viz,Mitman:2020bjf} rigorously establish the existence of gravitational memory and determine the final offset, they do not uniquely isolate the time-dependent signal to be searched for in data, since the corresponding expressions generally also contain oscillatory contributions. In particular, the commonly used memory model defined by the $(2,0)$ harmonic mode of the null part of the balance laws relies on additional assumptions, which we will discuss in the main text. On the other hand, an instantaneous step function model is unphysical, as it introduces arbitrarily high-frequency content.}

The central question of this work is therefore:
\begin{quote}
\emph{Can one identify a fundamental signature of gravitational memory in the frequency-band-limited response of a gravitational-wave detector that is unique to the defining memory offset and clearly distinguishable from purely oscillatory radiation?}
\end{quote}
This question is particularly timely in light of the prospect of a first detection of gravitational memory
\cite{Lasky:2016knh,McNeill:2017uvq,Johnson:2018xly,Yang:2018ceq,Hubner:2019sly,Islo:2019qht,NANOGrav:2019vto,Burko:2020gse,Boersma:2020gxx,Islam:2021old,Hubner:2021amk,Grant:2022bla,Sun:2022pvh,Gasparotto:2023fcg,Ghosh:2023rbe,Goncharov:2023woe,NANOGrav:2023vfo,Cheung:2024zow,Sun:2024nut,Hou:2024rgo,Inchauspe:2024ibs,Agazie:2025oug}.
While a statistical detection through the stacking of binary black hole (BBH) mergers may become possible in future observing runs of current ground-based detectors~\cite{Hubner:2019sly,Hubner:2021amk}, the space-based gravitational-wave observatory LISA (Laser Interferometer Space Antenna)~\cite{LISA,LISA:2024hlh,LISA_RosettaStone} is a prime candidate to achieve the first single-event detection, as demonstrated in Ref.~\cite{Inchauspe:2024ibs}.
Indeed, LISA is expected to observe mergers of massive black-hole binaries (MBHBs) in the frequency range $10^{-4}$--$10^{-1}\,\mathrm{Hz}$ with extraordinarily high signal-to-noise ratios (SNRs), particularly during merger, when most of the memory is generated.
For such events, the memory signal can reach sufficiently high SNR to allow for a detection, although predicted detection rates depend strongly on the underlying astrophysical population models~\cite{Barausse_2020,Barausse_Lapi_2021}.
Motivated by this prospect, the present work focuses on the detection of gravitational memory with LISA.
Nevertheless, the theoretical derivations and models developed here apply generally to gravitational memory in radiative processes and are therefore also relevant for ground-based detectors
\cite{LIGOScientific:2014pky,VIRGO:2014yos,KAGRA:2020tym,Reitze:2019iox,ET:2025xjr}
as well as Pulsar Timing Arrays (PTAs)~\cite{NANOGrav:2023gor,EPTA:2023fyk}.

The primary aim of this paper is to take a concrete step toward a future claim of detection of gravitational memory by providing a principled answer to the central question posed above. 
Our approach explicitly distinguishes between the defining net memory offset and the measurable time-dependent memory rise that can serve as the basis of a detection claim.
In this context, the core modeling insight developed in this work is to utilize a separation of scales between the high-frequency gravitational waves and the lower-frequency buildup of memory as an additional physical ingredient. 
We argue that this separation is naturally provided by the Isaacson description of gravitational-wave energy-momentum~\cite{Isaacson_PhysRev.166.1263,Isaacson_PhysRev.166.1272,Heisenberg:2023prj,Zosso:2025ffy}, which yields the coarse-grained flux entering the time-dependent memory signal.
On this basis, we introduce a robust theoretical framework for defining the time-dependent memory rise. 
This framework clarifies and systematizes the assumptions underlying existing models used for quasi-circular compact-binary coalescences (CBCs), while at the same time providing a principled route to extend memory modeling beyond this specific scenario.

To this end, we present an extensive and self-contained summary of the theory of gravitational memory and use it to arrive at an internally consistent definition of a memory model, in particular applicable to CBCs.
This framework allows us to address common misconceptions and criticisms surrounding the program of measuring gravitational memory and to articulate the theoretical significance of a potential upcoming detection.
Building on this theoretical foundation, a companion paper computes Bayes factors associated with the response of LISA to memory-free and memory-full radiation~\cite{Cogez:2025memoryLISA}, enabling statistically meaningful assessments of detection prospects.
With a dedicated Appendix, the present work also serves as a reference in which a comprehensive set of formulas and derivations relevant to gravitational memory, often scattered throughout the literature, are collected in a unified notation. Furthermore, we present gravitational memory in several complementary theoretical settings, including asymptotic BMS balance laws, the direct Isaacson description of radiative energy-momentum backreaction, and the post-Newtonian (PN) regime, and provide explicit cross-links and comparisons between these approaches.

This paper is organized as follows.
In Sec.~\ref{Sec:MemoryTheory}, we introduce a general and physically motivated definition of a time-dependent memory model.
More specifically, Sec.~\ref{sSec:Radiation} offers a concise but comprehensive summary of the theory of gravitational memory, while in Sec.~\ref{sSec:DefMemModel} we identify the characteristic definition of the memory rise that provides a clear theoretical link between the measured time-dependent signal and the defining property of gravitational memory as a permanent offset in the asymptotic spacetime metric.
We present the full analytic expression in a spin-weighted spherical-harmonic expansion and discuss the intrinsic limitations of the theoretical framework in uniquely defining a memory rise.
Section~\ref{sSec:Practical Approximation} applies this definition to the specific case of nonlinear memory from binary black hole mergers and discusses the nature of different gravitational memory models.
Furthermore, we analyze the memory model for quasi-circular binary black holes in both the time and frequency domains.
This discussion goes beyond the standard use of BMS balance laws and clarifies the regime in which approximations based on the null part of the flux-balance laws are valid.
In Sec.~\ref{Sec:claim of detection with LISA}, we specialize to the response of the LISA detector to gravitational radiation including a memory contribution.
In Sec.~\ref{sSec:CharacteristicProperties}, we discuss key properties of the LISA detector that are relevant for understanding the behavior of the memory response and address common misconceptions. In particular, we demonstrate that residual oscillatory components in a memory model can significantly compromise the memory response and lead to biased estimates of the associated SNR.\footnote{In particular, if one identifies memory directly with the full null part of the balance laws or with unaveraged flux-based constructions, the resulting signal can retain small oscillatory contributions at merger and ringdown frequencies, which may artificially enhance detectability estimates without being uniquely tied to the final memory offset.}
Finally, in Sec.~\ref{sSec:BayesFactor}, we present a robust Bayes factor analysis for a claim of detection of gravitational memory with LISA, summarizing the main results of the Bayesian analysis carried out in the companion paper~\cite{Cogez:2025memoryLISA}.
We emphasize the implications of the memory model developed in this work for hypothesis testing and claims of detection and provide updated forecasts for memory detectability.

\section{The theory of gravitational displacement memory}\label{Sec:MemoryTheory}

\subsection{Gravitational memory in asymptotic radiation}\label{sSec:Radiation}

\subsubsection{Definition}\label{ssSec:Definition}

A fundamental consequence of field theory in physics is the ability of fields to irreversibly transport energy away from a localized source, a process known as \emph{radiation}. For instance, in GR, radiation emitted from a localized source is, to a first approximation, described through two $\mathcal{O}(1/r)$ transverse traceless ($TT$) perturbations\footnote{The expressions for the polarization tensors are given in Eq.~\eqref{eq:DefsPolarizationTensors}.}
\begin{equation}\label{eq:gravitationalradiationGen}
    h^{TT}_{ij}=h_+ e^+_{ij}+h_\times e^\times_{ij}\,,
\end{equation}
of the metric on top of an asymptotically flat Minkowski solution described by a source-centered coordinate system $\{t,x,y,z\}$~\cite{misner_gravitation_1973,Thorne:1980ru} with $r\equiv\sqrt{x^2+y^2+z^2}$.
More precisely, variations in the quadrupole and higher moments of a localized source result in the creation of spacetime perturbations that decouple from the source and radiate away energy-momentum content described by the two polarizations $h_+$ and $h_\times$ as a function of asymptotic retarded time $u\equiv t-r$.


Note that this definition consists of two main ingredients. First, considering perturbations in the flat asymptotics of a spacetime captures the fact that, in general, radiation can only be identified far from the source at infinity, where one can clearly distinguish the fields that carry the energy lost forever to the localized system~\cite{DAmbrosio:2022clk}. This region of spacetime is generally known as \emph{radiation zone}. Second, in contrast to Coulombic long range potentials, radiation is intimately connected to the concept of \emph{dynamical degrees of freedom} of a theory. In other words, in GR, radiation is only associated with its two degrees of freedom of the asymptotic metric perturbations that represent the dynamical part of the field.\footnote{To be precise, perturbations on top of a given solution ought to be described through a pullback from a perturbed manifold to the manifold of the exact solution, involving a gauge freedom that can be associated with a choice of coordinates within a particular foliation. This redundancy in description divides metric perturbations into unphysical gauge modes and the physical degrees of freedom. Among the latter, the equations of motion of the theory dictate the amount of dynamical degrees of freedom. We refer to the following reviews for a more careful treatment of these statements~\cite{Weinberg1972,misner_gravitation_1973,WaldBook,Flanagan:2005yc,carroll2019spacetime,maggiore2008gravitational,Zosso:2024xgy}.}

Due to their dynamics and $1/r$ falloff, the radiative degrees of freedom carry information about the source that is radiated away through energy loss and can be extracted far from it. In particular, in the radiation zone, the $h^{TT}_{ij}$ metric perturbation components will dominate any measurement of changes in proper distance between freely falling test masses, as governed by the geodesic deviation equation~\cite{misner_gravitation_1973,carroll2019spacetime,maggiore2008gravitational}
\begin{equation}\label{eq:GeodesicDeviation}
    \ddot{s}_i=\frac{1}{2}\ddot h^{TT}_{ij} \,s^j +\mathcal{O}\left( \frac{1}{r^2}\right),
\end{equation}
with $s_i$ the spatial proper distance between two such test masses. The measurement of gravitational radiation is therefore captured by the time evolution
\begin{align}\label{eq:integratedGeodesicDeviation}
    \Delta s_i(u)  = \frac{1}{2} \Delta h^{TT}_{ij}(u)\, s_0^j  \,,
\end{align}
where we fixed an initial time $u_0$ and for any operator $\mathcal O$ we define
\begin{equation}\label{eq:DefDelta}
    \Delta \mathcal{O}(u)\equiv \mathcal O (u)- \mathcal O_0\,, \quad \text{with}\quad \mathcal O_0\equiv \mathcal O(u_0)\,,
\end{equation}
such that $s^0_j$ defines the initial proper spatial separation. A change in a particular direction $s_i=s\,\hat s_i$ therefore defines the fractional strain
\begin{align}\label{eq:integratedGeodesicDeviation2}
    \Delta s(u)  = \frac{s_0}{2} \Delta h^{TT}_{ij}(u)\, \hat s^i \hat s^j  \,.
\end{align}

Such radiation typically manifests as waves. Indeed, a primary mechanism for producing radiation is a periodic motion of a source, which generates oscillatory disturbances that propagate outward. In the case of gravitational waves (GWs), Eq.~\eqref{eq:integratedGeodesicDeviation} implies that the proper distance between freely falling test masses undergoes corresponding periodic variations in response to the passing wave, giving rise to the characteristic signals measured by GW observatories.
However, on a fundamental level, any such emission of gravitational radiation will not only contain wave-like oscillatory perturbations but will always also come with a non-oscillatory component, which permanently alters the structure of spacetime after its passage. Indeed, very generally, the asymptotic strain does not settle at its initial value after a burst of radiation~\cite{Christodoulou:1991cr}
\begin{equation}\label{eq:DefiningMemoryOffset}
\lim_{u\rightarrow\infty}\Delta h^{TT}_{ij}(u)\equiv \Delta h^{TT}_{ij}\neq 0\,,
\end{equation}
resulting in a nonzero and constant offset within any detector response
\begin{equation}\label{eq:memorydef}
    \lim_{u\rightarrow\infty}\Delta s(u)\neq 0\,.
\end{equation}
The part of the radiation responsible for such a permanent shift in proper distances is what we will call \emph{gravitational memory}~\cite{Zeldovich:1974gvh,Turner:1977gvh,Braginsky:1985vlg,Braginsky:1987gvh,Christodoulou:1991cr,Ludvigsen:1989cr,Blanchet:1992br,Thorne:1992sdb,PhysRevD.44.R2945,Favata:2008yd,Favata:2009ii,Favata:2010zu,Barnich:2009se,Bieri:2013ada,Pasterski:2015tva,Nichols:2017rqr,Garfinkle:2022dnm,Zosso:2024xgy,Zosso:2025ffy}.

As mentioned in the introduction, this statement has profound implications for the infrared structure of field theories. As we will understand below, the zero-frequency offset of the radiation in Eq.~\eqref{eq:DefiningMemoryOffset} encodes a universal imprint of the angular distribution of the net change in energy-momentum of a localized source, representing a universal structure of soft radiation. In turn, this universal behavior is closely connected to the asymptotic symmetries of spacetime.

\subsubsection{Proof of existence through balance laws}\label{ssSec:ProofofMemory}

\begin{figure}[h]
    \centering
    \includegraphics[width=0.44\textwidth]{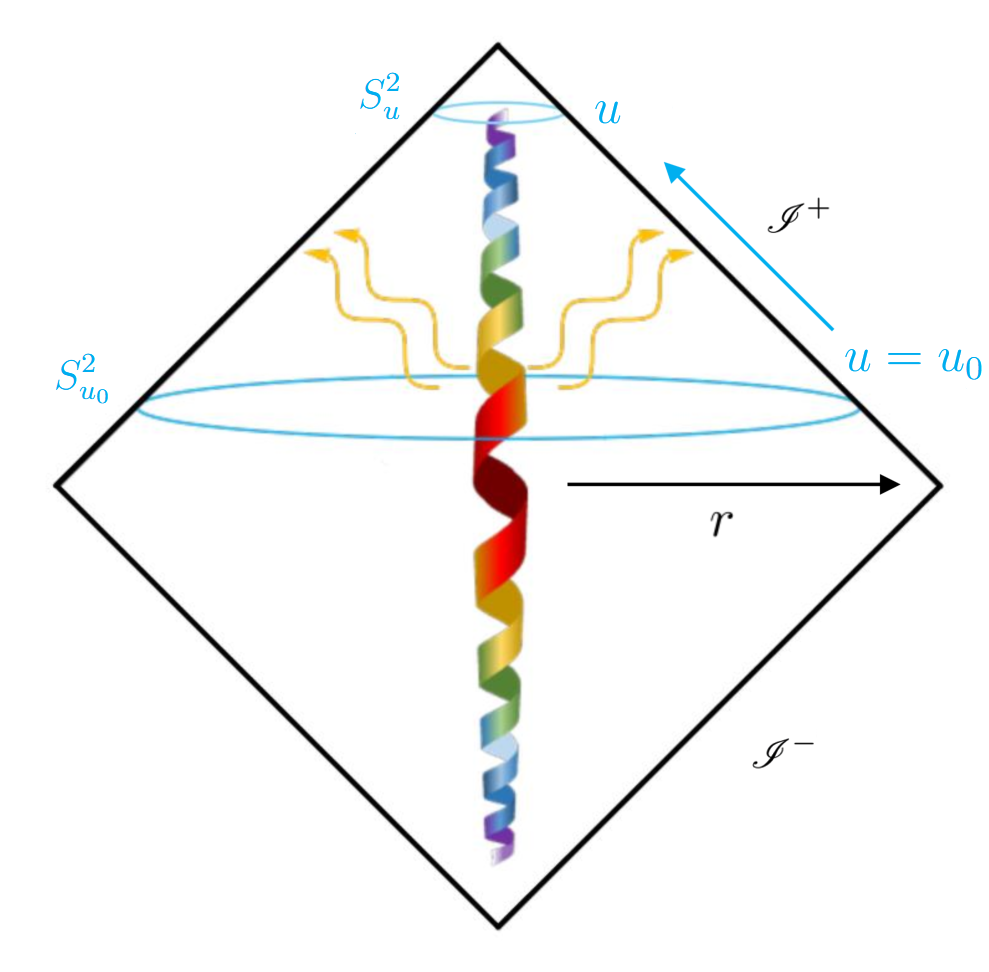}
    \caption{Penrose diagram of a conformally compactified asymptotically flat spacetime in asymptotic light-cone coordinates $\{u,r,\theta,\phi\}$, with $u=t-r$ and where time $t$ flows vertically. A localized source emits null radiation (yellow) toward future null infinity $\scri^+$, defined as the $r\to\infty$ limit at fixed retarded time $u$. The angular coordinates are not shown, but the asymptotic two-spheres at retarded times $u_0$ and $u$ are depicted schematically as blue circles. The BMS balance laws state that the supermomentum flux reaching $\scri^+$ between $S^2_{u_0}$ and $S^2_u$ is exactly balanced by the change in supermomentum charge between these two times. [Figure adapted from~\cite{DAmbrosio:2022clk,Zosso:2024xgy}.]}
    \label{fig:Asymptotically flat spacetime}
\end{figure}

A rigorous proof of the existence of gravitational-wave memory, as defined in Eq.~\eqref{eq:DefiningMemoryOffset}, is furnished by the BMS \emph{balance laws}~\cite{Bondi:1962px,Sachs:1962wk,Newman:1968,Geroch:1981ut,Ashtekar:1981,Ashtekar:1982,Wald:1999wa,Ashtekar:2014zsa,Flanagan:2015pxa,Strominger:2017zoo,Compere:2019gft,Ashtekar:2019viz,Mitman:2020bjf,DAmbrosio:2022clk,Mitman:2024uss}. These balance laws are associated with the exact asymptotic symmetries of spacetimes satisfying the mathematical assumptions of \emph{asymptotic flatness}~\cite{Penrose:1962ij,Geroch:1977big,WaldBook,GomezLopez:2017kcw}, which admit a well-defined notion of \emph{null infinity}.%
\footnote{The notion of asymptotic flatness can be formulated either geometrically, through the conformal compactification of spacetime first introduced by Penrose, or directly within the physical spacetime using a specific coordinate system, as originally done in Refs.~\cite{Bondi:1962px,Sachs:1962wk} in what is known as \emph{Bondi gauge}. For the purposes of this review, we will not delve into these technical constructions and instead present the relevant formulas in the source-centered Cartesian coordinates introduced above. For further details, see Refs.~\cite{Strominger:2017zoo,DAmbrosio:2022clk,Mitman:2024uss}.}
These mathematical structures are particularly valuable for sharply distinguishing the notion of Coulombic fields from radiation introduced above, despite the fact that both exhibit a leading-order asymptotic falloff proportional to $\sim 1/r$: Radiation is identified unambiguously as the component of the field that reaches future null infinity,
as illustrated in Fig.~\ref{fig:Asymptotically flat spacetime}.

At the same time, a concrete definition of an asymptotically flat structure of spacetime allows to inquire about the symmetries of spacetime, hence, the coordinate transformations that leave the asymptotic structure invariant. These symmetries are encoded in the BMS group~\cite{Bondi:1962px,Sachs:1962wk}, which constitutes an infinite-dimensional extension of the Poincar\'e group through the inclusion of angular-dependent translations known as \emph{supertranslations}. Heuristically, this enlargement reflects the fact that, at null infinity, neighboring points on the asymptotic two-sphere are no longer in direct causal contact, allowing independent translations along each asymptotic direction while still preserving the asymptotic structure of spacetime.

In close analogy with the Poincar\'e group of Minkowski spacetime, each generator of the BMS supertranslations is associated with a corresponding supermomentum charge $\mathcal{Q}_{(\alpha)}(u)$. At future null infinity, these charges are conserved modulo a supermomentum flux $\mathcal{F}^{\mathrm{R}}_{(\alpha)}(u)$ carried by asymptotic radiation, leading to the BMS balance law
\begin{equation}\label{eq:BMSBalanceLawI MT}
    \Delta\mathcal{Q}_{(\alpha)}(u) = - \mathcal{F}^{\mathrm{R}}_{(\alpha)}(u)\,.
\end{equation}
The geometric content of this relation is again visually represented in Fig.~\ref{fig:Asymptotically flat spacetime}.

The supermomentum charges associated with BMS supertranslations are defined in terms of the \emph{Bondi mass aspect} $M(u,\theta,\phi)$ as
\begin{equation}\label{eq:Supermomentum Charge MT}
    \mathcal{Q}_{(\alpha)}(u)
    =\frac{1}{4\pi G}\int_{S^2_u} d\Omega\,\alpha(\theta,\phi)\,M(u,\theta,\phi)\,,
\end{equation}
where the smooth function $\alpha(\theta,\phi)$ labels the infinite family of charges.
The corresponding supermomentum flux entering the BMS balance laws is given by
\begin{align}\label{BMSSupermomentumFlux MT}
    \mathcal{F}^\text{R}_{(\alpha)}(u)
    =\int_{u_0}^{u}du'\int_{S^2_{u'}} d\Omega\,
    \alpha(\theta,\phi)\bigl[F_\text{R}-F_{\rm sup}\bigr]\,,
\end{align}
where $F_\text{R}$ denotes the total null energy flux reaching future null infinity. This total energy flux decomposes into a gravitational and a matter contribution
\begin{equation}\label{eq:totalNullRadiation}
    F_\text{R}(u',\theta,\phi)=F_h+F_\text{m}\,,
\end{equation}
where $F_\text{m}$ represents the flux of any unbound null matter fields, hence electromagnetic radiation, while the gravitational radiation energy flux is
\begin{equation}\label{eq:GravitationaEF}
    F_h(u',\theta,\phi)
    =\frac{r^2}{32\pi G}\,\dot h^{TT}_{ij}\dot h_{TT}^{ij}\,,
\end{equation}
with the leading-order $r$ dependence canceling as expected at null infinity.
Moreover, there is an additional contribution depending solely on the radiative metric degrees of freedom 
\begin{equation}
    F_{\rm sup}(u',\theta,\phi)
    \equiv \frac{r}{16\pi G}\,\mathcal{D}^i\mathcal{D}^j \dot h^{TT}_{ij}\,,
\end{equation}
where $\mathcal{D}_i$ denotes the embedding of the covariant derivative on the 2-sphere. As discussed below, the presence of this term is a direct consequence of the enlargement of the asymptotic Poincar\'e group to include angle-dependent supertranslations.

We now briefly elaborate on the physical content of the BMS balance laws. For the interested reader, more information, including technical details, is gathered in Appendix~\ref{App:BMSFluxBalanceLaws}.
A particularly important case is obtained by choosing the supertranslation parameter $\alpha(\theta,\phi)=1$, for which the associated charge reduces to the \emph{Bondi mass}
\begin{equation}\label{eq:Supermomentum Charge E}
    m(u)\equiv \mathcal{Q}_{(1)}(u)
    =\frac{1}{4\pi G}\int_{S^2_u} d\Omega\,M(u,\theta,\phi)\,.
\end{equation}
Unlike the ADM mass $M_{\rm ADM}$~\cite{ADM:1959zz,Arnowitt:1962hi}, which is defined at spatial infinity and characterizes the total energy of an initial spacelike hypersurface, the Bondi mass is defined at future null infinity and measures the instantaneous energy of the isolated system on a hypersurface approaching constant retarded time $u$. 
The Bondi mass therefore explicitly depends on time $u$, encoding energy loss through outgoing radiation. Indeed, for $\alpha(\theta,\phi)=1$ the balance law [Eq.~\eqref{eq:BMSBalanceLawI MT}] simply reduces to a statement of conservation of total energy 
\begin{equation}\label{eq:BMSBalanceLawI 2s MT} 
\Delta m(u) =- \int_{u_0}^{u} d u'\int_{S^2_{u'}}d\Omega\,\left[F_h+F_\text{m}\right](u',\theta,\phi)\,, 
\end{equation}
corresponding to the famous Bondi mass loss formula.
In deriving Eq.~\eqref{eq:BMSBalanceLawI 2s MT}, we used the fact that the term $F_{\rm sup}$ in the supermomentum flux integrates to zero for $\alpha=1$ (see Appendix~\ref{App:BMSFluxBalanceLaws}), so that the right-hand side represents the total asymptotic energy flux. The same cancellation occurs for $\alpha(\theta,\phi)=Y_{1m}(\theta,\phi)$, corresponding to $\ell=1$ supertranslations, in which case the balance laws express conservation of linear momentum.

However, for supertranslations with angular structure beyond the first two harmonics, $\ell\geq 2$, the $F_\text{sup}$ term in Eq.~\eqref{BMSSupermomentumFlux MT} no longer vanishes. To understand the role of this additional contribution to the balance laws, it is instructive to isolate it and perform the trivial integration over retarded time. Moreover, since the vanishing of $F_{\rm sup}$ is no longer enforced by angular integration, we may now exchange the order of integration and obtain a local, angle-by-angle consistency relation between the asymptotic strain, the energy flux, and the Bondi mass aspect
\begin{align}\label{eq:BMSSupermomentumBalanceLaw each angle MT}
    \mathcal{D}^i\mathcal{D}^j\Delta h_{ij}^{TT}(u,\theta,\phi) =\,& \frac{16\pi G}{r}\int_{u_0}^{u}du'\, \left[F_h+F_\text{m}\right](u',\theta,\phi)\nonumber\\
    &-\frac{4}{r}\Delta M(u,\theta,\phi)\,.
\end{align}
This relation, which must hold pointwise on the asymptotic two-sphere, admits two complementary interpretations, which we refer to as the \emph{consistency relation} and the \emph{memory relation} of the BMS balance laws. For definiteness, and to simplify the discussion, we set $u_0\rightarrow -\infty$ in both cases.

The first viewpoint is provided directly by Eq.~\eqref{eq:BMSSupermomentumBalanceLaw each angle MT}, which may be interpreted as a consistency relation for the total asymptotic radiation $h^{TT}_{ij}(u)$ appearing on both sides of the equation. This interpretation becomes particularly transparent upon assuming
\begin{equation}
\lim_{u\rightarrow -\infty} h^{TT}_{ij}(u)=0 \,,
\end{equation}
in which case Eq.~\eqref{eq:BMSSupermomentumBalanceLaw each angle MT} manifestly represents a constraint on the total strain $h^{TT}_{ij}(u)$, relating it at each retarded time $u$ to the accumulated radiative energy flux and the Bondi mass aspect. In this form, the balance law acts as a nontrivial consistency relation that must be satisfied by any physically admissible radiation signal, and can therefore be used as a diagnostic for waveform models (see Appendix~\ref{sApp:BalanceLawAtEachAngle} and Refs.~\cite{Ashtekar:2019viz,Mitman:2020bjf,DAmbrosio:2024zok}).

The second interpretation, which is central to the present work, is obtained by taking the late-time limit $u\rightarrow +\infty$ and evaluating the net change 
\begin{equation}\label{eq:MemoryOffset2}
\Delta h^{TT}_{ij}=h^{TT}_{ij}(+\infty)-h^{TT}_{ij}(-\infty)\,,
\end{equation}
before and after a burst of radiation. In this limit, Eq.~\eqref{eq:BMSSupermomentumBalanceLaw each angle MT} becomes a memory relation in the sense that after inverting the angular derivatives on the left-hand side, it yields a direct expression for the permanent memory offset defined in Eq.~\eqref{eq:DefiningMemoryOffset} in terms of the total radiated energy flux and the change in the Bondi mass aspect. More precisely, the total memory amplitude associated with the \textit{zero-frequency} contribution to the asymptotic radiation is shown to only depend on the angular distribution of the total emission of radiation energy $\int_{-\infty}^\infty du' F_R$, as well as the total change $\Delta M$, independent of the details of the source. Gravitational memory thus emerges as a generic feature of gravitational radiation, expected to arise in a wide class of radiative processes. Put differently, gravitational memory is not an additional assumption or a model-dependent feature, but can be understood as a direct manifestation of the enlargement of the BMS symmetry to include supertranslations with modes $\ell \geq 2$.
In Appendix~\ref{App:BMSFluxBalanceLaws}, we explicitly show how the memory offset can be isolated from Eq.~\eqref{eq:BMSSupermomentumBalanceLaw each angle MT} and understand that the angular derivatives select out the electric-parity component of the strain.

The three terms on the right-hand side of the memory relation derived from Eq.~\eqref{eq:BMSSupermomentumBalanceLaw each angle MT} are commonly interpreted as distinct ``sources'' of gravitational memory. While the integral of $F_\text{R}(u',\theta,\phi)=F_h+F_\text{m}$ naturally correspond to the total radiative emission of null energy, a change in the Bondi mass aspect, $\Delta M(u,\theta,\phi)$, is both directly related to the presence of gravitationally unbound massive matter~\cite{Zeldovich:1974gvh,Turner:1977gvh,Braginsky:1985vlg,Braginsky:1987gvh,Epstein:1978gvh,Burrows:1995bb,Ott:2008wt,Murphy:2009,Sago:2004pn} and final remnant kick-velocities~\cite{Merritt:2004xa,Gonzalez:2006md,Favata:2008ti,Komossa:2012cy,Borchers:2022pah,DAmbrosio:2024zok}.
Traditionally, these contributions are given specific names. The memory due to the total emission of gravitational energy captured by the integral of $F_h$, is termed \emph{nonlinear memory}, reflecting the fact that the charge of gravity can be carried by the gravitational field itself. In contrast, the memory arising from radiative matter $F_\text{m}$ and the change in mass aspect $\Delta M$, is referred to as \emph{linear memory}.
Furthermore, the gravitational memory associated to the integral of the combined energy-flux of asymptotic null radiation $F_\text{R}$, is called \emph{null memory}, whereas the memory arising purely from $\Delta M$, which formally does not reach null infinity at leading $1/r$ order, is called \emph{ordinary memory}. Note that in the absence of unbound massless matter, null memory coincides with nonlinear memory and ordinary memory coincides with linear memory.

This classification reflects the viewpoint of the memory relation of the BMS balance laws. However, even when the balance laws are interpreted as a consistency relation for the total instantaneous strain at each $u$, it is customary to distinguish a \emph{null part} and an \emph{ordinary part} of the asymptotic waveform. We will adopt this nomenclature throughout the following discussion.

\subsubsection{Necessity to go beyond BMS balance laws}\label{ssSec:NecessityBeyondBMS}

In summary, BMS balance laws provide a consistency relation [Eq.~\eqref{eq:BMSSupermomentumBalanceLaw each angle MT}] that any radiation field at $\scri^+$ must satisfy. Most notably, in the infinite-time limit this relation generically implies the presence of a non-zero gravitational memory effect associated with any burst of gravitational radiation [Eq.~\eqref{eq:MemoryOffset2}]. By construction, these balance laws constrain the total DC memory offset, corresponding to the zero-frequency content of the signal.

However, as emphasized in the Introduction, interferometric detectors are intrinsically insensitive to this zero-frequency memory offset itself and can only probe the time-dependent transition between the initial and final states. Since detector sensitivity is limited to a finite frequency band, any measurable signal corresponds to a finite-frequency representation of the underlying memory effect. In other words, while gravitational memory is formally defined by its zero-frequency (soft) limit, its observational signature arises from the time-dependent evolution of the radiation within the detector band. 
A claim of detection therefore requires a definition of a time-dependent signal whose statistical significance directly tracks the magnitude of the underlying memory offset. This requirement generically excludes oscillatory features from a memory model: such components increase the signal-to-noise ratio while not contributing to the defining zero-frequency memory offset.\footnote{This issue is particularly relevant for nonlinear memory, since the same events that generate memory also emit oscillatory gravitational waves. As a result, both contributions are generically present in the data, making an unambiguous identification of the memory component nontrivial. This contrasts with pure linear memory, where the unbound energy-momentum sourcing the memory is not itself part of the gravitational radiation.}

Yet, while the balance laws determine the final memory offset, they do not uniquely specify the detailed time dependence of the transition between the initial and final states of the radiation field. This can also be understood from the fact that a BMS supertranslation captures only the constant and linearly growing components relating two distinct memory vacua, as shown explicitly in~\cite{DeLuca:2024cjl}. More generally, the time-dependent balance laws in Eq.~\eqref{eq:BMSSupermomentumBalanceLaw each angle MT} impose a constraint on the full asymptotic waveform, which generically contains both oscillatory gravitational waves and memory contributions.

These oscillatory contributions arise from both the ordinary and null sectors of the radiation. While for quasi-circular binary black hole mergers most oscillatory radiation resides in the ordinary sector, bursts of gravitational radiation generically also exhibit oscillatory features in the null sector of the time-dependent waveform~\cite{Mitman:2020bjf}.\footnote{The origin of these oscillatory features can be understood from the fact that the energy flux entering the balance laws provides only a notion of the \emph{total} radiated energy, $\int_{-\infty}^{\infty} du'\,F_h$, and does not define a localized energy density on scales shorter than the gravitational wavelength~\cite{Isaacson_PhysRev.166.1272,misner_gravitation_1973,maggiore2008gravitational,Flanagan:2005yc}. As a result, the partial time integral $\int_{-\infty}^{u} du'\,F_h$ need not evolve monotonically in retarded time and can contain oscillatory features.} Even when such oscillations are small in amplitude, they can influence the signal-to-noise ratio attributed to a given memory model, as we show explicitly in Sec.~\ref{ssSec:MemoryInLISA}.

A meaningful definition of the observable \emph{memory rise} therefore requires additional physical input beyond the BMS balance laws themselves. In the next section [Sec.~\ref{sSec:DefMemModel}], we therefore introduce a framework for defining a time-dependent gravitational memory signal that isolates the low-frequency memory rise from oscillatory gravitational waves while remaining directly connected to the underlying zero-frequency memory offset. From a modeling perspective, this approach yields a description applicable to a broad class of gravitational-wave sources. Finally, in Sec.~\ref{sSec:Practical Approximation}, we discuss practical approximations for nonlinear memory in quasi-circular binary black hole mergers and clarify in what sense BMS balance laws continue to inform such models.


\subsection{Defining a general memory signal model}\label{sSec:DefMemModel}

As anticipated, in this section, we address the problem of defining a unique time-dependent memory rise in an event of emission of gravitational waves from a localized source.

\subsubsection{An idealized but fundamental separation of scales}\label{ssSec:FundamentalMemoryModel}

To work toward a first-principles answer to this question, we begin with a thought experiment and consider an idealized emission of transient gravitational radiation from a simple localized source. We assume that this source produces gravitational waves with a clearly identifiable characteristic frequency scale $f_H$ over a finite duration $T_L$. This characteristic frequency is associated with the periodic variation of the source's quadrupole and higher multipole moments.
In this setting, gravitational displacement memory can be defined through the existence of a fundamental separation of scales between high ($H$) and low ($L$) frequencies, $f_H \gg f_L \sim 1/T_L$, separating the emitted oscillatory gravitational waves, also referred to as \emph{primary waves}, from a corresponding memory signal associated with a change in the background structure of the asymptotic spacetime. This separation of scales is practically implemented at the level of the asymptotic radiation through a spacetime averaging procedure
\cite{Isaacson_PhysRev.166.1263,Isaacson_PhysRev.166.1272,misner_gravitation_1973,Flanagan:2005yc,maggiore2008gravitational,Zalaletdinov:2004wd,Stein:2010pn,Favata:2011qi,Heisenberg:2023prj,Zosso:2024xgy,Zosso:2025ffy},
which we denote by $\langle\cdots\rangle$.
Choosing an averaging kernel adapted to the high-frequency scale $f_H$, the gravitational displacement memory of any leading-order asymptotic radiation $h_{ij}^{TT}$ is then encoded in its low-frequency ($f_L$) background component, defined schematically as
\begin{equation}\label{eq:MemLfromAverage}
    h_{ijL}^{TT} \equiv \langle h_{ij}^{TT} \rangle \, .
\end{equation}
By subtraction, this averaging procedure also identifies the corresponding high-frequency ($f_H$) gravitational-wave signal,
\begin{equation}\label{eq:MemHfromAverage}
    h_{ijH}^{TT} \equiv h_{ij}^{TT} - h_{ijL}^{TT} \, .
\end{equation}
In this thought experiment, Eqs.~\eqref{eq:MemLfromAverage} and \eqref{eq:MemHfromAverage} provide a first-principles separation of a leading order radiation into a gravitational displacement memory signal and a pure gravitational wave. While a definition of memory through a separation of frequency scales is obvious at the level of the soft limit characterizing the final memory offset, as we will now understand, this separation remains valid for the time-dependent interpolation of the memory.

The introduction of a fundamental separation of scales, together with an associated averaging operation, is not an ad-hoc trick for defining a memory model. Rather, it reflects a deeply rooted conceptual necessity in gravitational perturbation theory. In fact, these concepts were first described long before the theoretical discovery of the gravitational memory effect in an effort to provide an elementary definition of gravitational waves. The completion of this program is attributed to the work of Refs.~\cite{Isaacson_PhysRev.166.1263,Isaacson_PhysRev.166.1272} (see Refs.~\cite{misner_gravitation_1973,Flanagan:2005yc,maggiore2008gravitational} for reviews) and is known as the \emph{Isaacson approach} to gravitational waves.
Within this framework, it is emphasized that a physical definition of gravitational waves requires the existence of a clear separation between the scales of variation of the spacetime metric: a background spacetime at scales ($L$), defined through a spacetime average $\langle\cdots\rangle$, and high-frequency perturbations at characteristic scales ($H$), which describe the gravitational waves themselves. Crucially, the localized energy-momentum carried by gravitational waves is inherently only well-defined up to a characteristic coarse graining over the relevant scales of variation
\cite{Isaacson_PhysRev.166.1263,Isaacson_PhysRev.166.1272,maggiore2008gravitational}.\footnote{This limitation on energy localization can be understood both classically, as a basic property of the Fourier transform, and from a quantum-mechanical perspective through the Heisenberg uncertainty principle.}
For the interested reader, we provide additional details on the Isaacson approach, as well as explicit derivations of all equations appearing in this section, in Appendix~\ref{App:IssacsonMemory}.

It was recently noted in Ref.~\cite{Heisenberg:2023prj} that the solution to the leading order Isaacson equations of motion in the context of gravitational radiation from a localized source coincides in form with the general formula of gravitational displacement memory. Specifically, the assumed separation of scales within the Isaacson framework yields two sets of leading-order equations. The first is valid at high-frequency scales and represents a propagation equation for gravitational waves on top of an arbitrary low-frequency background spacetime. The second equation, on the other hand, can be viewed as a back-reaction equation of the coarse-grained energy-momentum carried by the gravitational waves onto the background spacetime. Indeed, the leading order low-frequency Isaacson equation can be viewed as the very definition of a well-behaved energy momentum tensor of gravitational waves [Eq.~\eqref{eq:DefPseudoEMTensorGR}], which can be shown to be generally gauge-invariant and conserved~\cite{Isaacson_PhysRev.166.1263,Isaacson_PhysRev.166.1272,maggiore2008gravitational}. In the asymptotic Minkowski limit, this Isaacson energy-momentum tensor is entirely characterized through the emitted energy flux, which can be written as~\cite{maggiore2008gravitational,Garfinkle:2022dnm}
\begin{align}\label{eq:EnergyFluxH}
    t_{ij}&=\frac{1}{r^2}\frac{dE}{dud\Omega}n_in_j=\frac{1}{32\pi G}\Big\langle \dot h_{ijH}^{TT} \dot h_{ijH}^{TT}\Big\rangle  \,n_in_j\nonumber\\
    &=\frac{1}{16\pi G}\Big\langle \dot h^2_{H+} +\dot h^2_{H\times}\Big\rangle \,n_in_j\,,
\end{align}
with $n_i\equiv n_i(\Omega)$ the unit radial direction defined in Eq.~\eqref{eq:Def Direction of Propagation n}. 

As we explicitly show in Appendix~\ref{sApp:Solving IsaacsonEquation}, the result of the back-reaction of the anisotropic emission of such energy content in gravitational waves $h_{ijH}^{TT}$, is the advent of an additional gauge-invariant low-frequency contribution within the asymptotic radiation $h_{ijL}^{TT}$, which reads [Eq.~\eqref{eq:memoryequationA}] 
\begin{equation}\label{eq:memoryequation}
    h_{ijL}^{TT}(u,r,\Omega)=\frac{4G}{r}\int_{-\infty}^u du'\int d\Omega' \frac{dE}{du'd\Omega'}\left[\frac{n'_in'_j}{1-\vec{n}'\cdot \vec n}\right]^{TT}\,,
\end{equation}
where the superscript $TT$ denotes a projection onto the physical $TT$ modes along the relevant radial direction $n_i(\Omega)$ defined in Eq.~\eqref{eq:Projectors}.

\subsubsection{Characteristics of the Isaacson memory formula}\label{sSec:CharacteristicsofIsaacsonMemory}
As promised, the solution in Eq.~\eqref{eq:memoryequation} indeed represents a displacement memory signal since the structure of the formula describes a non-zero final memory offset \begin{equation}
    \Delta h_{ij}^{TT}=\lim_{u\rightarrow\infty}h_{ijL}^{TT}(u)-h_{ijL}^{TT}(-u)\neq 0\,.
\end{equation}
Moreover, the contribution is a \emph{hereditary effect}, depending on the entire past history of the source. 
It is also worth noting that despite being sourced by the emitted energy-momentum, the memory component still scales as $1/r$ and is part of the radiation reaching null infinity. In terms of the asymptotic perturbative expansion described in Sec.~\ref{sSec:Radiation}, the nonlinear memory in Eq.~\eqref{eq:memoryequation} can therefore be understood as a leading order low-frequency effect.\footnote{This is in contrast to a particle scattering picture in which nonlinear memory is naturally described as a second order effect suppressed by an additional Planck mass factor. In the realm of radiation emission at large classical scales, this power-counting scheme therefore no longer holds [see Appendix~\ref{sApp:leadingOrdereqs in flat space} for more details].}
Up to subtle but important differences that we will discuss below, Eq.~\eqref{eq:memoryequation} coincides in its form with earlier versions of the nonlinear memory formula in the literature
\cite{Christodoulou:1991cr,PhysRevD.44.R2945,Thorne:1992sdb,Favata:2008yd,Favata:2009ii,Favata:2010zu,Garfinkle:2022dnm}. 

The most important distinction is the explicit averaging over oscillatory frequency scales appearing in Eq.~\eqref{eq:memoryequation}.\footnote{The usefulness of such an averaging was already noted in Ref.~\cite{Favata:2011qi}. In contrast to the ad hoc introduction adopted there, the present Isaacson-based treatment identifies the associated separation of scales as a fundamental ingredient in the derivation of the memory equation and the definition of its model. More recently, Refs.~\cite{Heisenberg:2023prj,Inchauspe:2024ibs,Zosso:2025ffy} of some of the authors have also employed the Isaacson definition of memory presented here.} 
Within the thought experiment introduced above, this implies that the emitted energy-momentum carried by primary waves at frequency scale $f_H$ sources a smooth memory buildup over the emission timescale $T_L$, corresponding to a characteristic low frequency scale $f_L\sim 1/T_L$. Assuming a sufficiently long emission, $T_L\gg 1/f_H$,\footnote{The emission must nevertheless remain short enough to ensure perturbative amplitudes. Otherwise, the memory contribution must be treated as part of the background spacetime~\cite{maggiore2008gravitational}.} the resulting low-frequency signal is physically distinct from the oscillatory gravitational waves and can be interpreted as part of the background scales of spacetime~\cite{Heisenberg:2023prj,Zosso:2025ffy}. 

Embedding Eq.~\eqref{eq:memoryequation} within the Isaacson framework offers important conceptual advantages, as detailed in Appendix~\ref{App:IssacsonMemory}. In particular, the equation is no longer viewed as a solution to an ad hoc subset of a second-order Landau-Lifshitz equation~\cite{PhysRevD.44.R2945,Favata:2010zu}, but instead as the unique leading-order low-frequency contribution in the asymptotically flat limit of the Einstein equations, with higher-order corrections systematically accessible. Consequently, the energy-momentum flux of gravitational waves is understood as the only nonlinear operator contributing at leading order. Moreover, unlike the BMS balance laws, the Isaacson approach does not rely on the rigid structure of exact asymptotic flatness.\footnote{In realistic cosmological settings, BMS symmetries can only be understood as approximate symmetries on localized patches where cosmic expansion is negligible.} 
Instead, as shown explicitly in Appendix~\ref{App:DerivationIsaacsonLow}, gravitational memory emerges directly from the sourced wave equation, with the source given by a general notion of coarse-grained, gauge-invariant energy-momentum~\cite{Isaacson_PhysRev.166.1263,Isaacson_PhysRev.166.1272,maggiore2008gravitational}.

More generally, the functional form of Eq.~\eqref{eq:memoryequation} characterizes gravitational memory in any local Lorentz-invariant theory as the universal solution to a sourced asymptotic wave equation~\cite{Heisenberg:2024cjk,Heisenberg:2023prj,Garfinkle:2022dnm,Zosso:2025ffy}, independent of the microscopic origin of the unbound energy-momentum.\footnote{For instance, replacing the gravitational-wave energy flux in Eq.~\eqref{eq:memoryequation} with the unbound energy-momentum of massive particles reproduces the standard expression for linear memory~\cite{PhysRevD.44.R2945,Thorne:1992sdb}.} 
This is a manifestation of the universality property of gravitational memory already commented on in Sec.~\ref{ssSec:ProofofMemory}: the effect is independent of the details of the local interaction and depends only on the angular and temporal distribution of outgoing (and incoming) unbound energy--momentum. 
In this work, we focus on the observationally most relevant case of gravitational memory sourced by gravitational-wave emission. In the nomenclature established in Sec.~\ref{ssSec:ProofofMemory} this corresponds to nonlinear memory, which is also part of the null memory.

At leading order, Eq.~\eqref{eq:MemLfromAverage} and Eq.~\eqref{eq:memoryequation} are formally equivalent. In practice, however, it is preferable to work with the direct memory formula Eq.~\eqref{eq:memoryequation}. This expression captures the conceptual origin of gravitational memory, closely tied to the universal soft structure of radiation theory. 
Moreover, the specification of the characteristic emission scale $f_H$ constitutes the essential physical input for defining a time-dependent memory model for both Eq.~\eqref{eq:MemLfromAverage} and Eq.~\eqref{eq:memoryequation}. In the context of binary black hole mergers, this is, however, a very natural starting-point, as we will discuss in Sec.~\ref{ssSec:Practical Approximation}. Furthermore, once a suitable $f_H$ scales is identified, Eq.~\eqref{eq:memoryequation} is robust against ambiguities in defining a high-frequency waveform model $h_H$ and provides a good model of gravitational memory, even when inserting the full asymptotic radiation.\footnote{This can readily be understood in terms of the bi-variate perturbation expansion of the Isaacson approach: cross-terms between high- and low-frequency fields average to zero, while the contribution of low-frequency fields to the right-hand side of Eq.~\eqref{eq:memoryequation}, known as the \emph{memory of the memory}, is parametrically suppressed by powers of $f_L/f_H$, even if the corresponding amplitudes are comparable. The Isaacson framework thus provides a systematic perturbative expansion [see Appendix~\ref{App:IssacsonMemory} for more details.]}

Finally, we emphasize that Eq.~\eqref{eq:memoryequation} is derived under the assumption that the energy flux is \emph{emitted} from a localized source (see Appendix~\ref{App:IssacsonMemory}). In this sense, gravitational memory is not directly associated with the propagation, but rather the acceleration of unbound energy-momentum and can thus be viewed as a gravitational analogue of internal bremsstrahlung~\cite{Jackson:1998nia,Zosso:2025orw}. Nevertheless, by virtue of the universal infrared structure of field theories, the effect depends only on the initial and final angular distributions of unbound energy-momentum, as encoded in Eq.~\eqref{eq:memoryequation}.

\subsubsection{Generalization to realistic sources}\label{ssSec:RealisticSources}

Although the identification of a displacement memory signal has so far been discussed in an idealized setting, the central claim is that, at leading order, Eq.~\eqref{eq:memoryequation} continues to provide a clearly distinguishable memory signal in more complex and realistic scenarios.
Naturally, as the source dynamics become more involved, the idealized separation of scales assumed in Sec.~\ref{ssSec:FundamentalMemoryModel} will eventually break down. For sufficiently complex sources with multiple or rapidly varying emission frequencies, a single, sharply defined characteristic frequency scale $f_H$ may no longer exist. Nevertheless, a generic feature of gravity is that gravitational waves are emitted phase-coherently by the bulk motion of their sources. As a result, in many realistic situations one can still associate, at each instant, a rather narrow band of frequencies to the dominant gravitational waves. Even if this scale evolves in time, the idealized analysis presented above remains applicable over successive time intervals, provided that the evolution of $f_H$ is sufficiently slow. In practice, this amounts to partitioning the emission into several time segments of duration $T_L$, within which the analysis holds for a quasi-stationary $f_H$. Combined with an appropriate averaging at each retarded time $u$, Eq.~\eqref{eq:memoryequation} thus provides the general nonlinear memory model proposed in this work.

From this perspective, it becomes clear that the Isaacson assumptions are notably pushed to their limits during the merger of a compact binary coalescence. While the inspiral phase of a binary system constitutes a textbook realization of the idealized scenario discussed in Sec.~\ref{ssSec:FundamentalMemoryModel}, the energy flux increases rapidly during merger, as illustrated, for example, in Fig.~\ref{fig:WaveformTypes}. A reliable definition of gravitational memory in this regime is nevertheless crucial, since binary black hole mergers are among the prime target sources for the first single-event detections of gravitational memory. A key result of this work is to demonstrate that even through the merger of typical binary black hole coalescences, the evolution of the high-frequency scales associated with the gravitational waves emitted by the movement of the binary source, remains sufficiently well-behaved to apply the Isaacson definition of an associated memory signal. These results are presented in Sec.~\ref{sSec:Practical Approximation} below. Consequently, Eq.~\eqref{eq:memoryequation} provides a well-motivated gravitational memory waveform model for searches in current and future black hole merger data. As a preparation for the specific discussion of binary black hole mergers, we conclude this section by expanding the general memory formula \eqref{eq:memoryequation} in a spin-weighted spherical harmonic basis.

\subsubsection{Expansion in spin-weighted spherical harmonics}\label{ssSec:SWSHExpansionMemoryModel}

Given a gravitational radiation signal defined in Eq.~\eqref{eq:gravitationalradiationGen},
it is convenient to introduce a complex scalar of definite spin-weight $s=-2$,
\begin{equation}\label{eq:SpinweightedScalar}
    h(u,r,\Omega)\equiv  h^{TT}_{ij} \bar{m}^i\bar{m}^j 
    =h_+-i h_\times\,,
\end{equation}
where the complex transverse basis vector of definite spin-weight $s=-1$ is defined as
\begin{equation}
    \bar m_i\equiv \frac{1}{\sqrt{2}}(\theta_i-i\phi_i)\,.
\end{equation}

For completeness, we briefly recall the notion of spin-weight (see also Refs.~\cite{Goldberg:1966uu,Newman:1966ub,DAmbrosio:2022clk}).
A complex function on the sphere with fixed direction $\bm{n}(\Omega)$ may still acquire a phase under rotations about $\bm{n}(\Omega)$. Functions of definite spin-weight $s$ are defined by their transformation under such $U(1)$ rotations
\begin{equation}\label{eq:SpinWeightTransfos}
    f_s(\theta,\phi)\rightarrow f_s(\theta,\phi)\,e^{is\psi}\,,
\end{equation}
where $\psi$ is the rotation angle. The spin-weight $s=-2$ of the scalar field in Eq.~\eqref{eq:SpinweightedScalar} therefore characterizes its transformation under rotations about the line of sight and is essential for a harmonic decomposition on the sphere. Accordingly, this radiation field can be expanded in spin-weighted spherical harmonics (SWSHs) [see Appendix~\ref{App:SWSH}] as
\begin{equation}\label{eq:MemoryPhasor}
    h (u,\theta,\phi)=\sum_{\ell m}  h_{\ell m}(u)\,_{\mys{-2}}Y_{\ell m}(\theta,\phi)\,.
\end{equation}

Applying this expansion to the low-frequency memory component in Eq.~\eqref{eq:memoryequation} yields an evolution equation for the individual SWSH modes of the memory signal. The explicit derivation is given in Appendix~\ref{App:SWSHExpansionMemory}, Eq.~\eqref{eq:NonLinDispMemoryModesGRApp} (see also Ref.~\cite{Heisenberg:2023prj}), with the result
\begin{equation}\label{eq:NonLinDispMemoryModesGR}
  h^L_{\ell m}(u)=\frac{1}{r}\sqrt{\frac{(\ell-2)!}{(\ell+2)!}}
  \int_{-\infty}^u du'\int_{S^2}d^2\Omega'\,\bar Y_{\ell m}\,
  r'^2\big\langle |\dot{h}_{H}|^2\big\rangle\,.
\end{equation}
Up to the explicit spacetime averaging, Eq.~\eqref{eq:NonLinDispMemoryModesGR} is equivalent in its form to the null part of the BMS balance laws in Eq.~\eqref{eq:BMSSupermomentumBalanceLaw each angle MT} (see Appendix~\ref{App:BMSFluxBalanceLaws}).

The angular integral in Eq.~\eqref{eq:NonLinDispMemoryModesGR} can be evaluated analytically by expanding the primary waves themselves in SWSHs, resulting in the final formula [Eq.~\eqref{eq:memory-pre-3j}; see also Refs.~\cite{Favata:2008yd,Zosso:2024xgy,Inchauspe:2024ibs}\footnote{This formula is for instance also equivalent to the implementation in the \texttt{GWMemory} package~\cite{Talbot:2018sgr}.}]
\begin{widetext}
\begin{align}
h_{\ell m}^{L}(u)&=\frac{1}{r}\sum_{\ell',\ell''\geq 2}\, \sum_{m',m''}
\Gamma^{\ell'm'm''\ell''}_{\ell m}
\int_{-\infty}^u du'\,r'^2
\langle\dot{h}^H_{\ell' m'} \dot{\bar h}^{H}_{\ell'' m''}\rangle\,, \label{eq:memorymodes}\\
\Gamma^{\ell'm'm''\ell''}_{\ell m}&\equiv (-1)^{m+m''}
\sqrt{\frac{(\ell-2)!}{(\ell+2)!}}
\sqrt{\frac{(2\ell'+1)(2\ell''+1)(2\ell+1)}{4\pi}}
\begin{pmatrix}
    \ell' & \ell'' & \ell\\
     m' & -m'' & -m
\end{pmatrix}
\begin{pmatrix}
    \ell' & \ell'' & \ell\\
    2 & -2 & 0
\end{pmatrix}.
\end{align}
\end{widetext}
Here, the big parentheses represent $3-j$ symbols, which give a nonzero value if 
\begin{equation}\label{eq:SelectionRule M}
    m=m'-m'' \quad \text{and} \quad |\ell'-\ell''|\leq \ell\leq \ell'+\ell''\,.
\end{equation}

\subsection{Practical approximations of nonlinear memory from BBH mergers}\label{sSec:Practical Approximation}

We will now turn our attention to the specific but observationally most important case of nonlinear memory from binary black hole mergers. In doing so, we want to provide a deeper understanding of popular nonlinear memory models used in the literature and provide an overview of their justification and limitations. A special focus will be given on the role of the Isaacson averaging introduced in the previous section~\ref{sSec:DefMemModel}, representing a main ingredient in identifying the general nonlinear memory model in Eq.~\eqref{eq:memoryequation} and its spin-weight expanded counterpart in Eq.~\eqref{eq:memorymodes}.

For arbitrary binary systems, one generally needs to resort to the full memory formula in Eq.~\eqref{eq:memorymodes}, since \textit{a priori} gravitational memory is not restricted to a specific spin-weighted mode. Moreover, in principle, also the linear memory from the remnant kick needs to be considered, although it is  generally found to be subdominant~\cite{Favata:2008ti,DAmbrosio:2024zok}. In this context, especially in the case of precessing or eccentric binaries~\cite{Favata:2011qi}, the spacetime averaging over oscillatory high-frequency scales is crucial to ensure the definition of a well-behaved memory model.

The most promising sources for a first detection of gravitational memory with LISA are, however, given by massive binary black hole systems close to equal mass~\cite{Inchauspe:2024ibs} which are expected to generally exhibit low precession and eccentricity~\cite{Porter:2010mb,ColemanMiller:2013jrk,Barausse_2020,Barausse_Lapi_2021}. Such quasi-circular binary systems allow for a simplified practical treatment of the memory model, as we will now elaborate. In Sec.~\ref{ssSec:meory is in the 20 mode} we identify the well-known corresponding dominant spin-weighted gravitational memory modes. In doing so, we will highlight the fact that, even in the quasi-circular case, the identification of isolated memory modes crucially relies on the Isaacson averaging. As a consequence, an averaging assumption, although often not explicit, was already used in most of the previously proposed gravitational memory models of CBCs. Based on this result, Sec.~\ref{ssSec:Practical Approximation} discusses the practical implementation of computing displacement memory in the case of BBH mergers. In particular, we will address the question of how numerical waveform models in the past decades were able to accurately compute an entire range of high-frequency gravitational wave modes while completely missing out on gravitational memory, which from an amplitude perspective represents one of the most important contributions next to the leading 
high-frequency mode. Section~\ref{ssSec:PropertiesMemoryModel} then gathers all relevant properties of the gravitational memory model of quasi-circular binary black hole mergers.

\subsubsection{The dominant $m=0$ memory modes}\label{ssSec:meory is in the 20 mode}

As anticipated, in practice, one does not have to compute all the modes within Eq.~\eqref{eq:memorymodes} to obtain an accurate memory signal model. In fact, in the case of very low precession and eccentricity BBH merger events, there is a single spin-weighted spherical harmonic of displacement memory that clearly dominates over all the others, namely the $(2,0)$ mode~\cite{Favata:2008yd,Favata:2009ii}. 
To gain a heuristic understanding of this statement consider the following: The leading order terms of the primary high-frequency modes $h^H_{\ell' m'}$ from a non-precessing CBC are proportional to~\cite{Boyle:2014ioa,CalderonBustillo:2015lrg,Varma:2018mmi,Barkett:2019tus}
\begin{equation}
    h^H_{\ell' m'}(u')\propto e^{-im'\phi(u')}\,,
\end{equation}
where $\phi(u')$ is the leading orbital phase. It is important to note, that this formula assumes a coordinate system in which the binary lies in the $x-y$ plane, such that an advancement in the orbital phase coincides with the polar angle of the associated spherical coordinate system. Although this approximation is strictly speaking only valid in the quasi-circular inspiral phase, it remains a good approximation throughout merger as well.

This condition, together with the selection rule in Eq.~\eqref{eq:SelectionRule M}, then, implies that the leading-order memory modes are proportional to
\begin{equation}\label{eq:phaseBehaviorM}
    h^L_{\ell m}\propto \big\langle e^{-i(m'-m'')\phi(u')}\big\rangle= \big\langle e^{-im\phi(u')}\big\rangle\,.
\end{equation}
Due to the averaging over orbital timescales, all modes with $m\neq 0$ will, therefore, be heavily suppressed and dominant memory modes will only be found at $m=0$.


Furthermore, also among the $m=0$ modes, there exists a clearly dominant contribution. First of all, the mirror symmetry across the $x-y$ plane present in non-precessing CBCs dictate that the high-frequency modes satisfy [see Appendix~\ref{App:DerivationReflectionIdentity} for a derivation]
\begin{equation}\label{eq:SymmOrbitPlaneModes}
        h^H_{\ell'm'}=(-1)^{\ell'}\,\bar h^{H}_{\ell'-m'}\,.
\end{equation}
As a consequence, any $(\ell,0)$ mode with odd $\ell$ will vanish identically~\cite{Favata:2009ii}. Moreover, since the dominant high-frequency harmonic of a quasi-circular non-precessing CBCs is the $(2,2)$ mode, the memory modes at $\ell\geq 4$ will be numerically suppressed, and we are left with the statement that the dominant gravitational memory mode is the $(2,0)$ mode. These effects can be seen in Fig.~\ref{fig:MultiModeMemory} of Appendix~\ref{App:SWSHExpansionMemory}. 

We emphasize that this statement relies crucially on the averaging over high-frequency scales within the emitted primary energy flux in Eq.~\eqref{eq:memorymodes}. Without such an averaging procedure, no clean mode selection occurs and Eq.~\eqref{eq:memorymodes} does not yield a purely monotonic memory contribution to the asymptotic radiation. Instead, even for quasi-circular compact binary coalescences with higher-harmonic waveform content, evaluating Eq.~\eqref{eq:memorymodes} without averaging generically produces oscillatory features (see Fig.~\ref{fig:MultiModeMemory}). As shown in Sec.~\ref{ssSec:MemoryInLISA}, these can have a non-negligible impact on SNR computations, thereby obstructing a clean identification of a memory signal associated with a monotonic offset.

An exclusive consideration of the $(2,0)$ mode for memory from non-precessing BBH mergers is well known and was in particular already argued for in Refs.~\cite{Favata:2008yd,Favata:2009ii}, although without invoking the precise notion of an Isaacson split in defining an averaging requirement. 
Yet, an averaging over oscillatory scales of variation appears to provide the only physically motivated input that can conservatively but robustly isolate a signal uniquely associated with the permanent memory offset within the total radiation, without introducing residual ambiguity at the level of detection.
In this sense, all models of nonlinear memory based on the $(2,0)$ mode are implicitly grounded in the general memory formula in Eq.~\eqref{eq:memoryequation}, including the Isaacson spacetime averaging.

Explicitly, in terms of the lowest order primary harmonics up to $\ell\leq 4$ that are taken into account in most up-to-date waveform models, the expression for the $(2,0)$ low-frequency memory mode in Eq.~\eqref{eq:memorymodes} is given by 
\begin{widetext}
\begin{align}
  h_{20}^{L} =\frac{1}{r}\int_{-\infty}^udu'&r'^2\Bigg\langle\frac{1}{7}\sqrt{\frac{5}{6\pi}}|\dot h^H_{22}|^2-\frac{1}{14}\sqrt{\frac{5}{6\pi}}|\dot h^H_{21}|^2-\frac{1}{14}\sqrt{\frac{5}{6\pi}}|\dot h^H_{20}|^2+\frac{5}{4\sqrt{42\pi}}\left(\dot h^{H}_{22}\dot{\bar h}^{H}_{32}+\dot{\bar h}^{H}_{22}\dot h^{H}_{32}\right)\nonumber\\
  &+\frac{5}{28\sqrt{6\pi}}\left(\dot h^{H}_{22}\dot{\bar h}^{H}_{42}+\dot{\bar h}^{H}_{22}\dot h^{H}_{42}\right)+\frac{1}{8}\sqrt{\frac{5}{21\pi}}\left(\dot h^{H}_{21}\dot{\bar h}^{H}_{31}+\dot{\bar h}^{H}_{21}\dot h^{H}_{31}\right)+\frac{5}{28\sqrt{3\pi}}\left(\dot h^{H}_{21}\dot{\bar h}^{H}_{41}+\dot{\bar h}^{H}_{21}\dot h^{H}_{41}\right)\nonumber\\
  &+\frac{1}{28}\sqrt{\frac{5}{2\pi}}(\dot h^H_{20}\dot{\bar h}^{H}_{40}+\dot{\bar h}^{H}_{20}\dot h^{H}_{40})+\frac{1}{2\sqrt{10\pi}}\left(\dot h^{H}_{33}\dot{\bar h}^{H}_{43}+\dot{\bar h}^{H}_{33}\dot h^{H}_{43}\right)+\sqrt{\frac{2}{105\pi}}\left(\dot h^{H}_{32}\dot{\bar h}^{H}_{42}+\dot{\bar h}^{H}_{32}\dot h^{H}_{42}\right)\nonumber\\
   &+\frac{1}{2\sqrt{42\pi}}\left(\dot h^{H}_{31}\dot{\bar h}^{H}_{41}+\dot{\bar h}^{H}_{31}\dot h^{H}_{41}\right)-\frac{2}{11}\sqrt{\frac{2}{15\pi}}|\dot h^H_{44}|^2-\frac{1}{11\sqrt{30\pi}}|\dot h^H_{43}|^2+\frac{1}{77}\sqrt{\frac{10}{3\pi}}|\dot h^H_{40}|^2\nonumber\\
   &-\frac{17}{77\sqrt{30\pi}}|\dot h^H_{41}|^2+\frac{4}{77}\sqrt{\frac{2}{15\pi}}|\dot h^H_{42}|^2\Bigg\rangle\,,
\label{eq:memorymodes20}
\end{align}
\end{widetext}
where we have used Eq.~\eqref{eq:SymmOrbitPlaneModes} to write all modes in terms of positive $m'$. For completeness we have also included the terms sourced by the high-frequency $h^H_{20}$ and $h^H_{40}$ modes, although their contribution will clearly be negligible.\footnote{However, as we will further discuss below, the (2,0) mode in general also contains a high-frequency contribution associated to the ringdown.}
Note that, as already mentioned, in more general situations of precessing and eccentric binaries, the arguments above break down and one has to rely on the more general memory formula given in Eq.~\eqref{eq:memorymodes}. Moreover, the restriction of the displacement memory to the $m = 0$ modes is in fact also a particular consequence of choosing the particular coordinate system, in which the $x-y$ plane coincides with the plane defined by the binary system~\cite{Favata:2008yd}. Another choice of coordinates would lead to mode-mixing and memory contributions with $m\neq 0$.

\subsubsection{The practical computation of memory models}\label{ssSec:Practical Approximation}

\begin{figure}
    \centering
    \includegraphics[width=\linewidth]{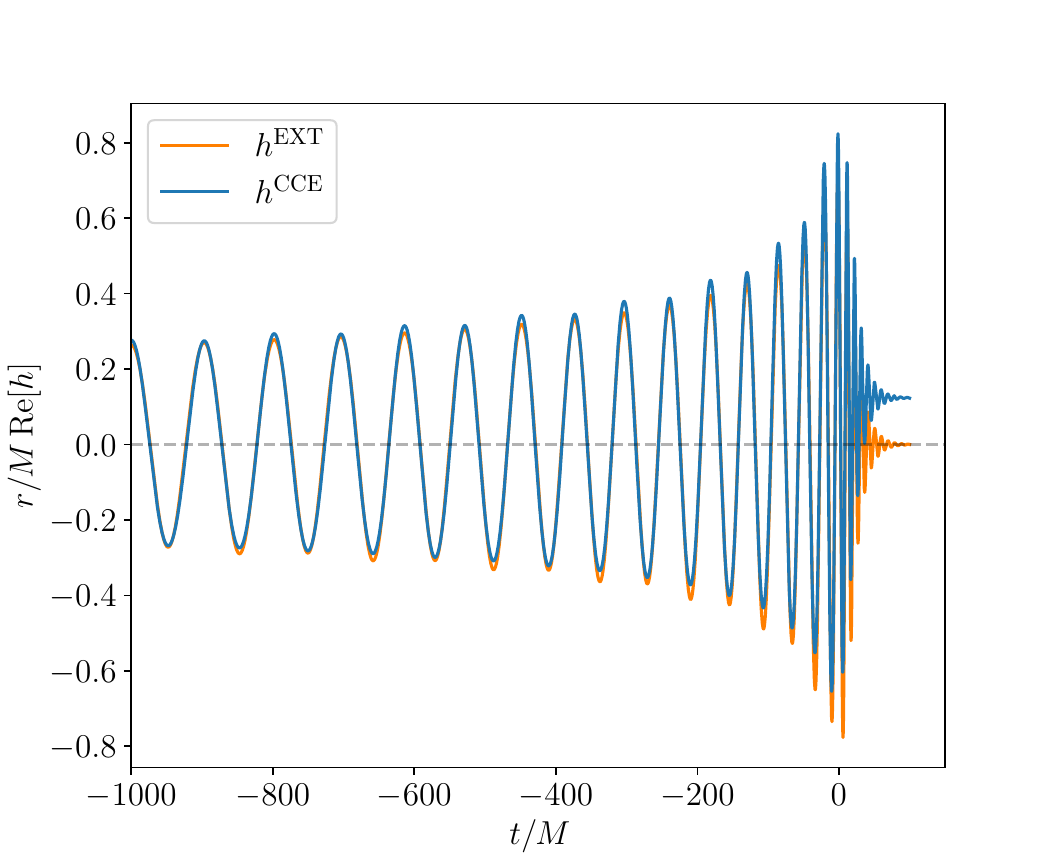}
    \caption{
    Illustration of type $(a)$ and type $(b)$ waveform models. The type $a$ waveform (orange) includes no memory and is based off the extrapolation method (EXT) while the CCE waveform (blue) is type $b$ and includes memory.
    }
    \label{fig:WaveformTypes}
\end{figure}

There are in principle two distinct ways of computing and isolating a gravitational memory model for quasi-circular BBH mergers, depending on the presence of memory within the given asymptotic waveform family:
\begin{enumerate}[$(a)$]
    \item Waveform models that do not include any gravitational memory yet.
    \item Full waveform models, which already include the memory information.
\end{enumerate}
A visualization of the difference between both types of waveform models is shown in Fig.~\ref{fig:WaveformTypes}. Throughout this work, we will be using surrogate models \texttt{NRHybSur3dq8} for type $(a)$ waveforms and \texttt{NRHybSur3dq8\_CCE} for type $(b)$ waveforms~\cite{Varma:2018mmi,Yoo:2023spi}. At the end of this section, we will also comment on the validity and limitations of defining a memory model through the BMS balance laws.

\paragraph{} 

Historically, asymptotic waveform models for BBH mergers did not include gravitational memory contributions. This is because, for a long time, numerical relativity simulations based on the extraction scheme of \textit{extrapolation}
(EXT) were unable to capture the memory component of the radiation due to technical limitations~\cite{Favata:2008yd,Favata:2010zu}. These limitations originate primarily from the finite extraction radius, which generically contaminates the low-frequency content of NR waveform data. In this context, defining a viable memory model is not a matter of isolating the memory signal from the full radiation, but rather of computing the nonlinear memory associated with a given high-frequency waveform. Without explicit intention, such waveform models have effectively already performed the \textit{a priori} challenging task of isolating a purely high-frequency waveform [cf.~Eqs.~\eqref{eq:MemLfromAverage} and \eqref{eq:MemHfromAverage}] within the multi-scale problem of BBH mergers. In this case, one can straightforwardly employ Eq.~\eqref{eq:memorymodes20} to compute the missing nonlinear memory contribution.

\begin{figure*}
    \centering
    \includegraphics[width=\linewidth]{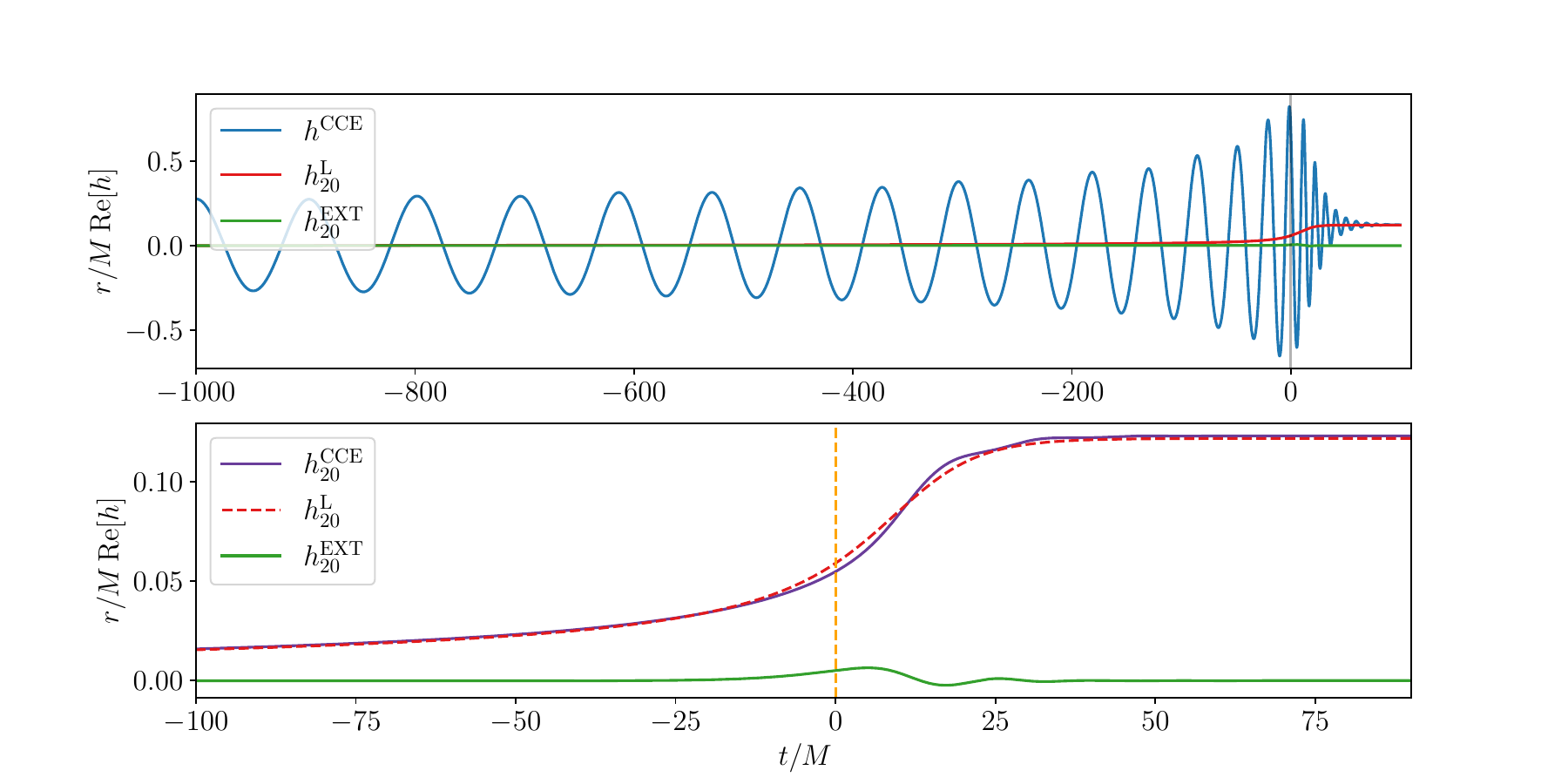}
    \caption{\textit{Top panel:} We show the full CCE waveform strain (blue, waveform type $b$), the $(2,0)$ mode of the extrapolated surrogate model (green, waveform type $a$), and the memory content (red) of the $(2,0)$ mode as calculated from the extrapolated waveform through Eq.~\eqref{eq:memorymodes20}. \textit{Bottom panel:} A close-up around the merger (orange dashed line) of the $(2,0)$ mode used to compare the CCE (purple, waveform type $b$), EXT (green, waveform type $a$), and memory calculation (dashed red). \textit{Parameters:} $Q=1.5$, $\chi = 0.6$. 
    }
    \label{fig:WaveformWithMem}
\end{figure*}

One may nevertheless wonder how such NR simulations, and the waveform models calibrated to them, can accurately reproduce the high-frequency gravitational-wave signal while simultaneously missing the memory contribution in the $(2,0)$ mode, whose amplitude can become comparable to that of the dominant $(2,2)$ oscillatory mode.\footnote{See also Fig.~2 of Ref.~\cite{Favata:2008yd} for a quantitative comparison of mode amplitudes.} The resolution lies in the assumed Isaacson separation of physical scales between the oscillatory and memory components of the radiation, presented in Sec.~\ref{ssSec:FundamentalMemoryModel}. A parametric separation in frequency between memory and gravitational waves allows NR simulations to accurately extract oscillatory modes while effectively eliminating the memory contribution. This can be achieved as follows~\cite{Favata:2008yd,Favata:2010zu}. Many of the most reliable NR waveforms are extracted using the Weyl scalar $\Psi_4$~\cite{Newman:1961qr,GomezLopez:2017kcw,Ashtekar:2019viz,DAmbrosio:2022clk}, whose spin-weighted spherical-harmonic modes are related to the strain via $\psi_{\ell m} \propto \ddot h_{\ell m}$. At the level of $\psi_{\ell m}$, the memory contribution is therefore parametrically suppressed relative to the high-frequency oscillatory signal by a factor of order $(f_L/f_H)^2$, and $\Psi_4$ contains no net memory offset. To recover the strain $h_{\ell m}$, the NR data for $\psi_{\ell m}$ must then be integrated twice in time. 
At this stage, ambiguities associated with poorly modeled low-frequency content from finite-radius extraction are absorbed into the choice of the two integration constants, which are fixed to remove any spurious secular drift and to set the late-time strain offset to zero~\cite{Berti:2007fi}. 
These choices effectively filter out the nonlinear memory while leaving the accurately resolved oscillatory modes unaffected. Finally, the lower-frequency behavior associated with the early inspiral, where the characteristic oscillatory frequency falls below the reliable range of the NR simulation, is restored by matching the waveform to analytic PN results. In this way, NR simulations and waveform models achieve an accurate representation of the oscillatory gravitational-wave signal while unintentionally, but systematically, discarding the nonlinear memory contribution. In particular, it remains possible to accurately model the oscillatory high-frequency content of the $(2,0)$ mode associated with the ringdown phase~\cite{Mitman:2020pbt,Mitman:2020bjf,Mitman:2024uss}, as shown in Fig.~\ref{fig:WaveformWithMem}.

As already stated, given such an exclusively high-frequency waveform model, the corresponding nonlinear memory in the $(2,0)$ mode is computed by inserting the mode decomposition of the waveform into Eq.~\eqref{eq:memorymodes20} and performing a numerical time integration. The result for a representative event is again displayed in Fig.~\ref{fig:WaveformWithMem}.
Furthermore, once the projection onto the $(2,0)$ harmonic is carried out, one can effectively disregard the explicit spacetime averaging in Eq.~\eqref{eq:memorymodes20}. Indeed, while the spacetime averaging appearing in Eqs.~\eqref{eq:memoryequation} and \eqref{eq:memorymodes} is essential to define an isolated memory signal and to establish that the gravitational memory of non-precessing BBH mergers resides in the $(2,0)$ harmonic, an explicit coarse-graining of Eq.~\eqref{eq:memorymodes20} is no longer required at this stage. The reason is that, to good approximation, any residual wave-like behavior characterized by the complex exponential in Eq.~\eqref{eq:phaseBehaviorM} is absent for $m=0$. Concretely, the memory formula employed in the companion Bayesian analysis of the LISA memory signal~\cite{Cogez:2025memoryLISA}, which accounts for the $(2,2)$, $(2,1)$, $(3,3)$, $(3,2)$, $(4,4)$, and $(4,3)$ high-frequency modes, reads
\begin{widetext}
\begin{align}
  h_{20}^{L} =\frac{1}{r}\int_{-\infty}^udu'r'^2&\Bigg[\frac{1}{7}\sqrt{\frac{5}{6\pi}}|\dot h^H_{22}|^2-\frac{1}{14}\sqrt{\frac{5}{6\pi}}|\dot h^H_{21}|^2+\frac{5}{4\sqrt{42\pi}}\left(\dot h^{H}_{22}\dot{\bar h}^{H}_{32}+\dot{\bar h}^{H}_{22}\dot h^{H}_{32}\right)\nonumber\\
  &+\frac{1}{2\sqrt{10\pi}}\left(\dot h^{H}_{33}\dot{\bar h}^{H}_{43}+\dot{\bar h}^{H}_{33}\dot h^{H}_{43}\right)-\frac{2}{11}\sqrt{\frac{2}{15\pi}}|\dot h^H_{44}|^2\Bigg]\,.
\label{eq:memorymodes20Selection}
\end{align}
\end{widetext}


\paragraph{} On the other hand, it is at present possible to implement NR simulations that are able to fully resolve the entire frequency spectrum of the asymptotic radiation of BBH mergers, including gravitational memory~\cite{Pollney:2010hs,Reisswig:2009rx,Babiuc:2010ze,Handmer:2014qha,Handmer:2015dsa,Handmer:2016mls,Mitman:2020pbt,Mitman:2024uss}. These results are based on an alternative extraction scheme known as Cauchy characteristic extraction (CCE)~\cite{Bishop:1996gt}, which takes the Cauchy evolution of an NR simulation at finite radius and propagates it to null infinity by solving the Einstein equations on outgoing null hypersurfaces. The resulting strain is free from low-frequency ambiguities and is able to fully resolve gravitational memory, including the late-time displacement memory offset, as seen in Fig.~\ref{fig:WaveformTypes}. This extraction scheme is indeed accurate as validated through a comparison to the direct computation of nonlinear memory and the BMS balance laws~\cite{Mitman:2020pbt,Mitman:2020bjf}, and most up-to-date simulations have recently been incorporated within surrogate waveform models, such as \texttt{NRHybSur3dq8\_CCE}~\cite{Yoo:2023spi} and \texttt{NRSur3dq8\_RD}~\cite{MaganaZertuche:2024ajz}.\footnote{There also already exist phenomenological models that incorporate the full $(2,0)$ mode, for instance as an extension of the computationally efficient waveform models \texttt{IMRPhenomTHM} ~\cite{Rossello-Sastre:2024zlr} and \texttt{IMRPhenomTPHM}~\cite{Rossello-Sastre:2025dep}, as well as extensions to the EOB models TEOBResumS-Dalí~\cite{Grilli:2024lfh} and TEOBResumS-GIOTTO~\cite{Albanesi:2024fts}.}


However, even for BBH mergers in which gravitational displacement memory is entirely found in the $(2,0)$ mode, it is not possible to directly use the $(2,0)$ of these CCE waveforms as a model for pure nonlinear memory. This is because, as anticipated above, the $(2,0)$ mode typically also contains an oscillatory part associated to the ringdown of a coalescence, shown explicitly in Fig.~\ref{fig:WaveformWithMem}. Therefore, also for waveforms of type $(b)$ an additional step is required to define isolated low-frequency gravitational memory and high-frequency gravitational wave models. In this context, it is in general not a good idea to use simple subtraction techniques using waveform models of type $(a)$ due to the occurrence of uncontrollable errors in particular at higher mass ratio, as examined in Ref.~\cite{Inchauspe:2024ibs}.\footnote{The main errors in such a comparison arise since the waveforms used in building the surrogates are in different frames. In other words, the extrapolated waveforms use a naive Newtonian-based center-of-mass correction while CCE uses a BMS charge-based method. Smaller, but still important, errors to account for are the intrinsic machine-learning modeling error and subsequent upgrades to the code.} A viable strategy in the particular case of quasi-circular CBCs is in fact to use the separation of the waveform into its null and ordinary parts as defined through the BMS balance laws [Eq.~\eqref{eq:BMSSupermomentumBalanceLaw each angle MT}]. Indeed, for such sources the oscillatory component within the $(2,0)$ mode is entirely attributed to the ordinary waveform, whereas the null part solely describes the monotonic memory rise~\cite{Mitman:2020bjf,Mitman:2024uss}. However, such a strategy requires access to the full CCE extraction data, which is generally not freely available. Instead, the use of an averaging procedure at the frequency scales of the ringdown contribution of the form of Eqs.~\eqref{eq:MemLfromAverage} and \eqref{eq:MemHfromAverage} could be envisaged. An efficient way of imposing such a separation of scales is in fact the direct use of Eq.~\eqref{eq:memorymodes20} where the $h^H$ are replaced by the memory-full total asymptotic strain. As already mentioned in Sec.~\ref{sSec:CharacteristicsofIsaacsonMemory}, the low-frequency characteristics of memory ensures that its contribution to the memory modes within Eq.~\eqref{eq:memorymodes} is negligible. At present, in most cases the direct use of Eq.~\eqref{eq:memorymodes20} for a full waveform model or waveform models of type $(a)$ is the most practical definition of a memory signal model~\cite{Inchauspe:2024ibs,Cogez:2025memoryLISA}.

\paragraph{} Alternatively, a number of studies construct memory models directly from the null component of the BMS balance laws discussed in Sec.~\ref{ssSec:ProofofMemory} (see, e.g., Refs.~\cite{Mitman:2020bjf,Grant:2022bla}). 
For quasi-circular CBCs, this approach is justified because, once the physical input of the Isaacson framework is adopted, namely that the dominant memory contribution resides in the $(2,0)$ mode, an explicit averaging procedure is no longer required, as explained in Sec.~\ref{ssSec:Practical Approximation}$a$. 
At the level of this $m=0$ mode, the memory obtained from the BMS balance laws is therefore identical to the construction using waveform models $(a)$, since, up to averaging over high-frequency scales, Eq.~\eqref{eq:NonLinDispMemoryModesGR} is formally equivalent to the nonlinear part of the BMS supermomentum balance law, a correspondence we derive explicitly in Appendix~\ref{App:BMSFluxBalanceLaws} [see Eq.~\eqref{BMSSupermomentumBalanceLawBD5 App}].\footnote{In fact, as discussed above, for BBH systems with negligible recoil, the balance laws further provide a natural decomposition of the waveform into null and ordinary contributions within the $(2,0)$ mode.}
It is, however, important to remember that the BMS balance laws are not \textit{per se} a tool to discriminate between wave-like and memory-like signatures within the asymptotic radiation, as elaborated on in Sec.~\ref{ssSec:NecessityBeyondBMS} above. Moreover, in Sec.~\ref{ssSec:MemoryInLISA} below, we illustrate the impact on SNR of directly using a memory model based on the null part of the BMS balance laws instead of Eq.~\eqref{eq:NonLinDispMemoryModesGR} including the spacetime averaging.

\subsubsection{Properties of the nonlinear memory of BBH mergers}\label{ssSec:PropertiesMemoryModel}

Consider a quasi-circular black hole binary of total mass $M$, mass ratio $Q$, and aligned spins $\chi$. We first establish the characteristics of the memory signal for the canonical case of an edge-on ($\theta=\pi/2$), equal-mass ($Q=1$), non-spinning ($\chi=0$) system in the time, frequency, and time-frequency domains, and then discuss how these features generalize across the parameter space.

\begin{figure}
    \centering
    \includegraphics[width=1\linewidth]{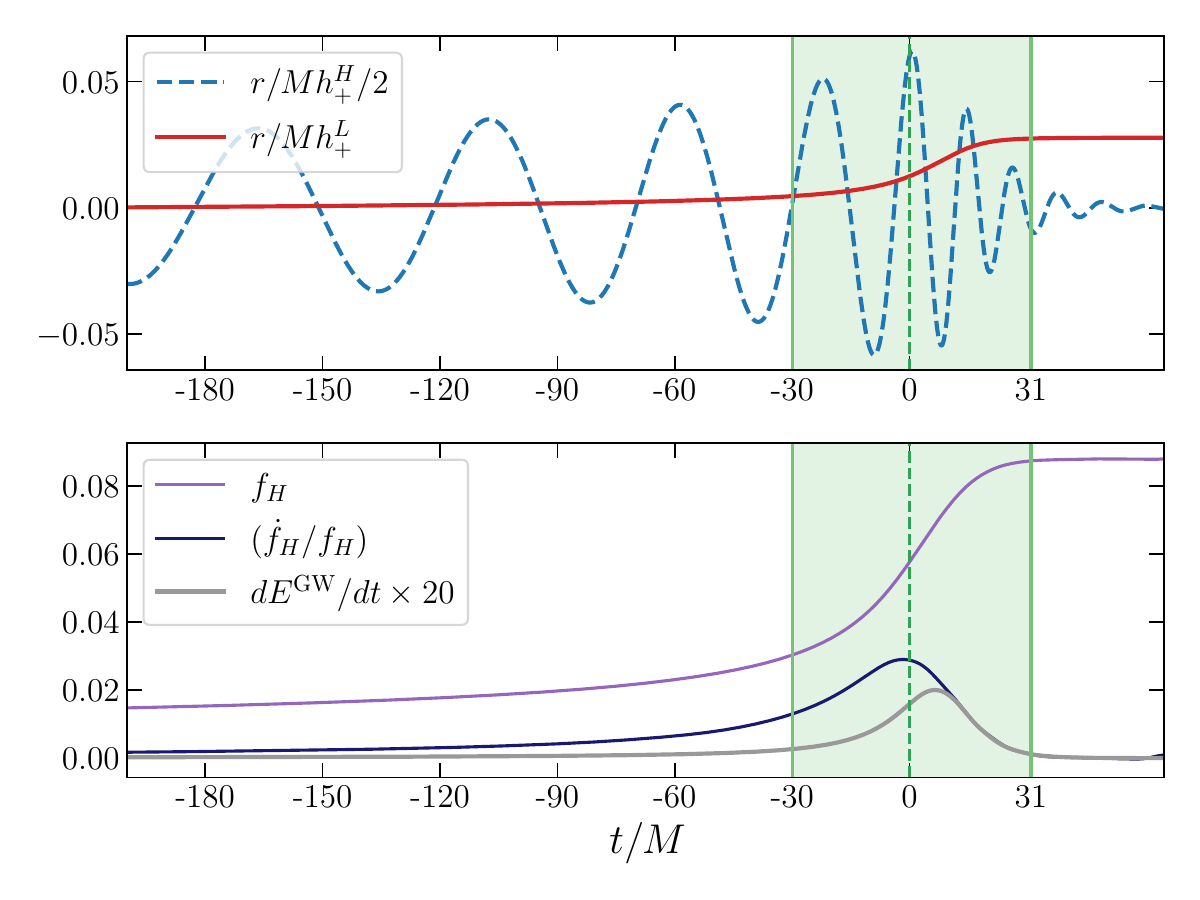}
    \caption{
    \textit{Top:} Dominant oscillatory waveform and memory for an equal-mass, nonspinning binary (edge-on).  
    The green band marks the interval $t \in [-30M,\,30M]$, during which approximately $66\%$ of the total radiated energy is emitted.
    \textit{Bottom:} Instantaneous gravitational-wave frequency $f_{\rm GW}$ of the dominant $(2,2)$ mode, together with the characteristic 
    memory-growth timescale $\dot{f}_{\rm H}/f_{\rm H}$, and the corresponding energy flux $dE^{\rm GW}/dt$.
 }
    \label{fig:memwith freq}
\end{figure}

\paragraph{General time-domain properties}
As seen in Fig.~\ref{fig:memwith freq}, the three stages of a binary evolution of the \emph{inspiral}, \emph{merger} and \emph{ringdown} (IMR)~\cite{Buonanno:2006ui} also result in three distinct types of memory signals:
\begin{enumerate}
    \item A very slow rise during the inspiral.
    \item A rather sudden jump around merger.
    \item A settling down at a final value during ringdown.
\end{enumerate}
More precisely, the growth of the memory signal is directly associated to the radiation reaction timescale of the binary, which in terms of the language introduced in Sec.~\ref{sSec:DefMemModel}, defines separate stages of emission time $T_L$ over which the emission of radiative energy does not significantly alter the binary.

\begin{enumerate}
\item During the \textit{inspiral} phase, the gravitational displacement memory grows steadily in amplitude over a potentially very long emission time $T_L^{\mathrm{I}}$, while the characteristic GW frequency $f_H^{\mathrm{I}}$, associated with the orbital timescale, evolves slowly. In Appendix~\ref{App:PNresults}, we review the leading PN calculation of the memory. In the PN framework, the memory arises from the quadrupole-quadrupole interaction term, which formally enters at 2.5PN order. However, because it is integrated over the entire evolutionary history of the binary, it contributes to the waveform amplitude at effectively 0PN order. Moreover, we show that the memory evolves on the radiation-reaction timescale
$f_H/\dot{f}_H$.
This instantaneous memory timescale is compared with the instantaneous GW frequency $f_H$ in Fig.~\ref{fig:memwith freq}, where the separation between the two is manifest at all times.
Throughout the inspiral, the cumulative GW energy emitted is insufficient to produce a memory contribution comparable in amplitude to the oscillatory waveform, whose magnitude remains dominant until the late inspiral.
\item
On the other hand, during the \textit{merger} phase, the large GW energy flux induces a rapid increase in the memory amplitude. Throughout this work, the merger time is defined at $t=0$ and corresponds to the peak of the $L^2$ norm of the surrogate waveform \texttt{NRHybSur3dq8\_CCE}~\cite{Yoo:2023spi}. As shown in Fig.~\ref{fig:memwith freq}, the step-like growth of the memory is largely concentrated within the green area, corresponding to the interval $[-30M,\,30M]$ around the merger time.\footnote{
The energy radiated within this time window accounts for approximately $66\%$ of the total emitted energy. Beyond $t \simeq 30M$, the memory saturates and reaches its final value with a relative accuracy of $\mathcal{O}(10^{-2})$.
}
During this interval, the instantaneous growth timescale of the memory becomes much shorter than during the inspiral.
Nevertheless, it remains substantially longer than the inverse of the characteristic GW frequency $f_H^{-1}$, so that a clear separation between the memory timescale and the dominant GW oscillation timescale persists even through merger. Finally, we note that the merger time coincides with the peak of the instantaneous memory growth rate, which precedes both the peak of the GW energy flux ($t\simeq 6M$) and the saturation of the memory signal ($t\simeq 30M$). 
\item Lastly, the relatively short and low-energy \textit{ringdown} will only marginally contribute to any measurable gravitational memory, depending on where the distinction between merger and ringdown phase is drawn~\cite{Buonanno:2006ui,Bhagwat:2017tkm}. Choosing the start of ringdown to be in the range $t=[10M,20M]$ leaves no more than $\sim 20 \%$ of the memory present in the signal as seen in Fig.~\ref{fig:memwith freq}. Nevertheless, it completes the memory accumulation in the waveform, yielding the late-time offset

\end{enumerate}

Hence, the amplitude of gravitational memory from such a binary black hole coalescence attains an appreciable value only due to the very energetic merger. In fact, for the example system discussed above, the memory signal amplitude around the merger becomes comparable in scale to that of the high-frequency oscillatory gravitational waves themselves. Thus, when referring to an observationally relevant memory signal, particularly in the context of a first detection, one is primarily referring to the merger-generated component of the memory, as we will further discuss in Sec.~\ref{Sec:claim of detection with LISA} below.

\paragraph{General frequency-domain properties}

\begin{figure}
    \centering
    \includegraphics[width=\linewidth]{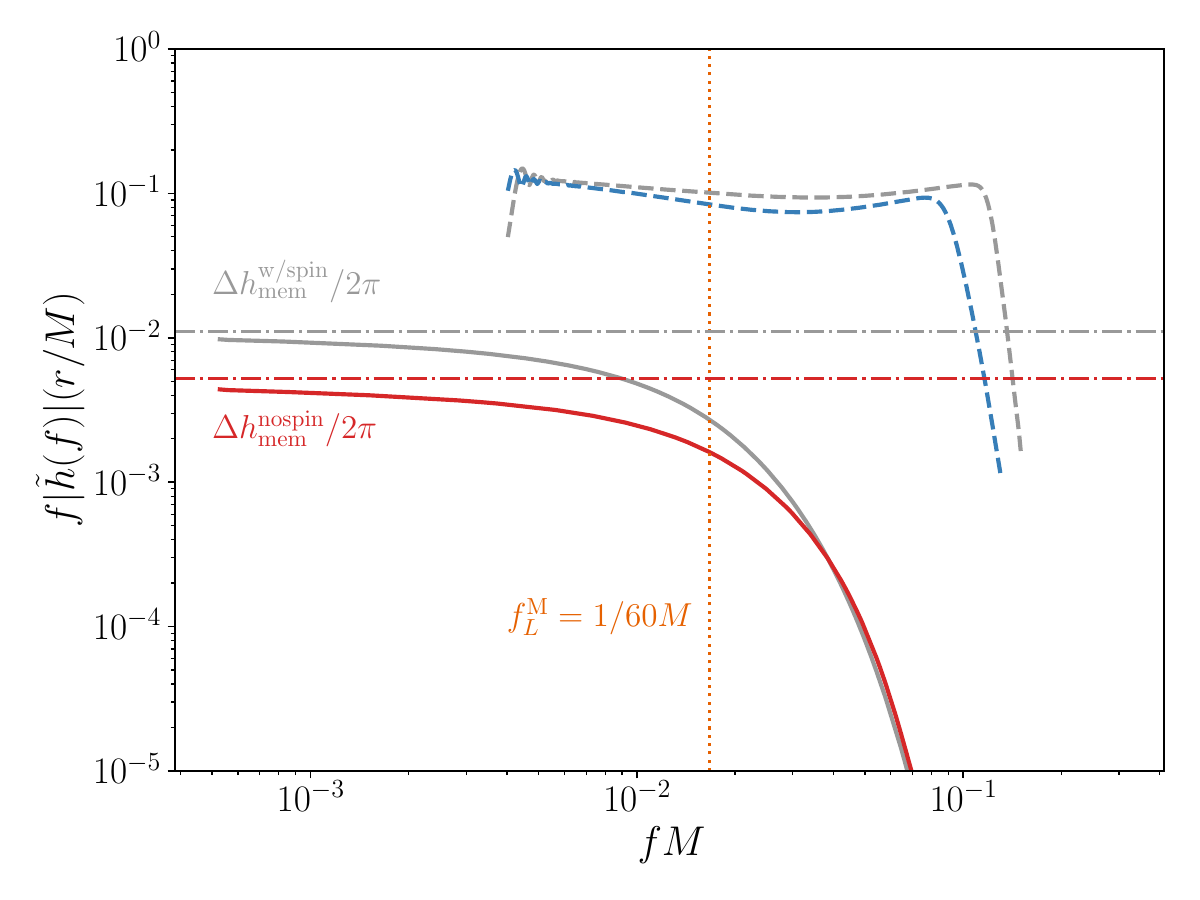}
    \caption{Characteristic strain in frequency space of the memory (solid) and of the dominant GW signal (dashed) the system in Fig.~\ref{fig:memwith freq} with zero spin (blue) and with spin $\chi=0.8$ (grey). The signal of the oscillatory GWs follows the typical shape of an IMR event, with a powerlaw increase in frequency during inspiral of $|\tilde h|\sim f^{-7/6}$, followed by a merger feature that ends in a sharp, damped ringdown at the highest frequencies. The memory signal on the other hand approaches a constant value $\Delta h_{\rm mem}/(2\pi)$ (dot-dashed) at low frequencies and decays at frequencies higher than $\sim f_L^{\rm M}$, as explained in the main text. }
    \label{fig:FTmemory}
\end{figure}

The gravitational memory associated with a BBH merger dominates the signal in the frequency domain, as illustrated in Fig.~\ref{fig:FTmemory}. To understand the shape of the memory signal in frequency domain, it is useful to consider the approximation of modeling the memory rise as a Heaviside step function $\Theta(t)$ with amplitude given by the memory jump at merger, $\Delta h_{\rm mem}$, smoothed over the memory rise time $T_L^{\rm M}\simeq 60M$. The Fourier transform of a step function reads
\begin{equation}
\tilde{\Theta}(t)
=-\frac{i}{2\pi f}+\frac{\delta(f)}{2},
\end{equation}
exhibiting the characteristic $1/f$ behavior at low frequencies, together with an unobservable zero-frequency component. This scaling explains the constant low-frequency plateau observed in terms of the characteristic strain $f|\tilde h(f)|$ in Fig.~\ref{fig:FTmemory}, with an amplitude proportional to $\Delta h_{\rm mem}/2\pi$ (dot-dashed lines). At higher frequencies, the spectrum rolls off around the cutoff $f_L^{\rm M}\sim 1/T_L^{\rm M}$.\footnote{The analyticity of the memory and the vanishing of its time derivative after merger guarantee the decay of its Fourier transform at high frequencies via the Paley–Wiener theorem~\cite{Mukhopadhyay:2021zbt}.}
Although frequencies below $f_L^{\rm M}$ are also generated before merger, they are non-observable, as the memory detectability is primarily driven by the burst-like growth of the memory around merger.

While the amplitude of the memory rise is the primary feature governing its detectability, we emphasize that a detector sensitive to the merger of the primary GW signal becomes increasingly responsive to the high-frequency decay of the memory near $f_L^{\rm M}$. This regime is not directly associated with the long-lasting, step-like growth of the memory, but rather reflects the non-trivial time dependence of the memory rise. This underscores the importance of defining a memory model via a unique interpolating function connecting the memory offset before and after merger.  
Furthermore, while Fig.~\ref{fig:FTmemory} may suggest that the separation of frequency scales between the oscillatory signal and the memory becomes blurred at the high-frequency end of the memory spectrum, this is not the case when focusing on the observationally relevant merger portion, where a clear separation still prevails. This point is further illustrated in the following paragraph through an explicit time-frequency analysis.

By contrast, if the detector bandwidth is sufficiently broad and/or the merger occurs at high frequencies, potentially outside the detector band, the response is dominated by the $1/f$ behavior of the memory. In this case, the frequency separation between the primary signal and the memory becomes clearer, while the dependence on the precise form of the memory-rise is reduced. This consideration motivates searches for memory from out-of-band events as a sensitive probe of the intrinsic $1/f$ scaling of the memory signal directly connected to BMS symmetries and soft theorems.

\paragraph{General time-frequency-domain properties}
As anticipated in Sec.~\ref{ssSec:RealisticSources}, the time-frequency representation is particularly well suited to separating the memory signal from the sourcing oscillatory field. By tracing the time evolution of the frequency content of the signal over the binary evolution, one can identify the characteristic scales associated with each component.
\begin{figure}
\centering
\includegraphics[width=\linewidth]{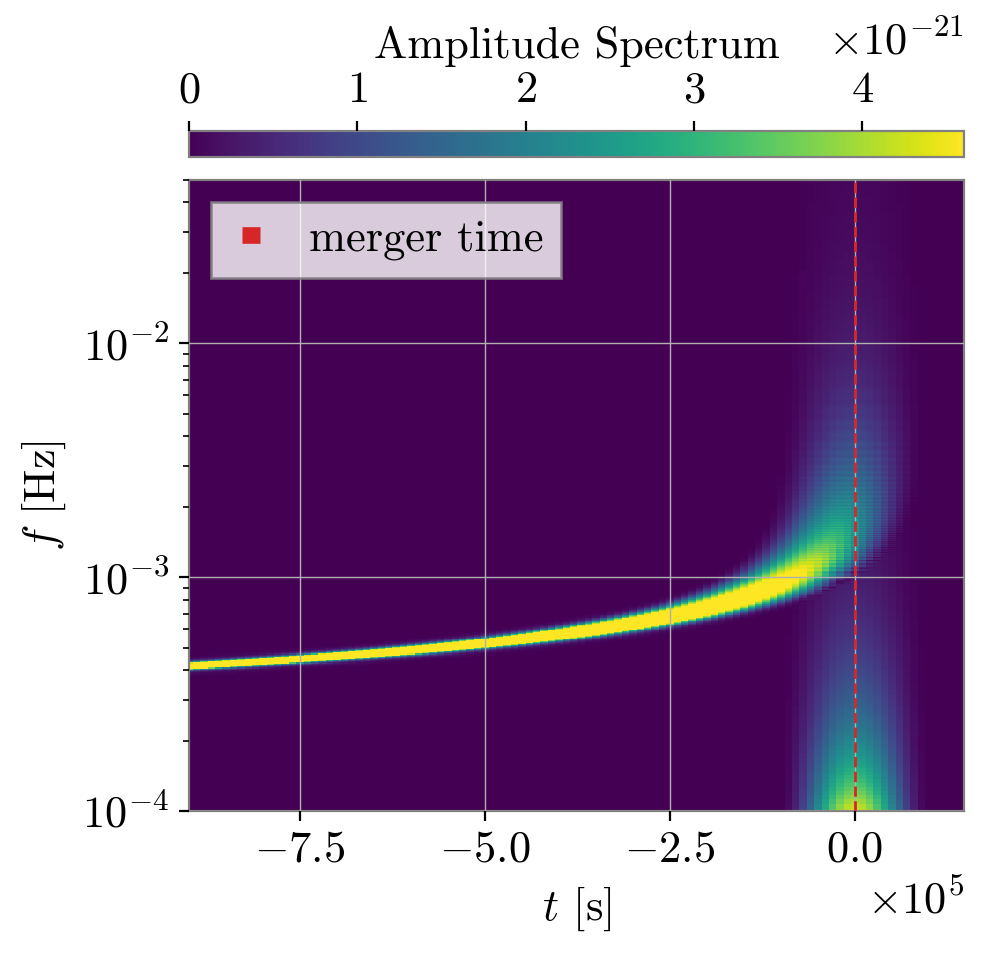}
\caption{Spectrogram of a radiation signal composed of the dominant $(2,2)$ oscillatory mode and the $(2,0)$ nonlinear memory contribution, generated by an aligned-spin MBHB system. Two distinct power blobs are distinguishable: one exhibiting the typical chirp-like behavior of the oscillatory $(2,2)$ mode, spread in time but narrow in frequency; the second, the memory piece, well localized at merger time, but spread in frequency with the expected $1/f$ behavior and high-frequency damping. This plot was made using the waveform model $(a)$ with Eq.~\eqref{eq:memorymodes20} for its memory. \textit{Parameters:} $M=5\times10^5 ~\mathrm{M}_\odot$, $Q=1$, $\chi = 0.95$, $\theta=\tfrac{\pi}{2}$. [The spectrogram is computed from \texttt{scipy.signal.spectrogram}, based on a short-time Fourier transform, and was optimized for visualization of the scale separations (e.g. time-frequency settings, colorbar saturation). It is using $N=5$ Hanning windows with $95 \%$ overlap, hence tolerating high correlations between pixels and tending to broaden the time localization of the signals power; see also~\cite{Inchauspe:2024ibs}]}
\label{fig:mem_spec}
\end{figure}
Fig.~\ref{fig:mem_spec} shows the short-time Fourier transform–based spectrogram of the gravitational radiation signal of an equal mass BBH merger, including the dominant $(2,2)$ GW mode and its nonlinear memory counterpart. Two distinct power regions are clearly visible. The brightest, associated with the $(2,2)$ mode, exhibits the characteristic chirp-like time-frequency behavior, with a slowly increasing frequency and amplitude leading up to merger, followed by a rapid rise and a sharp end. In contrast, the memory signal is primarily localized around the merger time, in agreement with the analysis in Sec.~\ref{ssSec:PropertiesMemoryModel}.\footnote{Due to high correlation between pixels in Fig.~\ref{fig:mem_spec} the memory feature appears much broader than it is in reality.}

Most importantly, the memory signal displays a broad frequency content clearly distinct from the oscillatory GWs, which follows the typical $1/f$ scaling at frequencies below the merger timescale $f_L^\text{M}$. This provides a confirmation of the claim in Sec.~\ref{ssSec:RealisticSources}, that the merger of quasi-circular binary black holes is generally slow enough to admit a characteristic high frequency $f_H$ with gravitational memory effectively acting as a change of background on which the high-frequency waves propagate. At the same time, it illustrates the potential to separate the two components in a model-agnostic manner, as already pointed out in~\cite{Inchauspe:2024ibs}.

\paragraph{Variations across parameter space}
Being entirely described by the $(2,0)$ SWSH mode, the dependence of the memory signal on the anlges of emission is given by the scaling of 
\begin{equation}
    _{\mys{-2}}Y_{20}(\theta,\phi)\propto \sin^2\theta\,.
\end{equation}
Hence, gravitational memory is maximal for edge-on systems $\theta=\pi/2$, while vanishes in the face-on scenario. 

As the mass ratio of a non-spinning binary black hole increases, the overall amplitude of the gravitational memory decreases, together with a corresponding reduction in the amplitude of the dominant $(2,2)$ oscillatory mode. While the qualitative structure of the memory signal remains the same, increasing mass asymmetry leads to a more gradual memory buildup, reflecting the longer merger timescale, thus shifting the characteristic frequency content. Increasing asymmetry further enhances the relative contribution of higher-order modes in both the oscillatory waveform and the memory signal, although the $(2,0)$ mode continues to dominate for quasi-circular binaries.

On the other hand, an increase in black hole spins generally amplifies the memory amplitude, both in absolute terms and relative to the peak of the dominant oscillatory mode. For example, for an edge-on configuration, the final memory amplitude increases from $\Delta h_L^{\rm nospin} \simeq 0.03\,M/r$ for non-spinning binaries to $\Delta h_L^{\rm w/spin} \simeq 0.07\,M/r$ for binaries with aligned spins of magnitude $\chi = 0.8$.
 These correspond to relative amplitudes with respect to the peak $(2,2)$ mode of approximately $\sim 0.15$ and $\sim 0.32$, respectively. For very high spin values, the memory amplitude can even entirely dominate the oscillatory signal.

\begin{figure}
    \centering
    \includegraphics[width=1\linewidth]{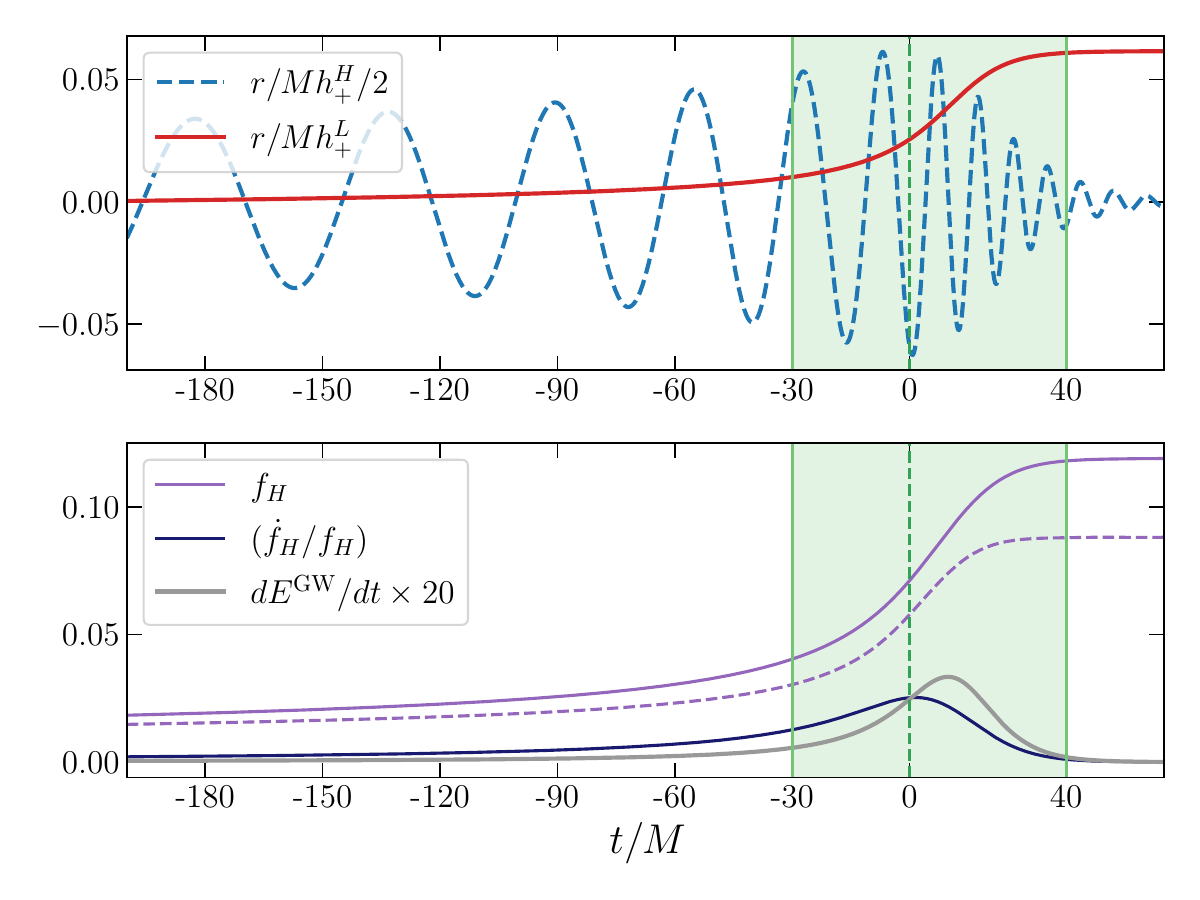}
    \caption{
    As in Fig.~\ref{fig:memwith freq}, but for a system with aligned spins $\chi=0.8$. For reference, we show the instantaneous frequency evolution for the zero spin case in a dashed line.
 }
    \label{fig:memwithfreqSPIN}
\end{figure}

A comparison with the dominant oscillatory waveform, together with the instantaneous GW frequency and energy flux, is shown in Fig.~\ref{fig:memwithfreqSPIN} for the aligned-spin case with $\chi = 0.8$. The enhancement of the memory can be attributed to the larger GW energy flux emitted around merger compared to the non-spinning case~\cite{Reisswig:2009vc, Barausse:2012qz}. Physically, this is a consequence of the orbital hang-up effect, which allows the spinning binary to reach closer separations and higher velocities before merging. In particular, aligned spins prolong the merger phase and reach higher instantaneous frequencies, leading to a larger saturation value of the memory. In this configuration, the memory reaches its maximum by $t \simeq 40 M$ with a relative accuracy of $\sim 10^{-2}$. 
This trend is also evident in the Fourier-domain representation (gray) in Fig.~\ref{fig:FTmemory}, where the increased merger frequency and larger amplitude shift the low-frequency plateau upward when comparing the $\chi = 0$ and $\chi = 0.8$ cases.  

By contrast, binaries with anti-aligned spins experience a shorter and less energetic merger phase, resulting in a reduced amplitude of the dominant $(2,2)$ oscillatory mode. The final memory amplitude is $\Delta h_L^{\rm w/aspin}\simeq 0.02 M/r$ and a relative amplitude to the $(2,2)$ mode of $\sim 0.16$ for a system with $\chi =-0.8$ anti-aligned with the binary angular momentum.

In summary, the systems exhibiting the largest gravitational memory are equal-mass binaries with aligned spins, which also undergo the most energetic merger phases.

\subsubsection{The propagation on cosmological scales}\label{ssSec:CosmologyModel}

We want to end this section with a quick note on the propagation of the memory signal over a cosmological background. So far, we have only talked about memory in asymptotically flat spacetime. However, real signals will propagate over the cosmic background observed in the universe~\cite{Weinberg2008Cosmology,dodelson2020modern,maggiore2008gravitational} 
\begin{equation}
    ds^2=-d t^2+ a^2(t)d\mathbf{x}\,,
\end{equation}
with $a(t)$ the evolving scale factor. This can be accounted for by using the asymptotically flat result as a proxy for the signal close enough to the source where the cosmic evolution is still negligible ($s$) and the universe had a given scale factor $a(t_)=a_s$, and subsequently evolving it over cosmic distances to the final observer ($o$) with current scale factor $a_o$. 

While this is a well-known problem addressed in the literature~\cite{thorne_2024_q890b-aem17,maggiore2008gravitational,Tolish:2016ggo,Bieri:2017vni}, we offer here a particularly concise derivation. The general result of cosmic propagation is that signals well inside the cosmic horizon, will continue propagating luminally, while its amplitude is being damped due to the cosmic expansion~\cite{dodelson2020modern}. More precisely, the memory of the form
\begin{equation}
    _sh^{TT}_{ijL}=\frac{1}{a_sr}A_s
\end{equation}
computed at ($s$), evolves in the cosmic flow through a typical amplitude suppression 
\begin{equation}
    _oh^{TT}_{ijL}=\frac{a_s}{a_o}\phantom{}_sh^{TT}_{ijL} = \frac{1}{a_or}A_s\,.
\end{equation}

However, one should not forget that the relevant distance measurement for radiation is the luminosity distance~\cite{Sachs:1962wk}. Indeed, while the Minkowski coordinate distance $a_sr$ corresponds to the luminosity distance at fixed scale factor $a_s$, the luminosity distance over the cosmic history generalizes to 
\cite{Weinberg2008Cosmology,maggiore2008gravitational}
\begin{equation}\label{eq:LuminosityDistance}
    _od_l\equiv (1+z) a_o r\,,
\end{equation}
with 
the redshift defined as
\begin{equation}\label{eq:redshiftDef}
    z\equiv \frac{a_o}{a_s}-1\,.
\end{equation}
Hence, in terms of redshift, the memory signal received by the observer reads
\begin{equation}
    _oh^{TT}_{ijL}= \frac{1+z}{d_l}A_s\,,
\end{equation}
and is thus enhanced by a factor of $1+z$ compared to the initial Minkowski result~\cite{Tolish:2016ggo,Bieri:2017vni}. This enhancement factor will result in a redefinition of the source-frame mass scales into a detector-frame, or redshifted mass $M_z$. Note that by redshifting the mass also the time dependence of the memory, depending on $Mt$, gets automatically redshifted as it should be.
In the companion paper~\cite{Cogez:2025memoryLISA}, these considerations were directly implemented in the defining memory formula.

Note, however, that in discussing this cosmic evolution for the specific case of the memory signal, the subhorizon assumption becomes of importance. This is because the zero frequency net offset of the memory signal will always lie outside the horizon and thus formally does not propagate. However, the frequency content of the memory signals accessible by current ground and space-based detectors always lies well inside the horizon. 

\section{Claiming a detection of gravitational memory with LISA}\label{Sec:claim of detection with LISA}

As previously emphasized, the careful construction of a conservative gravitational memory signal model that aims at a unique association with the permanent memory offset by incorporating its characteristic time-dependent rise in Sec.~\ref{Sec:MemoryTheory}, forms an essential foundation for any future claim of memory detection in gravitational-wave (GW) data. This necessity arises from a fundamental limitation: current and planned GW observatories are intrinsically unable to measure the net, DC component of the memory offset. Their band-limited sensitivity, confined to narrow frequency ranges analogous to the spectral filters used in electromagnetic astronomy, renders them effectively blind to DC signals, where instrumental noise grows steeply toward low frequencies. Specifically, ground-based interferometers such as LVK~\cite{LIGOScientific:2021djp} and future third-generation detectors~\cite{Maggiore:2019uih,Reitze:2019iox} operate in the kilohertz band, while the space-based LISA mission~\cite{LISA} targets millihertz frequencies, and PTA~\cite{NANOGrav:2023gor,EPTA:2023fyk,Reardon:2023gzh} probe the nanohertz regime. Consequently, detectors can only respond to those spectral components of the memory waveform that fall within their sensitivity band.

The frequency content of the memory signal is therefore a decisive factor for its observability: it must extend to sufficiently high frequencies for a detectable imprint to appear in the accessible band of a given instrument. In practice, this means that in the context of quasi-circular binary black holes discussed in Sec.~\ref{ssSec:PropertiesMemoryModel}, the memory signal targeted in GW data analysis is the merger-driven memory generated during the highly dynamical late stages of the compact-binary coalescence. Only this regime produces a rise time rapid enough to yield a cutoff frequency $f^\text{M}_L \sim 1/T^\text{M}_L$ that is both sufficiently high and accompanied by a large enough amplitude to be measurable.

Furthermore, interferometric detectors and the established search methods introduce additional obstacles to memory detection. Even though the gravitational-memory contribution is often comparable in amplitude to the high-frequency oscillatory modes of the radiation as described in Sec.~\ref{ssSec:PropertiesMemoryModel}, its observed imprint in GW data is significantly reduced in comparison to the high-frequency GW content, as we will discuss in detail in Sec.~\ref{ssSec:MemoryInLISA} below. 

Nevertheless, for massive black-hole binary sources whose memory rise timescales lie well within the LISA sensitivity band, the accumulated SNR can be substantial, reaching values up to $50$ for particularly bright and optimally inclined systems~\cite{Inchauspe:2024ibs, Gasparotto:2023fcg}. LISA therefore stands as a prime candidate to achieve the first single-source detection of gravitational memory. In what follows, we therefore focus specifically on the LISA mission, describe its response to memory signals, and, in conjunction with the companion paper to this work~\cite{Cogez:2025memoryLISA}, evaluate LISA's ability to robustly claim a detection of gravitational memory.

\subsection{Properties of the LISA response}\label{sSec:CharacteristicProperties}

\subsubsection{Characteristic of LISA response}

As discussed above, the imprint of the radiative memory signal within observed data is significantly affected by the properties of the detection instrument.
For completeness, we will shortly review here the characteristics of the LISA specific response to gravitational radiation (see also Ref.~\cite{Inchauspe:2024ibs}). The corresponding imprint of memory will essentially be characterized by the fact that at the low-frequency part of the spectrum, LISA's response to gravitational radiation effectively behaves as a third-order high-pass filter. 

To understand this statement, consider first the detector response of a single inter-spacecraft link in the triangular constellation of LISA to gravitational radiation as defined in Sec.~\ref{ssSec:Definition}. This response is fundamentally dictated by the geodesic deviation induced between the two freely falling test masses. More precisely, for wavelengths greater than the separation between the freely falling test masses, which corresponds to the low-frequency limit of the detector, the strain defined in Eq.~\eqref{eq:integratedGeodesicDeviation2} is proportional to the change in phase of laser light of frequency $\nu_0$ that travels between the spacecrafts, compared to the local phase-locked laser
\begin{equation}\label{eq:PhaseRelationToStrain}
    \Delta \varphi(u) \propto \nu_0s_0\Delta s(u)\,.
\end{equation}
This change in the phase of the laser light effectively measures a change in light-travel time.
In practice, however, single-link measurements are unusable, as they are overwhelmed by laser noise that is many orders of magnitude larger than the GW signal. This limitation is overcome in typical interferometers by exploiting Michelson-like multiple arm configurations that automatically cancel common-mode laser noise between pairs of arms. Due to LISA's, unequal and slowly time-varying armlengths, this cancellation needs to be achieved by an on-ground post-processing of the raw data known as time-delay interferometry (TDI).

Following the seminal work of Estabrook and Wahlquist~\cite{Estabrook:1975jtn}, the final output of the LISA response can be described in terms of Doppler change in frequencies of a laser light emitted and received between two freely falling test masses
\begin{equation}\label{eq:PhaseRelationToStrain2}
    \Delta \nu \propto \Delta \dot\varphi\,.
\end{equation}
More specifically, the laser frequency modulation induced by the passage of a gravitational radiation,
written as $\Delta \nu/\nu_0= y_{ab}$ for the link between spacecraft $a$ and $b$, is proportional to the difference between the radiative strain evaluated at the emission and reception events at time $u$ and $u+(1-\hat k\cdot \hat s)s_0$, with $s_0$ the initial separation of the spacecrafts~\cite{Burke:1975zz,Estabrook:1975jtn,Tinto_2010}
\begin{equation}\label{eq:OneWayDoppler}
y_{ab}
= \frac{1}{2}\frac{\hat{s}^i \hat{s}^j}{1-\hat{k}\cdot\hat{s}}
\left[  h^{TT}_{ij}(u) -h^{TT}_{ij}(u+(1-\hat k\cdot \hat s)s_0) \right].
\end{equation}
This is the so-called one-way Doppler response\footnote{In the community, the one-way Doppler response is also known as the ``two-pulse response'' or as the ``single-link'' response.} of LISA. As we see by combining Eqs.~\eqref{eq:PhaseRelationToStrain} and \eqref{eq:PhaseRelationToStrain2}, in the low-frequency limit, in which the GW wavelength largely exceeds the constellation armlength 
, the Doppler response can be approximated as being proportional to the time derivative of the strain
\begin{equation}
    \Delta \nu/\nu_0= y_{ab}\propto \Delta \dot s\,.
\end{equation}
Indeed, schematically, the one-way Doppler response in Eq.~\eqref{eq:OneWayDoppler} is obtained by applying a delay operator $\vb{D}_{ab}(S)$ of duration $S_{ab}=(1-\hat k\cdot\hat s)s_0$ accounting for the light-travel time between spacecrafts $a$ and $b$ onto the contracted radiation perturbation $h\equiv\hat{s}^i \hat{s}^j h^{TT}_{ij}/2(1-\hat{k}\cdot \hat s) $
\begin{equation}
y_{ab}(u) = \bigl( 1 - \vb{D}_{ab} \bigr)\, h(u),
\end{equation}
which, for slowly varying signals $(\dot h\,S \ll h)$ corresponds to a time derivative
\begin{equation}\label{eq:LowFreqExp}
y_{ab}= h(u)- h(u-S_{ab}) \sim S\,\dot h\,.
\end{equation}

Within TDI, the inter-spacecraft Doppler data streams $y_{ab}$ are then combined in a post processing to construct combinations of time-delayed measurements to cancel laser frequency noise~\cite{Vallisneri:2005ji,bayle_unified_2023}. In simple terms, this is achieved in two steps: A first-generation Michelson TDI observable $X_1$ cancels the noise assuming unequal but constant arm lengths by taking suitable, time-delayed combinations of the raw unequal-arm Michelson combination of a two-arm round trip 
\begin{align}\label{eq:MichelsonTwoArm}
    y(u)&=y_{13}+\vb{D}_{13}y_{31}-(y_{12}+\vb{D}_{12}y_{21})\nonumber\\
    &\sim 2\delta S\,\dot h+\mathcal{O}(S\delta S\,\ddot h)\,.
\end{align}
In the second line we have estimated here its contribution in the low frequency limit as in Eq.~\eqref{eq:LowFreqExp} with $S_{13}-S_{12}\sim \delta S$. Schematically, the cancellation of laser noise at leading order is achieved at this stage via a delay symmetrizing operation of the form
\begin{equation}
    X_1\sim(1-\vb{D}^2)\,y\sim  S\,\dot y\sim S\delta S\,\ddot h\,.
\end{equation}
Hence, in TDI, the leading order contribution of the naive two-arm Michelson output is canceled exactly, and in the low frequency limit, we are left with an effective second derivative of the strain.
Second-generation TDI $X_2$ introduces one additional set of delay-difference operators $1-\vb{D}^4$ to compensate for time-dependent arm lengths. Again, the combinations are such that in the low frequency regime, only the highest derivative terms survive
\begin{equation}
X_2 \;\sim\; S^2\delta S\, \dddot h,
\end{equation}
Hence, the second-generation Michelson TDI scales as the \emph{third time derivative} of $h$ in the low frequency detector limit~\cite{Babak:2021mhe}.

In total, there are three independent TDI channels traditionally denoted as $X$, $Y$, and $Z$, each constructed using one of the three spacecrafts as the central vertex. These three channels are in principle all sensitive to gravitational radiation, responding differently depending on their sky location and polarization. The $X$, $Y$, and $Z$ channels can be further combined into the orthogonal $A$, $E$, and $T$ channels, where in a first approximation the $A$ and $E$ channels are mostly sensitive to gravitational waves, providing independent information on polarization and source direction.
The T channel, by contrast, has a strongly suppressed response to gravitational waves in the low-frequency regime of the detector and is therefore largely insensitive to the memory signals considered here.

\begin{figure*}
    \centering
    \includegraphics[width=0.9\linewidth]{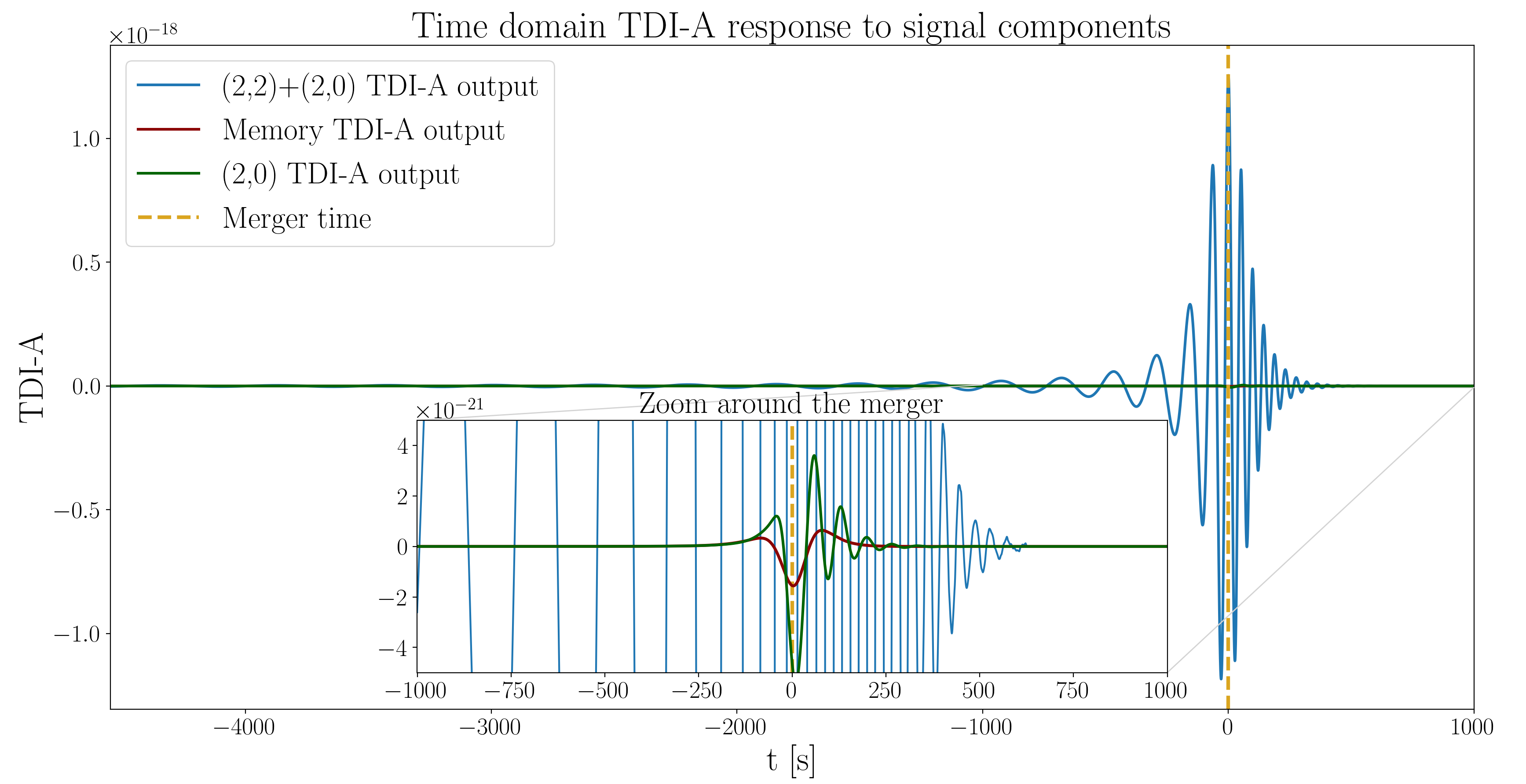}
    \caption{Time domain TDI-A channel obtained, after the response of the links to the radiation signals in Fig.~\ref{fig:WaveformWithMem}. The total waveform is shown in blue together with the $(2,0)$ SWSH mode split into the pure memory component $h^L_{20}$ in red and the purely oscillatory part $h^\text{EXT}_{20}$ in green.
    The TDI-A time domain channel shows that the memory component is suppressed by the LISA response and TDI post-treatment compared to the oscillatory signals.
    The parameters are the same as in Fig.~\ref{fig:WaveformWithMem} $Q=1.5$, $\chi=0.6$, with additional physical parameters adapted to the LISA scale $M_z = 10^6 \mathrm{M}_\odot$, $d_l = 10^4 \text{Mpc}$, $\theta = \pi/2$, $\psi=0$, and sky coordinates $\alpha = 0.74$ (right ascension), $\delta = 0.29$ (declination) as defined for instance in~\cite{Cogez:2025memoryLISA,LISA_RosettaStone}.}
    \label{fig:Comparing20Components}
\end{figure*}

\subsubsection{Memory in LISA}\label{ssSec:MemoryInLISA}

As shown in the previous section, collectively, the two-stage transfer function of TDI applied on the single-link Doppler response approximates to a third-order time differentiation for low-frequency signals, which is exactly the nature of the GW memory. This explains the characteristic signature of memory within the time domain TDI output of LISA shown in Fig.~\ref{fig:Comparing20Components} in the TDI-A channel. The memory component of the radiation from a quasi-circular BBH merger identified in Sec.~\ref{ssSec:PropertiesMemoryModel} as a continuous rise of the background spacetime [Fig.~\ref{fig:WaveformWithMem}] appears in its third order differentiated form as a burst like signal about merger. 

\begin{figure}
    \centering
    \includegraphics[width=1\linewidth]{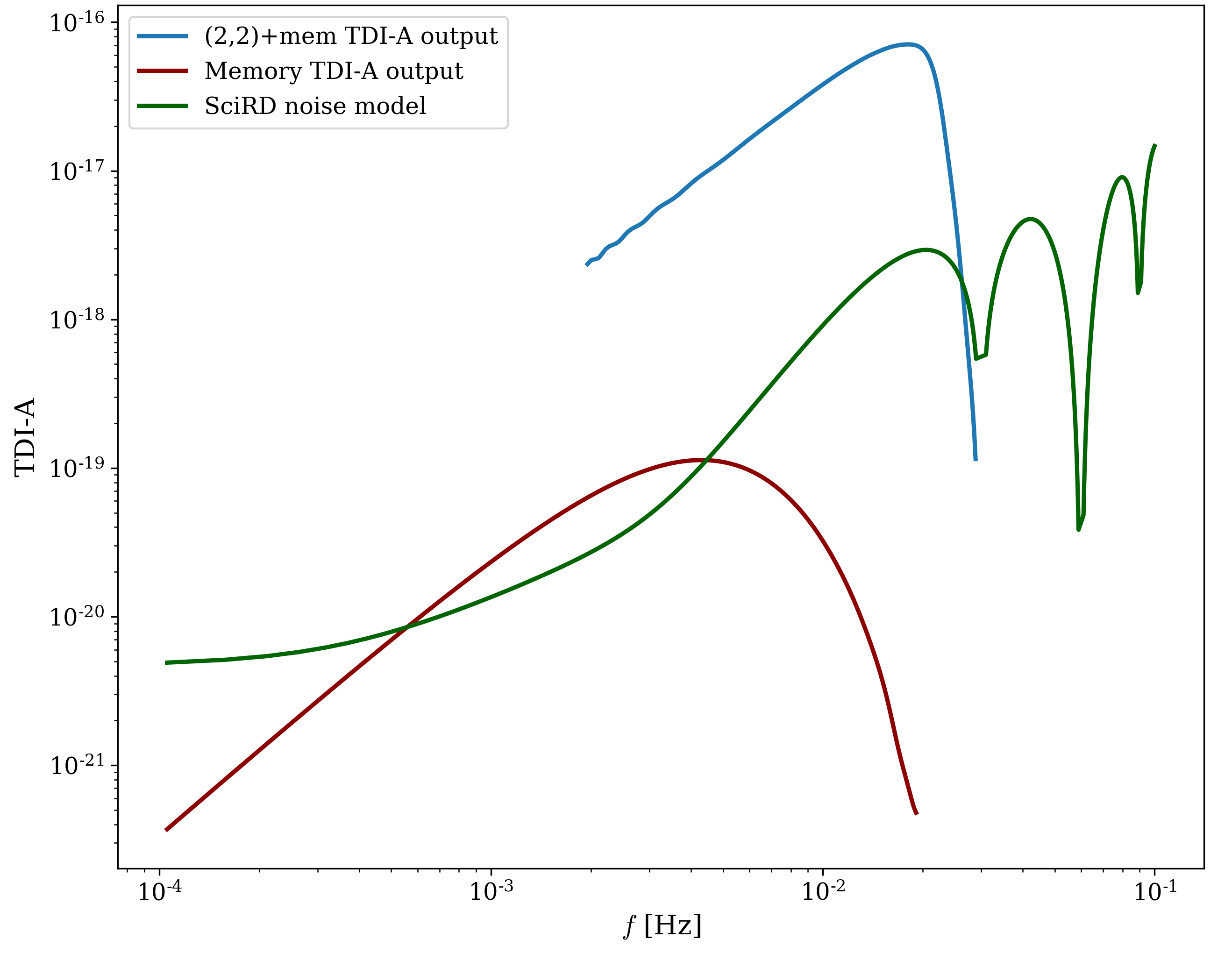}
    \caption{Frequency domain TDI-A channel obtained, after the response of the links, from Fig.~\ref{fig:WaveformWithMem}. The (2,2)+mem waveform is in blue and the memory component alone in red. The analytical PSD of the Science Requirements Document (SciRD)~\cite{LISA_SciRD} noise model was added in green, as a reference. This highlight the possible visibility of the memory component, here in the mHz region. The parameters are the same as in Fig.~\ref{fig:Comparing20Components}.}
    \label{fig:FreqTDIA}
\end{figure}

\paragraph{In-band mergers.}
As anticipated, the response of a detector to gravitational memory considerably alters the pure radiation signal from its initial shape in Fig.~\ref{fig:WaveformWithMem}. This is in particular relevant for events, in which both the oscillatory and the memory signal are in-band during merger, since compared to oscillatory high-frequency waves, gravitational memory is strongly attenuated in the TDI data streams. Indeed, within frequency space, the $1/f_L$ low frequency behavior of the gravitational memory signal within the radiation described in Fig.~\ref{fig:memwithfreqSPIN}, is converted into a $f_L^2$ dependence in the TDI output channels shown in Fig.~\ref{fig:FreqTDIA}. As a result, the memory contribution is effectively suppressed by a factor of $f_L^3 \ll f_H^3$ relative to the primary oscillatory signal.

Let us elaborate on this point. Ultimately, a judgment of detectability should not be made based on the signature of a particular signal within the response, but always needs a comparison to the detector noise. Indeed, the TDI filtering affects the noise in the same way and the LISA response could equivalently be described in terms of strain $[s]$ or phase-shift $[\varphi]$ units, Doppler frequency units $[y]$, or TDI streams $[X]$. What ultimately counts in the context of a matched filtering search strategy\footnote{Matched filtering is the optimal template-based technique for extracting coherent power from the data [see e.g. Ref.~\cite{maggiore2008gravitational}.} is the assessment of the SNR, reflecting the fact that noise is the main limiting factor for detectability.

Nevertheless, a characterization of the LISA TDI response is useful for an intuitive understanding of the relevance of the detector in measuring gravitational memory compared to oscillatory signals.
This is because the noise power spectral density (PSD) within the TDI data is rather well distributed over the entire frequency range, as shown in Fig.~\ref{fig:FreqTDIA}, with a tilt towards higher frequencies. Including the fact that as a very short burst like signal, the matched filtering of memory accumulates less coherent signal to noise compared to the oscillatory signals, it turns out that the visual comparison of amplitudes in terms of TDI channel outputs gives a much better intuitive estimate of detectability within the particular detector of LISA than a comparison at the level of the strain.
Moreover, it is important to note that no matter how the LISA detector output is represented, the strain, encoding the projection onto the frequency band limited detector, is not equivalent to the initial radiation. In particular, for memory, the low-frequency net-offset is simply not captured by the data.

Thus, in general, the impact of the modes within the radiation signal in Fig.~\ref{fig:WaveformWithMem} is very different from its amplitude, explaining the difficulty in detecting memory. From that perspective, the absence of a gravitational memory detection to date~\cite{Tiwari:2021gfl,Cheung:2024zow} is therefore not due to an intrinsic smallness of the memory effect itself, but rather a direct consequence of the design of frequency-band-limited detectors, which explicitly target oscillatory features at finite frequencies while inherently suppressing the slowly rising component characteristic of memory.
Furthermore, the discrepancy between the amplitude scale in the emitted radiation and the prominence of a signal in the detector SNR and parameter estimation has in particular two important consequences in the context of gravitational memory. Both highlight the necessity of a careful and unambiguous definition of a proper memory model, including its time-dependent rise property, as carried out in Sec.~\ref{Sec:MemoryTheory}. Although the discussion here is framed in the context of LISA, the same considerations apply to ground-based detectors.

\begin{figure*}
    \centering
    \includegraphics[width=1\linewidth]{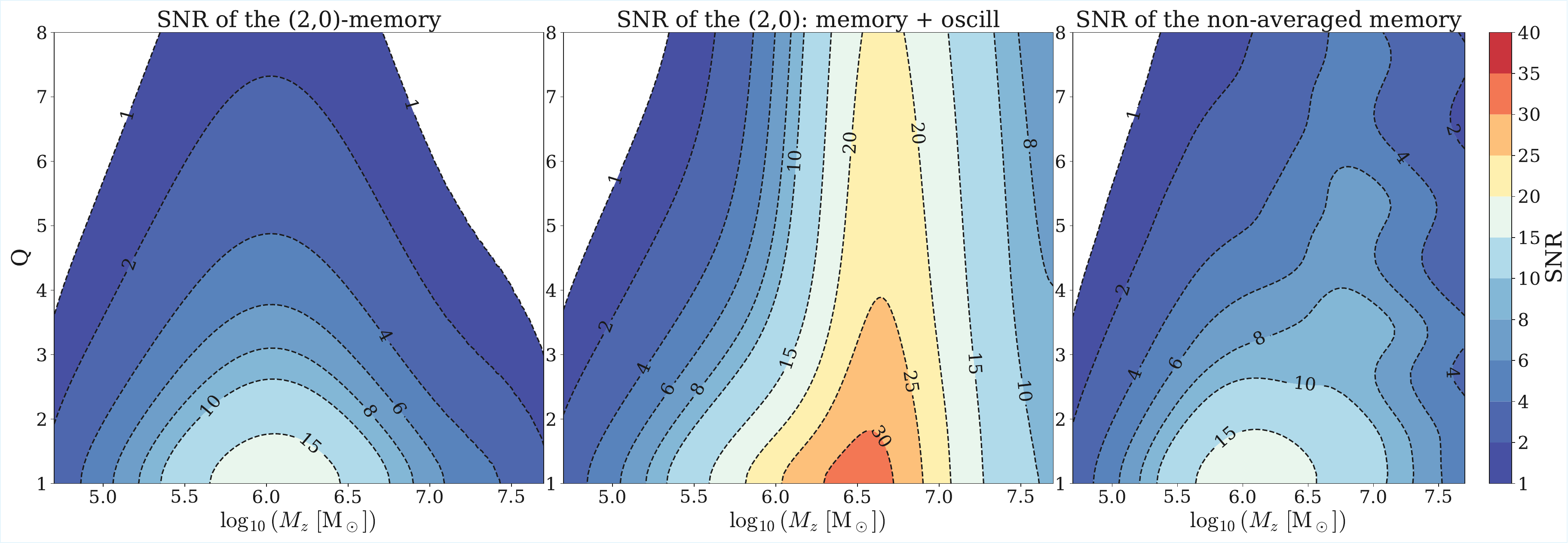}
    \caption{\small Memory SNR waterfall of the total binary  redshifted mass against the mass ratio for three different ``memory models'', showcasing the importance of a well-defined signal of the memory rise. \textit{Left:} Memory defined as in Eq.~\eqref{eq:memorymodes}, where the spacetime averaging selects out the $(2,0)$ mode. \textit{Middle:} Memory defined as the full $(2,0)$ mode within the waveform model $(b)$, including the oscillatory feature associated to the ringdown. \textit{Right:} Memory defined as Eq.~\eqref{eq:memorymodes}, including $m\neq0$ modes, without any averaging over high-frequency scales [Fig.~\ref{fig:MultiModeMemory}], which corresponds to a memory model defined as the null part of the BMS balance laws. \textit{Parameters:}  $\chi=0.4$, $\iota = \pi/3$, $d_l = 10^4 \textrm{Mpc}$, $\varphi_{ref} = 1$, $\psi = 0$, $\alpha = 0.74$, $\delta = 0.29$.}
    \label{fig:SNRMemOther}
\end{figure*}

The first concerns the oscillatory component that is generically present in the $(2,0)$ mode and is associated with the ringdown of the ordinary waveform discussed in Sec.~\ref{ssSec:PropertiesMemoryModel} above. In terms of the radiation modes shown in Fig.~\ref{fig:WaveformWithMem}, the memory contribution of a quasi-circular BBH merger appears to dominate the $(2,0)$ mode almost entirely. This makes it tempting to equate a detection of memory with a detection of the $(2,0)$ strain mode itself (see for instance Ref.~\cite{Rossello-Sastre:2025gtq}).  However, for the reasons discussed above, the oscillatory feature that is barely visible in Fig.~\ref{fig:WaveformWithMem} surpasses the memory in amplitude within the LISA response as shown in Fig.~\ref{fig:Comparing20Components}. As a consequence, this oscillatory feature has a significant impact on the computed SNR as we demonstrate in Fig.\ref{fig:SNRMemOther}. While a modification in SNR values across different masses is already present in the equal mass case, the importance of the oscillatory ringdown feature within the $(2,0)$ mode gains in significance as the mass ratio $q$ is increased. As shown in~\cite{Cogez:2025memoryLISA,Inchauspe:2024ibs}, this oscillatory contribution can moreover have a significant effect on parameter estimation.

The second consequence concerns additional modes entering the general expression for gravitational memory in Eq.~\eqref{eq:memorymodes}. As discussed above, in the absence of an explicit averaging over the high-frequency orbital timescale, the memory formula generically yields oscillatory contributions in modes with $m\neq 0$, as illustrated in Fig.~\ref{fig:MultiModeMemory}. In fact, as already mentioned in Sec.~\ref{ssSec:Practical Approximation}$c$, the null part of the BMS balance laws is equivalent to Eq.~\eqref{eq:memorymodes} without any spacetime averaging. While the amplitude of these oscillatory components is typically orders of magnitude smaller than the memory offset in the $(2,0)$ mode, their impact on the SNR is nevertheless non-negligible as also shown in Fig.~\ref{fig:SNRMemOther}, in particular as the mass ratio is increased. A proper memory model must therefore carefully isolate the genuinely non-oscillatory component of the radiation. This requirement becomes increasingly important once the non-precessing approximation is relaxed, in which case non-negligible gravitational memory can also arise in modes with $m\neq 0$.

\paragraph{Out-of-band mergers.}

\begin{figure}
    \centering
    \includegraphics[width=\linewidth]{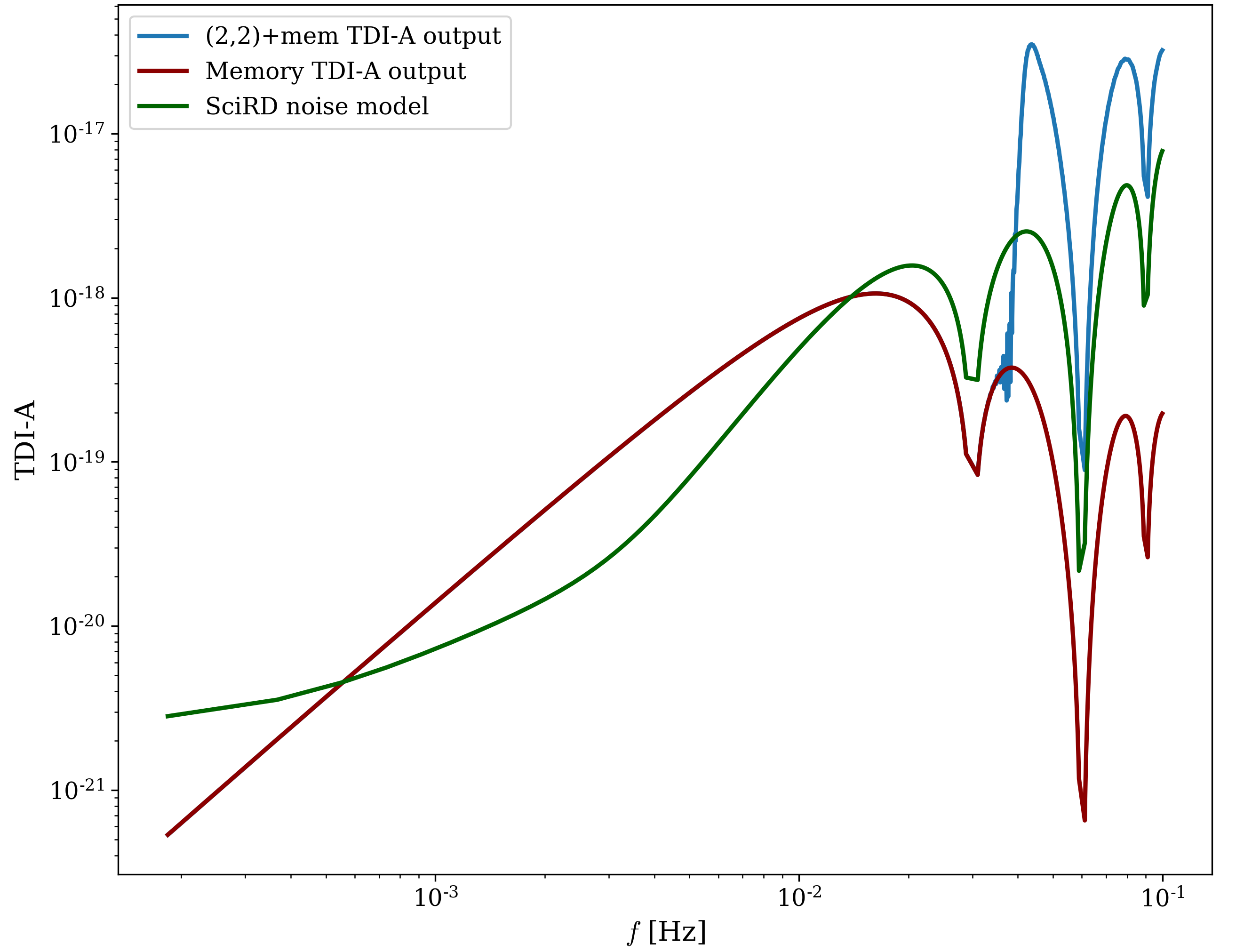}
    \caption{Example of the LISA frequency domain response for an out-of-band merger source. The inspiral at $t\simeq 1.10\,\mathrm{h}$ before merger (blue) and the memory signal (red) are shown. The SNR is $197$ in total, and $31$ for the memory alone. \textit{Parameters:} $\chi=0.7$, $\theta=\pi/2$, $d_l=10^2\,\mathrm{Mpc}$, $\varphi_{\rm ref}=0$, $\psi=0$, $\alpha=0.74$, $\delta=0.29$ and $M_z=10^4 \mathrm{M}_\odot$.}
    \label{fig:outofband}
\end{figure}

Before presenting our single-event memory detectability results, we discuss the LISA response to \emph{out-of-band} sources, i.e., binaries whose merger (and ringdown) frequencies lie above the detector's sensitive band. As a low frequency effect, gravitational memory may be observable even when the detector is effectively blind to the oscillatory merger signal. As discussed in Sec.~\ref{ssSec:PropertiesMemoryModel}b, in this regime, the separation of scales between the primary waveform and the memory contribution becomes particularly transparent, and in time domain, the signal is well approximated by a step-like profile whose parameterization involves only a few degrees of freedom~\cite{McNeill:2017uvq,Gasparotto:2025wok}. In this sense, out-of-band detections provide a particularly clean probe of the universal, non-oscillatory memory offset.

The frequency-domain memory signal for such an out-of band case is shown in Fig.~\ref{fig:outofband} for the merger of a binary with total mass $10^4\,\mathrm{M}_\odot$. At higher frequencies the spectrum is modulated by LISA's frequency-dependent response. At lower frequencies, $f\lesssim 10^{-2}\,\mathrm{Hz}$, 
the Fourier-domain signal of the memory effect exhibits the expected $f^2$ behaviour of the TDI output. While the merger of such an event would not be captured by LISA, the early inspiral portion is still in-band, information that can be used to estimate the expected memory signal together with its arrival time at merger. The inspiral that enters the LISA band at $t_{start}=-80000\,t/M$ ($1.10$ hours) prior to merger is explicitly shown in Fig.~\ref{fig:outofband}. For such an event, the observation of gravitational memory would not only be particularly clean, but would have great effect on the parameter estimation, as shown for instance in Refs~\cite{GasparottoGasparotto:2025adb, Gasparotto:2023fcg}. Yet, the detection of the memory burst becomes increasingly challenging as uncertainties in the measurement of the parent signal increase\footnote{A promising strategy in this case is \emph{multi-band} observation with detectors operating in complementary frequency ranges, e.g. memory in PTAs combined with merger/ringdown in LISA for SMBHBs, or LISA together with ground-based interferometers for intermediate- and stellar-mass binaries. Triggered searches for out-of-band memory in LISA using LIGO/Virgo detections have been proposed~\cite{Ghosh:2023rbe}. However, current population models suggest that individual detections are unlikely and may require stacking multiple events, with optimistic scenarios reaching detectability only over a $\sim6$-year LISA–Einstein Telescope mission~\cite{ET:2025xjr}.} and events generally needs to be too close to be realistically supported by current population models.

\subsection{Bayes factor analysis}\label{sSec:BayesFactor}

In the following, we want to assess the statistical significance required to claim a detection of gravitational memory with LISA, focusing on in band events. While previous SNR analysis~\cite{Inchauspe:2024ibs} provide a useful diagnostic for the strength of a signal compared to the assumed detector noise, it is not, by itself, a sufficient criterion for establishing a detection. A robust claim requires two ingredients: $(i)$ a coherent and physically well-motivated definition of the memory model, and $(ii)$ a statistical framework capable of quantifying how strongly the data favor the presence of memory over its absence. The Bayes factor provides exactly such a measure, comparing how likely the observed data are under a hypothesis that includes memory to a competing hypothesis that does not.

Ingredient $(i)$ implies that, to employ the Bayes factor meaningfully, a definition of waveform models that unambiguously isolate the memory contribution is essential. The LISA signal response attributed to memory must be explainable solely by the presence of a genuine memory offset, even though, as discussed above, LISA cannot directly measure its defining DC shift. The theoretical efforts in Sec.~\ref{Sec:MemoryTheory} aim precisely to equip the defined memory models with such a distinguishing characteristic. Based on the insight that gravitational memory is generated on radiation-reaction timescales and therefore occupies a distinct, lower-frequency regime than the oscillatory orbital and higher-harmonic content of the waveform, it is possible to attribute a unique time-dependent signature to gravitational memory that can be searched for in frequency-band-limited GW data. This separation between a low-frequency memory rise and the high-frequency oscillatory radiation is implemented via Isaacson averaging and ensures, for instance, that memory from quasi-circular BBH mergers is not confused with the oscillatory ringdown contribution to the $(2,0)$ mode, which is physically unrelated to the final DC memory component.

By combining a rigorously defined memory model with a full Bayesian model-selection analysis, we can therefore provide a statistically principled framework for evaluating whether LISA can make a definitive claim of detecting gravitational-wave memory. The companion paper~\cite{Cogez:2025memoryLISA} provides a detailed Bayesian analysis to determine the detectability of the memory effect in a single event using LISA. Here, we will provide a brief summary of the key findings regarding detectability.

\subsubsection{Definitions and summary of methodology}

To assess the detectability of gravitational memory for a given source, we perform Bayesian model selection by running two independent dynamical nested samplers~\cite{DynamicNestedSampling, Dynesty, Multinest, Skilling_2004, Skilling_2006, Speagle_2020}: one employing a waveform model that includes memory (mem) and one using an otherwise identical model without it (no mem). Thus, we compare two hypotheses that differ only by the presence of gravitational-wave memory as defined through the Isaacson model introduced in Sec.~\ref{Sec:MemoryTheory}.
In practice, we use the waveform type $(a)$ defined in Sec.~\ref{ssSec:Practical Approximation}. That is, we use waveforms that by construction do not contain the Isaacson memory rise memory as the hypothesis of radiation lacking memory, and construct the second hypothesis of memory-full radiation through a direct computation via the analytic memory formula in Eq.~\eqref{eq:memorymodes20Selection}, which is applicable to any quasi-circular binary black hole merger. Ensuring that both hypotheses employ the same oscillatory waveform model while differing only in the low-frequency Isaacson memory rise guarantees that modeling imperfections in the oscillatory signal do not influence the Bayes factor computation.
Concretely, we either focus solely on the dominant oscillatory mode $(2,2)$, or include several higher GW modes (HM), namely $(2,1)$, $(3,3)$, $(3,2)$, $(4,4)$, and $(4,3)$. Including higher modes improves accuracy at the cost of increased computational time. Within the Bayesian framework, the main waveform model used is \texttt{NRHybSur3dq8\_CCE}~\cite{GWSurrogate, Yoo:2023spi} --with both modes content--, along with the {\tt SEOBNRv5HM}~\cite{pySEOBNR, SEOBNRv5HM} model employed for validation in the HM case.\footnote{We do not use the $(2,0)$ mode provided by the \texttt{NRHybSur3dq8\_CCE} model, as it contains non-memory oscillatory contributions. Its full mode content corresponds to a type $(b)$ signal.}



Bayesian inference updates prior beliefs $p(\bm{\theta} | m)$ about parameters $\bm{\theta}$ of a model $m$ using data $d$ through Bayes' theorem
\begin{equation}
p(\bm{\theta} | d, m) = \frac{p(d | \bm{\theta}, m)\, p(\bm{\theta} | m)}{p(d | m)},
\end{equation}
where the denominator
\begin{equation}
\mathcal{Z}\equiv p(d | m) = \int p(d | \bm{\theta}, m)\, p(\bm{\theta} | m)\, d\bm{\theta},
\end{equation}
is the Bayesian evidence, or marginal likelihood. 
The evidence $\mathcal{Z}$ quantifies a model's overall predictive adequacy by averaging its likelihood $p(d \mid \bm{\theta}, m)$ over the full parameter space. This integral naturally penalizes unnecessary model complexity, because models that introduce large or weakly constrained parameter spaces dilute their average likelihood unless the data tightly support those additional degrees of freedom.

Each nested sampler yields the Bayesian evidence $\mathcal{Z}$ for its respective model. The preference for one model over the other is captured by the Bayes factor
\begin{equation}
\mathcal{B} = \frac{\mathcal{Z}_{\text{mem}}}{\mathcal{Z}_{\text{no~mem}}},
\end{equation}
which expresses how much more likely the data are under the memory model than under the no-memory model. For practical purposes, we work with the corresponding $\log_{10}$ Bayes factor. Following standard interpretive criteria provided by the Jeffreys scale~\cite{Jeffreys_1998}, we consider gravitational memory to be detected when the evidence for the memory model is decisive, i.e., when $\log_{10}\mathcal{B} \geqslant 2$.

In practice, the concrete set up for the computation of the evidence and the Bayes factor is as follows. Focusing on quasicicular binary black hole mergers, the parameters required to obtain a signal template, along with the range of values considered for these parameters, are summarised in Table~\ref{tab:Parameters}.\footnote{The ranges considered here are only indicative and do not always represent the prior used for a given computation. However, as long as the prior is not overly constrained for certain parameters and remains consistent across both models in the Bayesian analysis, the resulting Bayes factor will be computed correctly. More details are found in the companion paper~\cite{Cogez:2025memoryLISA}.}
\begin{table}[H]
 \begin{ruledtabular}
    \begin{tabular}{ccc}
        Parameter & Considered range & Reason \\
        \hline
        Mass ratio $Q$ & $[1, 8]$ & Waveform limitation \\[0.4ex]
        Spin(s) along z-axis & \multirow{2}{*}{$[-0.8, 0.8]$} & \multirow{2}{*}{Waveform limitation} \\[-0.4ex]
        $\chi_{z}$ & & \\[0.4ex]
        Redshifted mass $M_z$ & $[10^4, 10^8]$ M$_\odot$ & Roughly LISA band \\[0.4ex]
        Luminosity & \multirow{2}{*}{$[10^2, 10^6]$ Mpc} & Indicative range \\[-0.4ex]
        distance $d_\mathrm{L}$ & & only \\[0.4ex]
        Inclination $\iota$ & $[0, \pi]$ & Physical range \\[0.4ex]
        Polarization $\psi$ & $[0, \pi]$ & Physical range \\[0.4ex]
        Right ascension $\alpha$ & $[0, 2\pi]$ & Physical range \\[0.4ex]
        Declination $\delta$ & $[0, \pi]$ & Physical range \\
    \end{tabular}
    \caption{Summary of parameters and generic priors}
    \label{tab:Parameters}
 \end{ruledtabular}
\end{table}
Based on the two hypotheses outlined above, the evidence computation is then performed using the dynamical nested sampling method~\cite{DynamicNestedSampling}, with the {\tt Dynesty}~\cite{Dynesty, DynamicNestedSampling, Skilling_2004, Skilling_2006, Speagle_2020, Multinest} software.


\subsubsection{Main results for the claim of detection}

\begin{figure}
    \centering
    \includegraphics[width=1\linewidth]{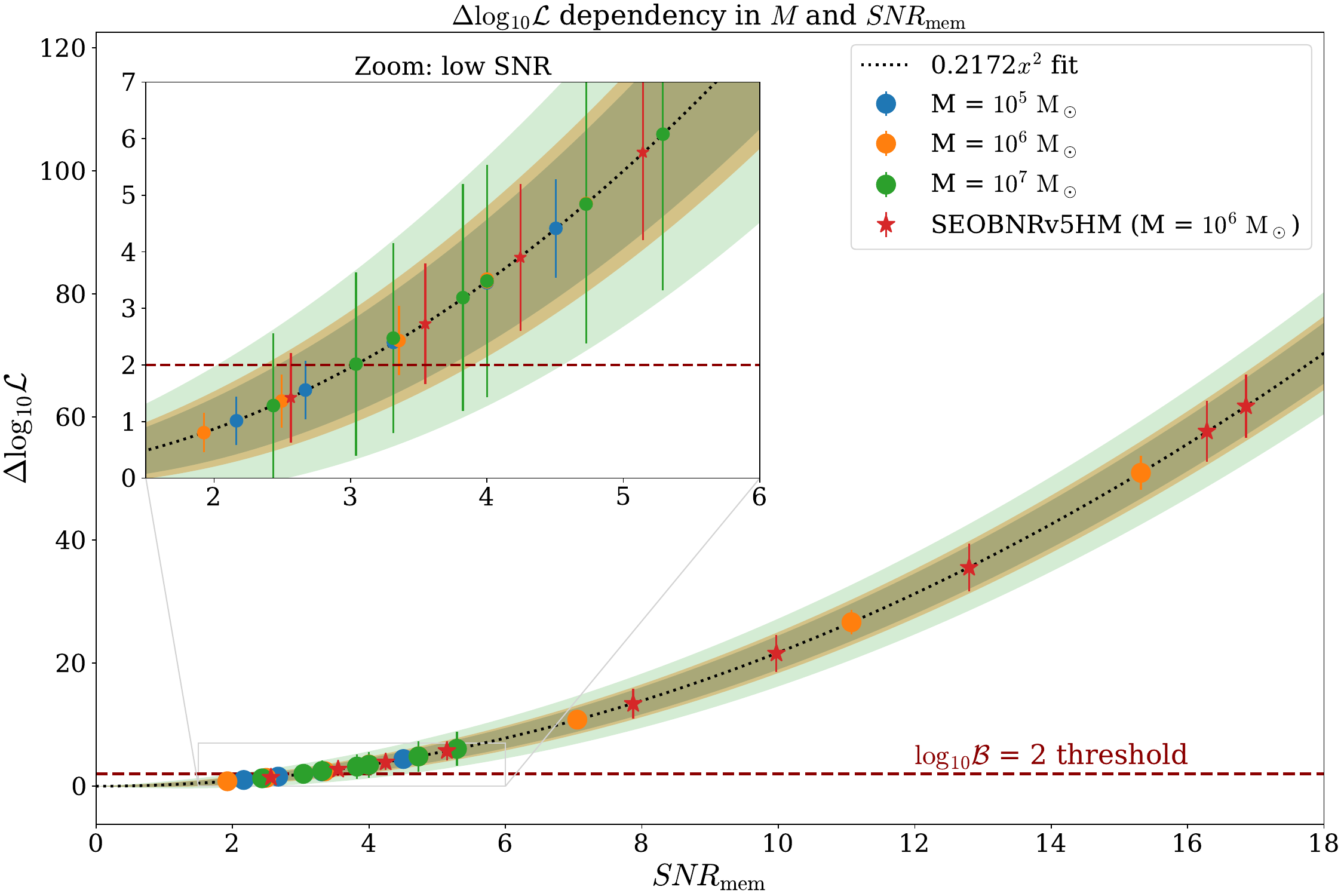}
    \caption{Mean and dispersion values of $\Delta \log_{10}\mathcal{L} \approx \log_{10}\mathcal{B}$ for various parameters sets. The black dotted line shows the fitted power law linking $\SNRmem$ and $\log_{10}\mathcal{B}$, and the red dashed line correspond to the $\log_{10}\mathcal{B} = 2$ threshold. Different total redshifted mass $M_z$ parameters are distinguished by different colors (blue for $M_z=10^5 \mathrm{M}_\odot$, orange for $M_z=10^6 \mathrm{M}_\odot$, and green for $M_z=10^7 \mathrm{M}_\odot$), both for computed points and the estimated dispersion, represented as colored areas. 
    The dot points, which are used to perform the fit, are computed using waveform model $(b)$ including only the (2,2)-mode and computing memory through Eq.~\eqref{eq:memorymodes20Selection}. The red star points correspond to additional runs computed using the {\tt SEOBNRv5HM} waveform, including higher SWSH modes to compute the memory contribution.}
    \label{fig:MeanAndSigmaValues}
\end{figure}

The main result of the Bayesian model selection analysis is the identification of a viable SNR threshold across the parameter space for which a full Bayes factor computation yields, with high probability, a justified claim of gravitational-memory detection. As shown in detail in the companion paper~\cite{Cogez:2025memoryLISA}, the signal-to-noise ratio of the memory component, $\SNRmem$, serves as a robust and largely parameter-independent detection metric, irrespective of the total SNR of the full waveform. Although $\SNRmem$ varies with intrinsic and extrinsic source parameters, such as mass ratio, spins, and inclination, its value alone provides a reliable predictor of the expected $\log_{10}\mathcal{B}$ for a given source. Fig.~\ref{fig:MeanAndSigmaValues} illustrates the power-law relationship between $\SNRmem$ and $\Delta \log_{10}\mathcal{L} \approx$ $\log_{10}\mathcal{B}$\footnote{Due to Bayes factor computation being time-expensive, the companion paper introduces an approximation to quickly estimate the value of $\log_{10}\mathcal{B}$ using the difference of $\log_{10}\mathcal{L(\bm{\theta}_{injection})}$ for the two models.}. This trend reveals a key operational conclusion: LISA should be capable of detecting gravitational memory whenever the expected memory SNR satisfies $\SNRmem \geqslant 3$, independent of the detailed source parameters. 

\begin{figure}
    \centering
    \includegraphics[width=1\linewidth]{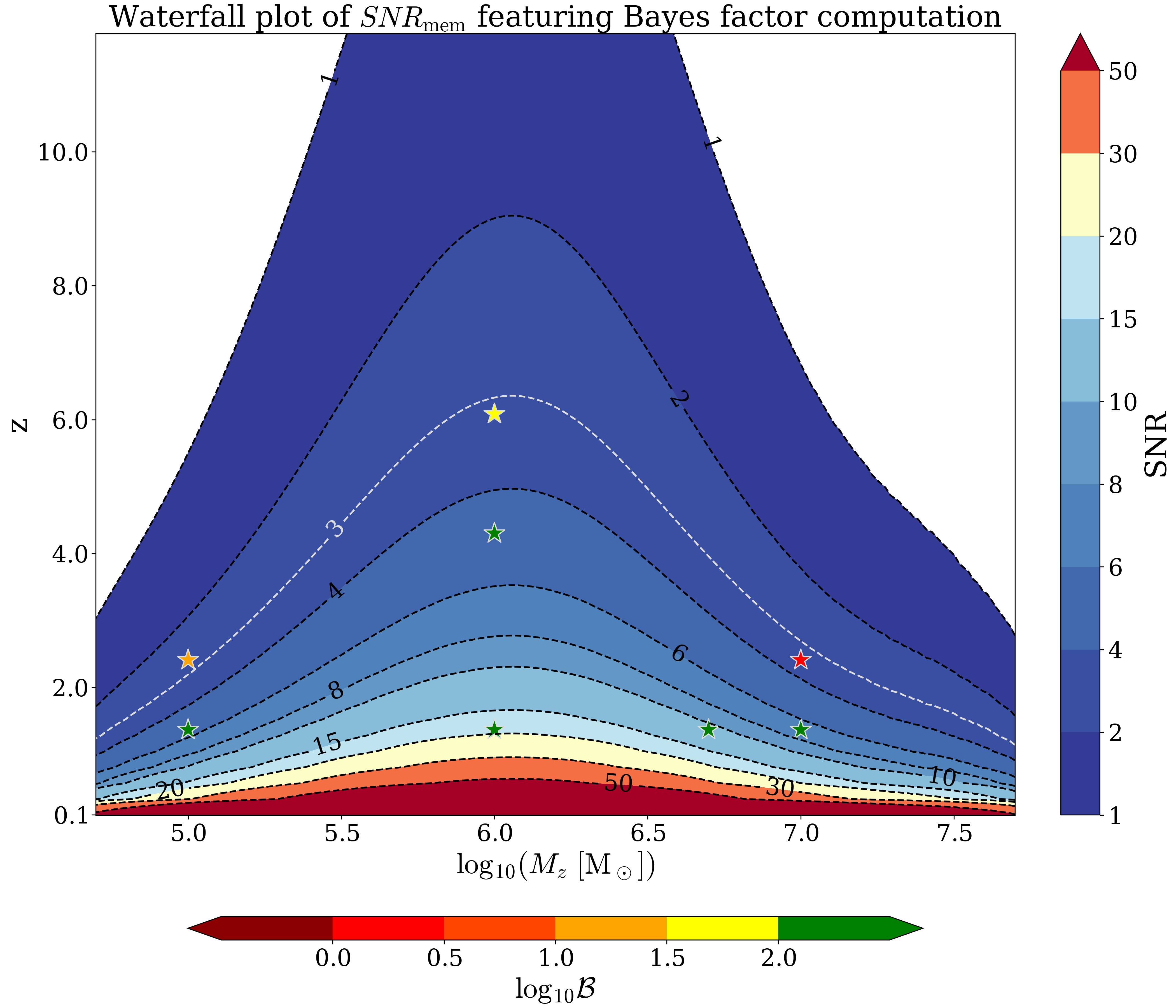}
    \caption{
    Memory waterfall plot with stars corresponding to the computed $\log_{10}$Bayes factor. 
    The color of the stars corresponds to the Jeffreys scale~\cite{Jeffreys_1998}, indicated by the colorbar under the figure. 
    The light gray dashed line represents the ISO-SNR contour $\SNRmem = 3$.
    This plot used the waveform model $(b)$, including only the (2,2)-mode and computing memory through Eq.~\eqref{eq:memorymodes20Selection}. \textit{Parameters:} $Q=1$, $\chi=0.4$, $\theta = \pi/3$, $\psi = 0$, $\alpha = 0.74$, $\delta = 0.29$.}
    \label{fig:MemoryWaterfallPlotWithBF}
\end{figure}

\begin{figure}
    \centering
    \includegraphics[width=1\linewidth]{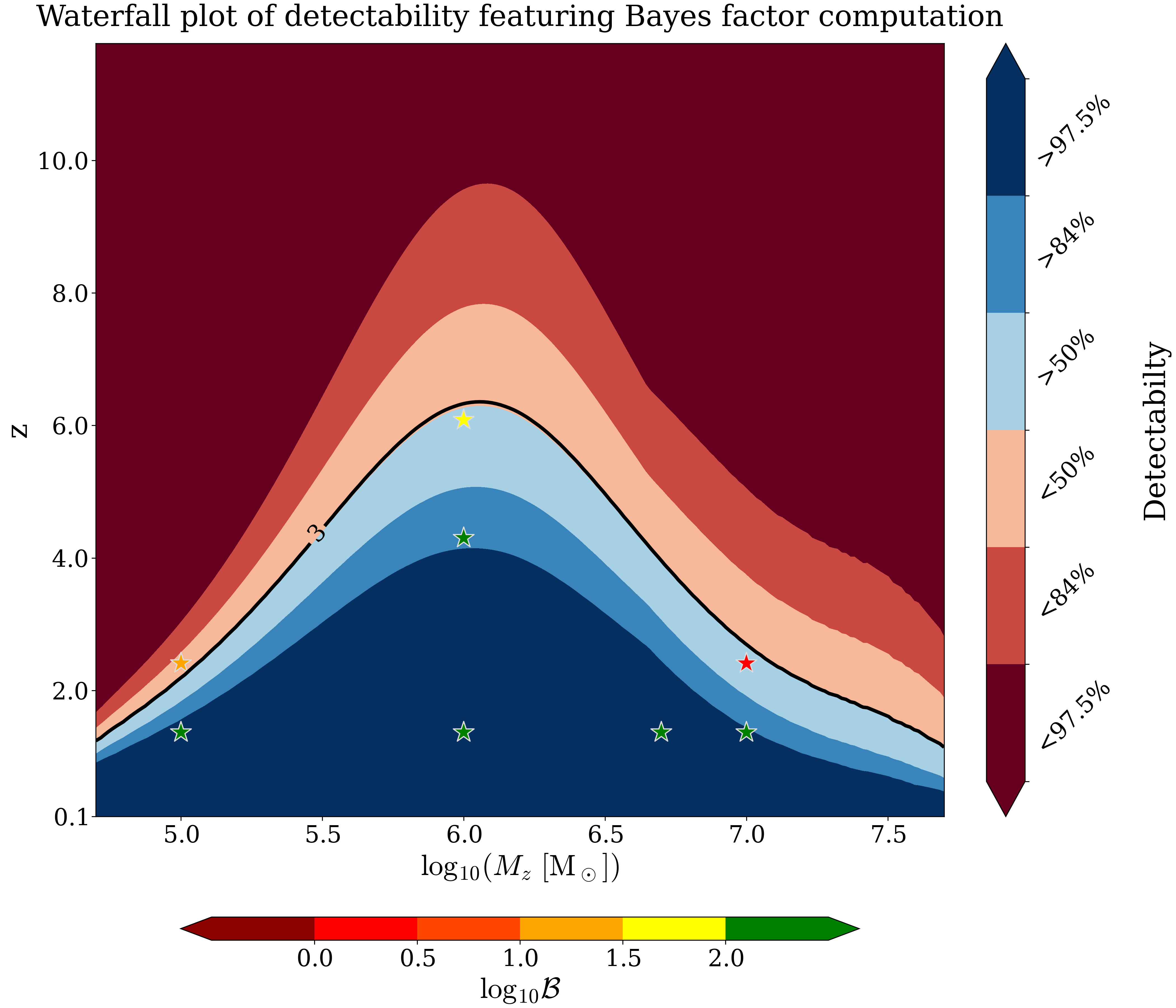}
    \caption{Conversion of the Fig.\ref{fig:MemoryWaterfallPlotWithBF} SNR waterfall plot into a detectability plot. The main colorbar (on the right) provides information on how likely we are to detect memory for a given set of parameters. The black line shows the $\SNRmem = 3$ threshold. We kept stars from the previous Bayes factor computations to compare with the prediction, using the same colorbar as in Fig.~\ref{fig:MemoryWaterfallPlotWithBF}.}
    \label{fig:DetectabilityMap}
\end{figure}

The Bayes factor computation, however, is subject to variations introduced by stochastic noise realizations in the data. Even for sources with identical physical parameters, different noise draws can shift the inferred $\log_{10}\mathcal{B}$ upward or downward relative to its expected value. This dispersion becomes especially relevant near the detection boundary, where modest fluctuations in the likelihood can alter the qualitative interpretation of the model comparison. Hence, a subset of sources with $\SNRmem \approx 3$ may occasionally yield Bayes factors below the decisive threshold purely due to unfavorable noise conditions. Accounting for this effect motivates the adoption of a more conservative criterion. Specifically, requiring $\SNRmem \geqslant 5$ ensures that the corresponding Bayes factor remains decisively in favor of the memory model across essentially all noise realizations, thereby providing a highly reliable and reproducible detection threshold. One can notice that the dispersion depends on the total redshifted mass $M_z$ of the source. This is explained by the connection between $M_z$ and the frequency range of the signal, depending on the mass the memory signal will be located in a more or less sensitive frequency range of LISA and therefore be more or less affected by the noise's dispersion. 

This result enables a systematic characterization of the region of the source-parameter space in which the memory signal becomes detectable. Fig.~\ref{fig:MemoryWaterfallPlotWithBF} provides an illustrative example in the form of an SNR waterfall plot, showing how $\SNRmem$ varies as a function of the total redshifted mass $M_z$ and redshift $z$. Superimposed on this plot are stars colored according to the Bayes factor associated with each parameter combination. Points indicated in green correspond to configurations for which a detection can be confidently claimed. Note that for simplicity, this plot only uses the (2,2) SWSH mode of the waveform model and its associated (2,0) memory component. Accounting for higher modes slightly enhance the values of $\SNRmem$ without modifying the relation between $\SNRmem$ and $\log_{10}\mathcal{B}$, as shown by the red stars on Fig.~\ref{fig:MeanAndSigmaValues}.
The power-law relationship between $\SNRmem$ and $\log_{10}\mathcal{B}$ then allows a conversion of the SNR computation into an estimation of the probability of distinguishing the memory signal across the parameter space. Concretely, the detectability map presented in Fig.~\ref{fig:DetectabilityMap} reformulates the information of Fig.~\ref{fig:MemoryWaterfallPlotWithBF} into a direct assessment of detection probability.

\begin{figure*}
    \centering
    \includegraphics[width=1\linewidth]{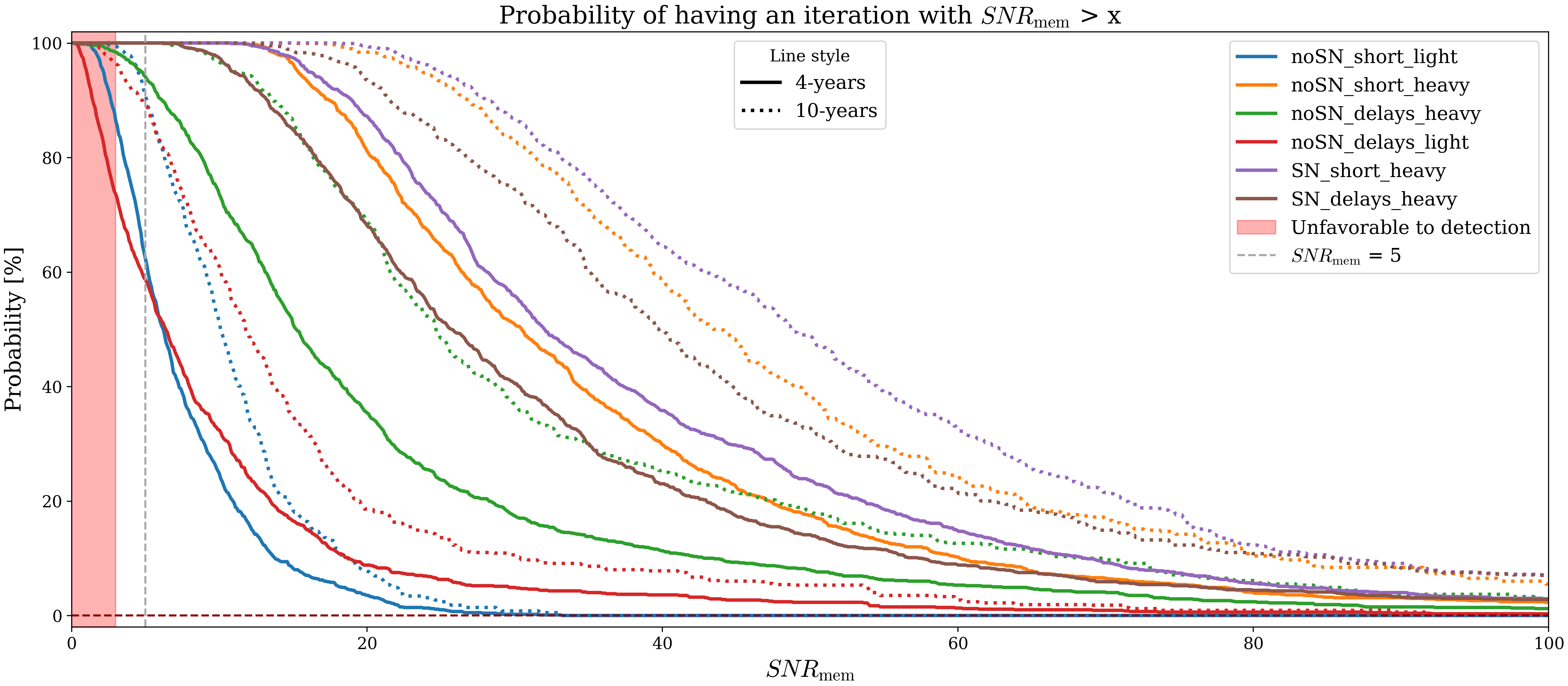}
    \caption{Probability of having an iteration with $\SNRmem$ greater than a given value (x-axis). Each color correspond to a population model in Ref.~\cite{Barausse_2020, Barausse_Lapi_2021}. Solid lines corresponds to 4-years iterations and dotted lines to 10-years. The red area covers the region where we are under the threshold $\SNRmem^{\textrm{thresh}} = 3$. The gray dashed line shows the value $\SNRmem = 5$ over which memory should be always detected. The $\SNRmem$ computed here include higher mode contribution.}
    \label{fig:Barausse_10yrs_ProbaOfMax}
\end{figure*}

\subsubsection{Expected memory events from MBHB}

As previously noted, one of the primary target sources for LISA are massive black hole binaries (MBHB), which are expected to produce the loudest GW signals of the mission~\cite{LISA:2024hlh}.
We therefore apply the considerations developed above to the latest MBHB population models of Barausse et al.~\cite{Barausse_2020,Barausse_Lapi_2021} and update previous forecasts for detection prospects~\cite{Gasparotto:2023fcg,Inchauspe:2024ibs}.
These population models comprise eight distinct scenarios, each corresponding to different choices for the main astrophysical uncertainties governing the evolution of massive black holes and their binaries, from their initial seeding to their growth via accretion, binary formation and mergers.
In particular, the models differ in the assumed nature of the initial black hole seeds, originating either from the collapse of first-generation stars (``light'') or from the direct collapse of protogalactic gas clouds (``heavy'').
They further account for uncertainties in the expected delay between galaxy mergers, which may be either prolonged (``delays'') or negligible (``short'') due to dynamical processes at parsec scales, as well as for the impact of supernova feedback (``SN'') or its absence (``noSN''), which can suppress accretion through the expulsion of gas from central regions and thereby affect massive black hole growth and merger rates. Detailed distributions for each population models are available in the companion paper~\cite{Cogez:2025memoryLISA}. 

Fig.~\ref{fig:Barausse_10yrs_ProbaOfMax} shows, for each population model, the probability of observing at least one event with a given $\SNRmem$ over 4 years (the nominal LISA mission lifetime) or 10 years (the maximal mission extension). All results are computed using the full multipolar content of the waveform, which maximizes sensitivity to the memory contribution.
For a 10-year LISA duration, nearly all scenarios shown in the figure are likely to produce at least one confident memory detection, defined as $\SNRmem \geqslant 5$. One can even hope for rather loud memory events, offering an opportunity to test the General Relativity. 
The only exceptions are the light-seed models including supernova feedback (SN), which yield no events detectable in the LISA band. 

The hierarchy of detection prospects in Fig.~\ref{fig:Barausse_10yrs_ProbaOfMax} is driven by the combined impact of seed mass, supernova feedback, and dynamical delays on both merger rates and merger frequency range. Short-delay heavy-seed scenarios yield the highest probabilities of memory detection, as massive seeds form binaries efficiently and merge in a mass and frequency range optimally matched to the LISA sensitivity band, largely independent of SN feedback. When delays are included, merger rates in heavy-seed models are reduced, as prolonged dynamical evolution suppresses the number of coalescences. However, in this case SN feedback has a noticeable impact by limiting early mass growth and spin-up, thereby reducing recoil kicks that would otherwise eject merger remnants from galactic centers. This explains why SN-delays-heavy outperforms noSN-delays-heavy scenarios. Light-seed models on the other hand are generally disfavored. Especially including SN feedback inhibits black hole growth, leading to lower-mass binaries that merge at higher frequencies where LISA is less sensitive.

\section{Conclusion}\label{Sec:Conclusion}

In this work, we have developed a principled theoretical framework for defining conservative gravitational memory models that can be used for observations with frequency-band-limited gravitational-wave detectors, with a particular focus on the future LISA mission. Our central goal has been to clarify what constitutes a physically meaningful and observationally relevant signature of gravitational memory in the absence of direct access to the defining memory offset, and thereby to provide the necessary conceptual and mathematical tools to support a future claim of detection. Such an observation would open a direct observational window onto the universal infrared structure of field theory, with direct links to questions in quantum gravity and physics beyond general relativity~\cite{Zosso:2024xgy,Weinberg_PhysRev.140.B516,Bondi:1962px,Sachs:1962wk,FrauendienerJ,Ashtekar:2014zsa,Compere:2019gft,DAmbrosio:2022clk,Mitman:2024uss,DeLuca:2024cjl,Strominger:2014pwa,Strominger:2017zoo,Goncharov:2023woe,Heisenberg:2023prj,Heisenberg:2024cjk,Heisenberg:2025tfh,Heisenberg:2025roe}.

In the case of ordinary memory associated with unbound matter, including linear memory from massive particles and null memory from electromagnetic radiation, a direct association between radiation and the resulting memory exists in principle. This is \textit{a priori} not the case for nonlinear memory sourced by gravitational waves themselves.
Concretely, the issue can be illustrated as follows. Consider the total radiation signal from a binary black hole merger, for instance the blue curve in Fig.~\ref{fig:WaveformTypes}, which induces a permanent shift in the proper distance between freely falling test masses. How can one claim the presence of this permanent offset in the response of a frequency-band-limited gravitational-wave detector? Since the detector is only sensitive to the transition between the initial and final values of the offset, it is necessary to define a time-dependent signal associated with this transition. While, in principle, a likelihood-based comparison naturally accounts for the detector response, a meaningful interpretation of such an analysis requires a physically motivated model of the memory rise that is distinguishable from purely oscillatory radiation and whose statistical significance directly reflects the underlying memory offset. Any viable definition must therefore separate the contribution responsible for the final memory offset from the oscillatory waveform while reproducing the correct asymptotic behavior.

The proposed solution is to introduce, as additional physical input, the knowledge of the characteristic frequencies of the emitted gravitational waves and to average the signal over these scales. Under the assumption that the evolution of these high frequencies is sufficiently slow, the gravitational memory rise can be identified with a smooth background change which is clearly distinct from the oscillatory signal.
This procedure is supported by the theoretical understanding of gravitational memory as arising from the backreaction of accelerated, unbound energy-momentum. In the case of memory sourced by gravitational waves, the definition of their energy-momentum inherently requires an assumed separation of scales, reflecting the fundamental nonlocality of gravitational energy-momentum. This naturally induces a corresponding separation between memory and what is identified as gravitational waves.

We, therefore, bring forward the Isaacson approach to gravitational-wave energy-momentum backreaction as a natural theoretical framework for implementing this physical input. Within this framework, the Isaacson averaging procedure provides the natural ingredient for defining a physically meaningful memory rise by explicitly separating the low-frequency memory component from the high-frequency gravitational waves in successive time windows of continuously evolving emission frequencies. While this separation is guaranteed in idealized scenarios, it becomes increasingly challenged as the source complexity grows. In particular, when multiple or strongly time-varying emission frequencies are present, maintaining a clearly identifiable separation between the gravitational-wave and memory scales may no longer be straightforward.

We have shown, however, that in the case of quasi-circular binary black hole mergers, the Isaacson-based memory model remains viable and that the required physical separation of scales is guaranteed. Through a secure separation of scales in time-frequency space, it becomes possible to understand how numerical-relativity-based waveform models faithfully capture the multimode content of emitted gravitational waves while entirely disregarding any memory contribution, thereby implicitly performing the crucial task of separating a purely oscillatory signal from the full radiation. Moreover, the Isaacson approach provides clean theoretical arguments for the standard practice of defining memory as the monotonically rising signal in the $(2,0)$ spin-weighted spherical-harmonic mode.

In this context, gravitational memory, although arising from a nonlinear backreaction, is understood as the leading-order low-frequency contribution to the radiation. It is instructive to compare this with other nonlinear predictions of GR. For instance, during the inspiral, the post-Newtonian expansion encodes a hierarchy of nonlinear interactions, including hereditary tail and ``tails-of-memory'' terms, which characterize the propagation of GWs on curved backgrounds~\cite{Blanchet:1992br}. While these PN corrections describe field self-interactions, their signatures can be degenerate with environmental effects, such as interactions with surrounding matter~\cite{Alnasheet:2025mtr}. Furthermore, as hereditary corrections to the oscillatory radiation that decay once emission ceases, they remain subdominant at high frequencies and are parametrically suppressed at low frequencies relative to the memory signal, making them difficult to isolate. By contrast, the characteristic time–frequency structure of the memory signal is clearly distinguishable from the dominant oscillatory features, as shown in Fig.~\ref{fig:mem_spec}. Another nonlinear signature arises in the ringdown, where mode–mode couplings generate second-order quasinormal modes~\cite{Lagos:2022otp}. These modify amplitudes and introduce overtones beyond the linear regime, which may dominate the high-frequency spectrum after merger~\cite{Cheung:2022rbm}. However, their short-lived nature and entanglement with merger dynamics make them intrinsically difficult to extract in a model-independent way. Moreover, the nonlinear ringdown examines strong-field, near-horizon dynamics, intrinsically different to the unique measure of the emission of radiative energy-momentum of the memory effect. Ultimately, these effects offer complementary windows into GR's nonlinear dynamics across the inspiral, merger, and ringdown phases, providing multiple avenues to test the theory where linearized predictions are insufficient.

We have also clarified how residual oscillatory components in a putative memory model can contaminate the inferred signal and artificially bias SNR estimates. This insight is essential for constructing reliable detection strategies and avoiding false positives driven by model-dependent artifacts. In particular, it is important to be aware of the non-negligible impact of small oscillatory signatures within the $(2,0)$ memory mode of quasi-circular binary black holes, which are associated with the ringdown. At the same time, we have explicitly demonstrated the danger of neglecting averaging over high-frequency scales when computing nonlinear memory. In particular, we have emphasized that while the BMS balance laws correctly encode the total radiated null energy, they do not by themselves provide a prescription for defining a local, instantaneous energy flux and hence a unique time-domain memory signal. This limitation underscores that the balance laws are not, by themselves, a tool for discriminating between wave-like and memory-like time-dependent components of the total radiation.

The clean definition of observable memory models developed in this work also enables systematic extensions beyond the quasi-circular approximation. Robust modeling of gravitational memory across the full parameter space will require generalizations to precessing and eccentric sources~\cite{Schmidt:2010it,Khan:2018fmp,Yu:2023lml,Hamilton:2023qkv,Thompson:2023ase,Arredondo:2024nsl,Hamilton:2025xru,Gupta:2024gun,Favata:2011qi,Talbot:2018sgr}. In these cases, a consistent multiscale treatment, including high-frequency averaging to isolate genuine memory growth, will be even more critical, as additional spin-weighted spherical-harmonic memory modes are generically excited. Isolating genuine memory growth in these cases is therefore expected to require an explicit implementation of the high-frequency averaging procedure. Addressing these challenges is left for future work.

In the present study, the theoretical results were directly applied to quasi-circular massive black hole binary mergers expected to be observed by LISA, for which the memory amplitude around merger can become comparable in scale to other radiative contributions. Building on the validated construction of two well-defined hypotheses of gravitational radiation with and without memory, we update previous SNR-based assessments with a statistically meaningful quantification of detection prospects via Bayes factor computations. We elaborate on this analysis in a companion paper~\cite{Cogez:2025memoryLISA}, which performs a Bayesian analysis of LISA's response to memory-free and memory-full radiation. These results indicate that LISA has strong potential to achieve the first single-event detection of gravitational memory, particularly for mergers of heavy-seed massive black hole binaries.

Finally, we highlight the importance of complementary validation strategies beyond matched filtering and Bayes factor assessments. While matched filtering provides optimal sensitivity under well-controlled noise assumptions, and the Bayes factor provides a statistically rigorous quantification of a detection claim, a convincing observation of gravitational memory must also account for non-stationary noise, or unforeseen physical effects. Specifically, a robust detection claim should be supported by a false alarm rate statistic from a frequentist perspective. In this context, it is instructive to compare to the current search for lensing of gravitational-wave~\cite{Li:2018prc,Grespan:2023cpa,Hannuksela:2025uol}, another solid prediction of general relativity that remains undetected. Although recent events have yielded log-Bayes factors exceeding the threshold of two~\cite{Goyal:2025eqo}, the community remains cautious in claiming a definitive detection~\cite{Keitel:2024brp,LIGOScientific:2025cwb,Chan:2025kyu,Shan:2025dcd}.

A primary difference between the search for lensed GW events and gravitational memory is, however, that while lensing is statistically expected in only a small fraction of events, determined by the presence of suitable lenses along the line of sight, general relativity predicts that nonlinear memory is present in every event of GW emission, with characteristics entirely determined by the source properties. At the same time, the distinctive features of gravitational memory identified in this work, most notably its specific timing relative to the oscillatory signal and its confinement to a low-frequency band, provide powerful consistency checks to rule out instrumental glitches or parameter uncertainties. In particular, the separation between the oscillatory chirp and the low-frequency memory signal can be understood as a concrete realization of the Isaacson separation of scales, allowing for a physically motivated disentangling of these components in the data. We therefore suggest leveraging this characteristic information within informed, unmodeled searches using existing algorithms~\cite{Cornish:2014kda,Cornish:2020dwh,Robson:2018jly,Gupta:2023jrn,Ebersold:2020zah,Drago:2020kic} to verify the presence of excess power at the expected time and frequency around merger, thereby significantly suppressing the false-alarm probability. A detailed quantitative assessment of these complementary strategies is left for future work. Moreover, it will be essential to generalize and stress-test memory detection pipelines within a full LISA global fit that accounts for all simultaneous GW signals and stochastic foregrounds~\cite{Robson:2017ayy,Littenberg:2023xpl,Katz:2024oqg,Strub:2024kbe,Johnson:2025oyu}.

\begin{acknowledgments}
JZ is supported by funding from the Swiss National Science Foundation
(Grant No. 222346) and the Janggen-Pöhn-Foundation. LMZ is supported through Research Grants No. VIL37766 and No. VIL53101 from Villum Fonden and the DNRF Chair Program Grant No. DNRF162 by the Danish National Research Foundation. The Center of Gravity is a Center of Excellence funded by the Danish National Research Foundation under Grant No. 184. AC acknowledges financial support from the Commissariat à l'Énergie Atomique (CEA) and from the Centre national d’études spatiales (CNES), within the framework of the LISA mission. HI thanks the Belgian Federal Science Policy Office (BELSPO) for the provision of financial support in the framework of the PRODEX Programme of the European Space Agency (ESA) under contract number PEA4000144253.
\end{acknowledgments}

\appendix

\section{Details on BMS balance laws}\label{App:BMSFluxBalanceLaws}

In Sec.~\ref{ssSec:ProofofMemory} we introduce the notion of asymptotic flatness, BMS symmetries, and the associated BMS balance laws. In this appendix we will present useful technical details and mathematical formulas in dealing with BMS balance laws omitted in the main text.

For convenience, we recall here that the balance laws can be written as a consistency relation of the asymptotic tensor radiation $h^{TT}_{ij}$ of GR [Eq.~\eqref{eq:BMSBalanceLawI MT}]
\begin{align}\label{eq:BMSSupermomentumBalanceLaw MT App}
   \int_{u_0}^{u} &du' \int d\Omega\,\alpha\,\mathcal{D}^i\mathcal{D}^j\dot h_{ij}^{TT}(u)=\\
   &=\frac{16\pi G}{r}\left\{ \int_{u_0}^{u}du'\int d\Omega\,\alpha\, \left[F_h+F_\text{m}\right]-\Delta \mathcal Q_{(\alpha)}(u)\right\}\,, \nonumber
\end{align}
where $\mathcal D_i$ defines the embedding of the covariant derivative on the 2-sphere.\footnote{We define $\mathcal D^i \equiv \mathcal D^A e_A^{\;i}$, where $\mathcal D_A$ is covariant derivative on the unit 2-sphere with capital Latin letters denoting indices on the 2-sphere $x^A=\{\theta,\phi\}$. Moreover, $e_A^{\;i}=\partial n^i/\partial x^A$ are the tangent vectors associated with the embedding of the sphere into the three-dimensional space orthogonal to the radial direction $n^i$. By construction $n_i e_A^{\;i}=0$, so that $\mathcal D^i$ acts intrinsically on transverse fields. For any transverse vector $v_i^T$ with $n^i v_i^T=0$, one has $\mathcal D^A v_A=\mathcal D^i v_i^T$, where $v_A=e_A^{\;i}v_i^T$.}
Here, the integral over $F_h$ defines the total flux of gravitational radiation [Eq.~\eqref{eq:GravitationaEF}]
\begin{equation}\label{eq:GravitationaEF App}
    F_h(u,\theta,\phi)=\frac{r^2}{32\pi G} \dot h^{TT}_{ij}\dot h_{TT}^{ij}\,,
\end{equation}
while $F_\text{m}$ capture the potential presence of matter radiation. Moreover, $\Delta \mathcal{Q}_{(\alpha)}$ denotes a change in supermomentum charge [Eq.~\eqref{eq:Supermomentum Charge MT}]
\begin{equation}\label{eq:Supermomentum Charge MT App}
    \mathcal{Q}_{(\alpha)}(u)=\frac{1}{4\pi G}\int_{S^2_u}d\Omega\,\alpha(\theta,\phi)\,M(u,\theta,\phi)\,,
\end{equation}
defined via the Bondi mass aspect $M(u,\theta,\phi)$. For simplicity of presentation, we will from now on neglect any presence of massless matter fluxes and set $F_\text{m}=0$.

First of all, note that 
in terms of the spin-weighted scalar of the strain defined in Eq.~\eqref{eq:SpinweightedScalar}\footnote{In the asymptotic limit, this quantity may be associated to the notion of a \emph{shear} $\sigma(u,\theta,\phi)$. The conversion to the commonly used definition of the shear is given by $h=2\bar \sigma$.}
\begin{equation}\label{eq:RadiationScalar}
    h(u,\theta,\phi)\equiv h_+-i h_\times\,,
\end{equation}
the BMS balance laws can be written as
\begin{align}\label{eq:BMSSupermomentumBalanceLaw MT scalar App}
    \int d^2\Omega\,&\alpha\,\Re[\eth^2\Delta h(u)]=\\
    &=\frac{1}{r}\left\{ \int_{u_0}^{u}du'\int d\Omega\,\alpha r^2\, |\dot h|^2-16\pi G\Delta \mathcal Q(u)\right\}\,. \nonumber
\end{align}
Here, we used that
\begin{equation}
    \dot h^{TT}_{ij}\dot h_{TT}^{ij} = 2|\dot h|^2\,,
\end{equation}
and 
\begin{equation}
    \mathcal D^i\mathcal D^j h^{TT}_{ij} = \frac{1}{2}\left(\eth^2 h+\bar{\eth}^2\bar{h}\right)=\Re[\eth^2 h]\,.
\end{equation}
In this last equality, we used the notion of angular derivative operator that we define in Eq.~\eqref{eth}. It is an operator that is used in the definition of spin-weighted spherical harmonics that raises and lowers the spin-weight of functions on the sphere, which is very useful in dealing with derivatives on the 2-sphere. Note that therefore the quantity $\eth^2\dot{h}$ is of spin-weight $0$, as it should. We further carried out its trivial $u'$ integral, where $\Delta{h}(u)\equiv h(u)-h(u_0)$. In the following, we will explicitly present commonly used reformulations of the BMS balance laws, which are especially useful in the context of gravitational memory.

\subsection{Spin-weighted expansion of memory}\label{sApp:SWSH expansion of memory in balance law}

A useful way to isolated the total memory offset $\Delta{h}$ on the left-hand-side of Eq.~\eqref{eq:BMSSupermomentumBalanceLaw MT scalar App} is by expanding the spin-weight $s=-2$ scalar $h$ as [Eq.~\eqref{eq:MemoryPhasor}]
\begin{equation}
   h(u,\theta,\phi)=\sum_{\ell'=2}^\infty \sum_{m'=-\ell}^\ell h_{\ell' m'}(u)\,_{\scriptscriptstyle-2}Y_{\ell' m'}(\theta,\phi)\,.
\end{equation}
Indeed, using the identity 
\begin{equation}
    \eth^2\phantom{}_{\scriptscriptstyle-2}Y_{\ell' m'}=\sqrt{\frac{(\ell'+2)!}{(\ell'-2)!}}\,Y_{\ell' m'}\,,
\end{equation}
one can analytically integrate the angular integral on the left hand side after parameterizing the supertranslations in terms of spherical harmonics $\alpha(\theta,\phi)=Y^*_{\ell m}(\theta,\phi)$.
We arrive at
\begin{widetext}
\begin{align}\label{BMSSupermomentumBalanceLawBD3 App}
\begin{split}
  \frac{1}{2}\big(\Delta h_{\ell m}(u)+(-1)^m&\Delta h^*_{\ell-m}(u)\big)=\sqrt{\frac{(\ell-2)!}{(\ell+2)!}}\frac{1}{r}\left[\int d\Omega\,Y^*_{\ell m} \int_{-u_0}^{u} d u'r^2\,|\dot{h}|^2-16\pi G\Delta\mathcal{Q}_{(Y^*_{\ell m})}(u)\right]\,.
\end{split}
\end{align}
\end{widetext}
The expression on the left hand side is not just any combination, but precisely corresponds to the to the electric-parity moment of the spin-weighted multipole (see for instance appendix D.1 in~\cite{Zosso:2024xgy})\footnote{Compared to standard notation, we have here defined $ h^E_{lm}=U_{\ell m}/\sqrt{2}$.}
\begin{align}
    h^E_{\ell m}&=\left[h_{\ell m}+(-1)^m \,h^*_{\ell -m}\right]\,,\label{eq:AUVlmToHlm}
\end{align}
such that we can write
\begin{align}\label{BMSSupermomentumBalanceLawBD4 App}
  \Delta &h^E_{\ell m}(u)= \sqrt{\frac{(\ell-2)!}{(\ell+2)!}}\times\\
  &\frac{1}{r}\left[\int_{S^2} d\Omega\,Y^*_{\ell m} \int_{-u_0}^{u} d u'r^2\,|\dot{h}|^2-16\pi G\Delta\mathcal{Q}_{(Y^*_{\ell m})}(u)\right]\,. \nonumber
\end{align}

As discussed in Sec.~\ref{ssSec:ProofofMemory} in the formal limit of $u\rightarrow\infty$ and $u_0\rightarrow-\infty$ the left-hand-side of this expression represents the total displacement memory offset within the electric-parity part of the strain of a given radiation event $\Delta h^E_{\ell m}$, whose value can be split on the right-hand side into a nonlinear null contribution of an integral representing the total gravitational energy flux and an ordinary component associated to a change in supermomentum charge. This implies that all gravitational radiation does not only contain a oscillatory gravitational waves, but always also includes a non-wave like memory component that leaves its traces on the asymptotic background spacetime geometry. 

As we will understand below, there is no displacement memory contribution in the associated magnetic part $\Delta h^M_{\ell m}=0$, such that Eq.~\eqref{BMSSupermomentumBalanceLawBD4 App} if fact simply represent the total memory. In its form, the null part of this expression corresponds to the memory formula which was obtained from the Isaacson approach in Eq.~\eqref{eq:NonLinDispMemoryModesGR}
\begin{align}\label{BMSSupermomentumBalanceLawBD5 App}
\begin{split}
  \Delta h^\text{null}_{\ell m}=r\sqrt{\frac{(\ell-2)!}{(\ell+2)!}}\int_{S^2} d\Omega\,Y^*_{\ell m} \int_{-\infty}^{\infty} d u'\,|\dot{h}|^2\,.
\end{split}
\end{align}
Indeed, in the limit $u\rightarrow\infty$ and $u_0\rightarrow-\infty$ of the time integral, all periodic oscillatory components are integrated zero, while only the averaged memory will contribute to the final offset. However, outside of this limit, the gradual $u'$ integral will not kill these oscillations in the time-dependent strain $h_{\ell m}(u)$ and the formulas are not equivalent anymore.

\subsection{A BMS balance law at each angle}\label{sApp:BalanceLawAtEachAngle}

Another way of rewriting the BMS balance laws is in terms of a relation that is valid at each angular direction as described in Sec.~\ref{ssSec:ProofofMemory}, such that [Eq.~\eqref{eq:BMSSupermomentumBalanceLaw each angle MT}]
\begin{equation}\label{eq:BMSSupermomentumBalanceLaw each angle MT App}
    \Re[\eth^2\Delta h(u)] =\frac{1}{r}\left\{16\pi G \int_{u_0}^{u}du'\, \left[F_h+F_\text{m}\right]-4\Delta M(u)\right\}\,.
\end{equation}
In vacuum, and in a post-Newtonian Bondi frame, in which 
\begin{equation}
    \lim_{u_0\rightarrow-\infty}h(u_0)=0
\end{equation}
this can be rewritten as
\begin{align}\label{eq:BMSSupermomentumBalanceLaw consistency App}
   \Re[\eth^2h(u)] &=\frac{1}{r}\bigg\{  \int_{-\infty}^{u}du'r^2\,|\dot h|^2\\
   &-4\Re\left[\Psi_2+\frac{1}{4}\bar h\dot{h}\right](u)-4M_\text{ADM}\bigg\}, \nonumber
\end{align}
where we have used the Einstein equations, which dictate the following relation 
\begin{equation}
    \partial_u M=-\partial_u\Re\left[\Psi_2+\frac{1}{4}\bar h\dot{h}\right]\,.
\end{equation}
Here, $\Psi_2$ represents the asymptotic Newman-Penrose scalar~\cite{Newman:1961qr} (see also e.g. Refs.~\cite{GomezLopez:2017kcw,Ashtekar:2019viz,DAmbrosio:2022clk}), which captures ``Coulombic'' information of the asymptotic gravitational field and satisfies 
\begin{equation}
    \lim_{u\rightarrow -\infty}\Re \Psi_2=- M_\text{ADM}\,.
\end{equation}

While one could now isolate the strain on the left-hand-side with similar computational techniques as in Sec.~\ref{sApp:SWSH expansion of memory in balance law} (see e.g.~\cite{Grant:2022bla}), it is alternatively possible to resort to a completion of the left-hand side to the full asymptotic waveform. Indeed, recall that the real part of the angular derivative acting on the strain in the expression $\Re[\eth^2h(u)]$ effectively selects out its electric parity component. In terms of obtaining a consistency relation for the entire asymptotic radiation, it is therefore useful 
to use the Bianchi relation~\cite{Mitman:2020bjf}
\begin{equation}
    \Im\left[\eth^2h\right]=-4 \Im\left[\Psi_2+\frac{1}{4}\dot{h} \bar h\right]\,,
\end{equation}
to write
\begin{align}\label{eq:BMSSupermomentumBalanceLaw consistency App2}
\eth^2 h(u)
&=
\frac{1}{r}\Bigg\{
\int_{-\infty}^{u} du'\, r^2 |\dot h|^2
-4 \left[\Psi_2+\frac{1}{4}\bar h \dot h\right]
-4 M_\text{ADM}
\Bigg\}\,.
\end{align}
Recall that the complex strain $h$ has spin weight $s=-2$. The operator $\eth^2$ acting on $h$ therefore produces a spin-weight zero quantity.

To invert this relation and recover the full strain, it is convenient to act with $\bar\eth^2$ on both sides.
One then obtains [Eq.~\eqref{eq:IdentityEth2}]
\begin{equation}\label{eq:EthBarEthIdentity}
\bar\eth^2\eth^2 h
=\mathcal D^2\bigl(\mathcal D^2-2\bigr) h\,,
\end{equation}
where $\mathcal D^2=\bar\eth\eth$ is the spin-weighted Laplacian on the unit two-sphere.
Using Eq.~\eqref{eq:Ethcommutation} one can then write
\begin{equation}\label{eq:FinalConsistencyA}
h(u)
=
\frac{1}{2r}\,
\bar\eth^2
\mathscr D^{-1}
\left\{
\frac{r^2}{4}\int_{-\infty}^{u} du'\,|\dot h|^2
-
\left(\Psi_2+\frac{1}{4}\bar h \dot h\right)
\right\}\,,
\end{equation}
where the scalar operator
\begin{equation}\label{eq:DefScrD}
\mathscr D
\equiv
\frac{1}{8}\,\mathcal D^2\bigl(\mathcal D^2+2\bigr)\,.
\end{equation}
annihilates all $\ell\le1$ spherical-harmonic modes. We therefore define its inverse $\mathscr D^{-1}$ as a pseudo-inverse acting only on $\ell\ge2$ modes, mapping all $\ell\le1$ components to zero~\cite{Mitman:2020bjf}.
Eq.~\eqref{eq:FinalConsistencyA} provides a consistency relation for the full asymptotic waveform at each retarded time $u$.
As already stated in the main text, this constraint equation can be used as a sanity check for state-of-the-art waveform models~\cite{Ashtekar:2019viz,Mitman:2020bjf,DAmbrosio:2024zok}.

\subsection{Gravitational memory as a finite BMS supertranslation}

Finally, we want to explicitly show that the permanent gravitational memory effect admits a natural interpretation in terms of the asymptotic symmetries of asymptotically flat spacetimes. In particular, the two vacuum configurations before and after the burst in a given angular direction are directly related by a finite supertranslation generated by a function $\alpha(\theta,\phi)$ on the 2-sphere, via 
\begin{align}\label{eq:AppElectricDecomposition} 
\Delta h^{TT}_{ij}(\theta,\phi)& = \frac{4G}{r}\, \perp_{ijab}(\theta,\phi)\, \mathcal{D}_a\mathcal{D}_b \,\alpha(\theta,\phi)\,, \\
&= \frac{8G}{r}\left(\mathcal{D}_i \mathcal{D}_j -\frac{1}{2}\,\perp_{ij}\,\mathcal{D}^2 \right)\,\alpha(\theta,\phi)\,,\nonumber
\end{align} where $\perp_{ijab}$ is the transverse-traceless projection operator defined in Eq.~\eqref{eq:ProjectorsA}.

Equation~\eqref{eq:AppElectricDecomposition} shows that the memory offset is purely electric parity and correspons to a metric shift generated by a finite supertranslation with parameter \(\alpha(\theta,\phi)\). 
The memory relation of the BMS balance laws [Eq.~\eqref{eq:BMSSupermomentumBalanceLaw each angle MT}] for modes with \(\ell\geq2\) define the explicit supertranslation parameter \(\alpha(\theta,\phi)\) required to satisfy this equation, namely
\begin{equation}\label{eq:AlphaDefinition}
\mathscr D \alpha(\theta,\phi)
=
\frac{\pi}{2}
\left[
\int_{-\infty}^{+\infty} du' \, F_{\rm R}(u',\theta,\phi)
-
\Delta M(\theta,\phi)
\right],
\end{equation}
with $\mathscr D$ defined in Eq.~\eqref{eq:DefScrD}.


\section{Details of the Isaacson approach to memory}\label{App:IssacsonMemory}

\subsection{The Isaacson assumptions.} 
The definition of gravitational waves on an arbitrary background spacetime requires two central assumptions: 
\begin{enumerate}[(1)]
    \item We require the existence of an exact solution $\bar{g}_{\mu\nu}$ about which we define perturbations.
    \item We assume the existence of a physical separation of characteristic scales of variation between a slowly varying background component ($L$) and a highly varying wave-perturbation ($H$).
\end{enumerate}
These two assumptions lead to the following leading-order decomposition of the metric\footnote{See Refs.~\cite{WaldBook,carroll2019spacetime,Zosso:2024xgy} and references therein for a mathematically more precise statement of such a decomposition of the metric.}
\begin{equation}\label{eq:IsaacsonDecompositionMetric}
    g_{\mu\nu}=\bar g_{\mu\nu}+h^L_{\mu\nu}+h^H_{\mu\nu}\,,
\end{equation}
where the exact solution $\bar g_{\mu\nu}$ is assumed to be static, or more generally to be part of the slowly-varying scales ($L$). The notion of gravitational waves can then be described by the gauge-invariant and dynamical degrees of freedom (dofs) within the high-frequency perturbations $h^H_{\mu\nu}$. This result is rather intuitive: the concept of waves as propagating perturbations only makes sense if we can ensure a physical separation between the periodic perturbations and the background. This is because the separation between a background and a perturbation is an arbitrary choice, which is not unique~\cite{Flanagan:2005yc,maggiore2008gravitational}. In particular, if the background is varying at the same scale as the perturbations, there is no way in uniquely distinguishing the oscillatory wave-part and the background~\cite{misner_gravitation_1973}. A physical definition of waves only exists if a clear separation between the scales of variation between the two, such that local physical experiments would be able to determine a unique separation.

Formally, a background spacetime can then be defined through a coarse-graining spacetime average $\langle ...\rangle$ over the short scales $(H)$, which in terms of the decomposition defined in Eq.~\eqref{eq:IsaacsonDecompositionMetric} reads
\begin{equation}
    \langle g_{\mu\nu}\rangle=\bar g_{\mu\nu}+\delta h_{\mu\nu}\,.
\end{equation}
While a concrete covariant definition of an averaging procedure was for instance given in Ref.~\cite{Brill:1964zz}, the exact averaging scheme is actually not important.  It only matters that the averaging operation satisfies the following set of leading order properties
\cite{Isaacson_PhysRev.166.1272,Brill:1964zz,misner_gravitation_1973,Flanagan:2005yc,Zalaletdinov:2004wd,maggiore2008gravitational,Stein:2010pn}:
\begin{enumerate}[(I)]
    \item the average of an odd number of $(H)$-operators vanishes;
    \item total covariant background derivatives of operators average out to zero;
    \item as a corollary of the above, integration by parts of covariant derivatives are allowed.
\end{enumerate}
With an appropriate averaging procedure at hand, the slowly varying contribution of any operator $O$ is defined as
\begin{equation}
    \left[O\right]^L\equiv\langle O\rangle\,,
\end{equation}
while the corresponding highly varying part simply reads 
\begin{equation}
    \left[O\right]^H\equiv O-\langle O\rangle\,.
\end{equation} 

A scale separation can be expressed through the characteristic frequency content of the fields, requiring a clear hierarchy
\begin{equation}\label{eq:SmallParam1GR}
f_{L}\ll f_{H},,
\end{equation}
between a slowly varying background ($f<f_L$) and perturbations with typical frequency $f_H$. In Eq.~\eqref{eq:IsaacsonDecompositionMetric}, this notation was already used within the labels $L$ and $H$ to indicate components associated with low or high frequencies.
Alternatively, one can formulate the separation in terms of spatial scales $L_L$ and $L_H$, demanding
\begin{equation}\label{eq:SmallParam1GRLength}
L_H\ll L_{L},,
\end{equation}
the so-called short-wave expansion, where $L_H$ is the wavelength of the perturbations.
While $L_H$ and $f_H$ are generally connected by the dispersion relation, $L_L$ and $f_L$ of the background are not necessarily related. Thus, slow variation in space and in time are conceptually distinct, though either condition can be used to separate background and perturbations. In practice, especially for terrestrial gravitational-wave detectors, the frequency hierarchy in Eq.~\eqref{eq:SmallParam1GR} is the most relevant~\cite{maggiore2008gravitational}.
In general, it is safest to assume both Eq.~\eqref{eq:SmallParam1GR} and Eq.~\eqref{eq:SmallParam1GRLength}. For simplicity, however, we focus here on the frequency condition. 

To summarize, the Isaacson scheme therefore relies on two small parameters in relation to the two Isaacson assumptions identified above:
\begin{enumerate}[(1)]
\item the perturbation amplitude: $|h_{\mu\nu}|\ll |g_{\mu\nu}|\sim 1$;
\item the ratio of background to perturbation frequency scales: $\frac{f_L}{f_H}\ll 1$.
\end{enumerate}
An important implication of the additional scale hierarchy is that perturbative equations cannot merely be expanded in the small amplitude. 
In the next subsection, we will discuss the corresponding leading-order Einstein equations. 

\subsection{Leading order equations of motion.}\label{sSec:LeadingOrderIsaacsonEquations}

Like in the main text, for simplicity, we will restrict ourselves to consider the vacuum Einstein equations. A more general treatment can be found in Refs.~\cite{maggiore2008gravitational,Zosso:2024xgy}. The assumed exact solution $\bar g_{\mu\nu}$ therefore satisfies
\begin{equation}
    G_{\mu\nu}[\bar g]=0\,,
\end{equation}
with $G_{\mu\nu}$ the Einstein tensor.
The goal is now to identify the leading order equations of motion about this exact solution.

Very generally, the assumed physical separation of scales will lead to two distinct sets of leading order equations. This is because the Isaacson split into low- and high-frequency parts ought to be applied directly to the equations as well, effectively implementing a multiple-scale analysis. For concreteness, we will denote the typical amplitude of high-frequency perturbations as
\begin{equation}\label{eq:SmallParam2GRH}
    |h^H_{\mu\nu}| = {\cal{O}}(\alpha)\,,
\end{equation}
whereas we write the amplitude of low-frequency perturbations as
\begin{equation}\label{eq:SmallParam2GRL}
    |h^L_{\mu\nu}| = {\cal{O}}(\beta)\,,
\end{equation}
with both $\beta,\alpha\ll 1$.
Schematically, the scale of derivatives acting on low- or high-frequency components can then be written as~\cite{Isaacson_PhysRev.166.1263,misner_gravitation_1973}
\begin{align}
\partial h^L_{\mu\nu}=\mathcal{O}(\beta f_L)\,,\quad
\partial h^H_{\mu\nu}= \mathcal{O}(\alpha f_H)\,.\label{eq:OrderDerivativeH}
\end{align}
Together with the knowledge that by dimensional analysis $G_{\mu\nu}$ must involve two derivative operators, it is then an easy task to estimate the order of each operator within a perturbation series
\begin{align}
  G^{\mys{(i)}}_{\mu\nu}[h_L]={\cal{O}}(\beta^i f_L^2)\,,\quad  G^{\mys{(i)}}_{\mu\nu}[h_H]&={\cal{O}}(\alpha^i f_H^2)\,,\label{eq:OrderCountingGRHigh1}
\end{align}
where $(i)$ denotes the order in perturbations. 
Lastly, while linear terms in the high-frequency fields vanish under the application of a space-time average $\langle...\rangle$ (recall property $(I)$ above), second order terms of $h_H$ will contribute both to the high-frequency and the low-frequency equations, since two high-frequency modes can interfere to produce a low-frequency component.

Taking all of the above into account it is rather straightforward to arrive at the following two leading order Isaacson equations\footnote{Note that here we adopt the convention used in Ref.~\cite{maggiore2008gravitational} and write $G^{\mys{(2)}}_{\mu\nu}[h_H]$ without explicitly factoring out the $1/2$ factor associated to the perturbative second-order nature of the operator. Compared to Ref.~\cite{Zosso:2024xgy} there is therefore a factor of $1/2$ difference to be aware of.}~\cite{misner_gravitation_1973,maggiore2008gravitational,Zosso:2024xgy}
\begin{align}
   G^{\mys{(1)}}_{\mu\nu}[h_H]&=0+{\cal{O}}(\alpha^2 f_H^2)\,,\label{eq:EOMIISGR}\\
   G^{\mys{(1)}}_{\mu\nu}[h_L]&=-\,\big\langle G^{\mys{(2)}}_{\mu\nu}[h_H]\big\rangle+{\cal{O}}(\alpha^3 f_H^2)+{\cal{O}}(\beta^2 f_L^2)\,.\label{eq:EOMISGR}
\end{align}
The first one, the equation at high-frequency scales straightforwardly describes a propagation equation of the gravitational waves on the non-trivial background spacetime. On the other hand, the leading order low-frequency scale equation naturally includes a coarse-grained contribution from the high-frequency fields at second order in perturbation theory. As such, its natural interpretation is to represent a back-reaction equation of the coarse-grained high-frequency operator giving rise to a perturbation of the background spacetime $h^L_{\mu\nu}$. In this sense, the right-hand side of Eq.~\eqref{eq:EOMISGR} can be identified with the gravitational-wave energy-momentum tensor\cite{Isaacson_PhysRev.166.1263,Isaacson_PhysRev.166.1272,misner_gravitation_1973,Flanagan:2005yc,maggiore2008gravitational}
\begin{equation}\label{eq:DefPseudoEMTensorGR}
t_{\mu\nu}[h_H]\equiv -\frac{1}{8\pi G}\big\langle G^{\mys{(2)}}_{\mu\nu}[h_H]\big\rangle,.
\end{equation}
Indeed, one of the main achievements of the Isaacson framework is the identification of a well-defined energy-momentum tensor of gravitational waves, which to leading order is gauge invariant under infinitesimal high-frequency transformations~\cite{Isaacson_PhysRev.166.1263,maggiore2008gravitational}. Moreover, because the averaging operation commutes with covariant differentiation, this energy-momentum tensor is also generally conserved
\begin{equation}
\bar\nabla^\mu t^L_{\mu\nu}[h_H]=0\,.
\end{equation}

Note that the back-reaction interpretation of Eq.~\eqref{eq:EOMISGR} naturally ties the low-frequency amplitude parameter $\beta$ to the amplitude of high-frequency waves $\alpha$ and the small dimensionless fraction $f_L/f_H$. Concretely, we have
\begin{equation}\label{eq:BetaParam}
\beta\sim \alpha^2\frac{f_H^2}{f_L^2}\,,
\end{equation}
so that requiring $\beta \ll 1$ enforces a hierarchy between the expansion parameters\footnote{We want to mention here that Isaacson's original analysis~\cite{Isaacson_PhysRev.166.1263} assumed $\alpha\sim f_L/f_H$, which misses out on the perturbative perturbative version of the low-frequency equation presented here.}
\begin{equation}
\label{eq:hierarchy-inequ}
\alpha\ll \frac{f_L}{f_H}\,.
\end{equation}
As we will now show, applied to the simplest scenario of an emission of gravitational waves from a localized source in asymptotically flat spacetime, the resulting radiative background perturbation $h^L_{\mu\nu}$ is the gravitational memory effect. 

\subsection{Leading order asymptotic equations}\label{sApp:leadingOrdereqs in flat space}

Within the Isaacson approach, assuming asymptotic flatness implies the choice the a particular exact solution of Minkowski spacetime, about which we are expanding on. More precisely, there exists a coordinate system, in which we can write $\bar g_{\mu\nu}=\eta_{\mu\nu}$, where $\eta_{\mu\nu}$ is in the familiar Minkowski form. The perturbations are then associated to ${\cal{O}}(1/r)$ corrections in the asymptotic limit. As an illustrative example, in the case of a standard binary black hole inspiral at mass scale $M$, the quadrupole formula implies that the dimensionless amplitude of high-frequency perturbations is $\alpha$ is given by 
\begin{equation}
    \alpha \sim \frac{GM}{r}(f_HGM)^{2/3}\,.
\end{equation}
Note that indeed, in this macroscopic scenario in which one expects $\alpha\sim 1$ very close to the source, the smallness of $h^H_{\mu\nu}$ is determined by the distance $r$ being large and not by the smallness of Newton's constant $G$. In other words, the counting in $\alpha$ is not equivalent to a counting in Plank mass factors, but reflects a counting in powers of $1/r$.


In order to determine the leading order equations of motion for the dynamical degrees of freedom of the theory from Eqs.~\eqref{eq:EOMIISGR} and \eqref{eq:EOMISGR} we now have to explicitly address the gauge ambiguity within the formulation of perturbation theory. As is widely known, this gauge redundancy is directly associated to infinitesimal coordinate transformations $x^\mu\rightarrow \tilde x^\mu=x^\mu-\xi^\mu$, with $|\xi^\mu|\ll 1$. 
Concretely, the gauge freedom of perturbation fields is written as
\begin{equation}\label{eq:CoordGaugeTransf}
  h_{\mu\nu}\rightarrow h_{\mu\nu}-(\partial_\mu \xi_\nu+\partial_\nu \xi_\mu)\,.
\end{equation}
By performing suitable coordinate transformations, it is well-known that the following transverse-traceless (TT) conditions can be imposed at the level of the equations of motion
\begin{equation}\label{eq:FirstOrderGaugeFixing}
    \partial^\mu h_{\mu\nu}=0\,,\quad \eta^{\mu\nu}h_{\mu\nu}\equiv h^t=0\,.
\end{equation}
These gauge conditions can be imposed for both the high-frequency perturbations $h^H_{\mu\nu}$ describing gravitational waves and the low-frequency $h^L_{\mu\nu}$.
Indeed, given the presence of a parametric separation of physical scales one should also separate infinitesimal coordinate transformations into two distinct classes $\xi^\mu_L(x)$ and $\xi^\mu_H(x)$ that match the characteristics of each scale~\cite{maggiore2008gravitational}, such that the gauge-freedom can be individually imposed on $ h^H_{\mu\nu}$ and $ h^L_{\mu\nu}$.

As concerns the high-frequency equation Eq.~\eqref{eq:EOMIISGR} the leading order propagation equations on the asymptotically flat background in the gauge in Eq.~\eqref{eq:FirstOrderGaugeFixing} reduce unsurprisingly to a simple wave equation
\begin{equation}\label{eq:EOMIISGR3}
    G^{\mys{(1)}}_{\mu\nu}[h^H]=-\frac{1}{2}\Box h^H_{\mu\nu}=0\,.
\end{equation}
On the other hand, applying the gauge conditions in Eq.~\eqref{eq:FirstOrderGaugeFixing} onto the low-frequency perturbations on the left-hand side of Eq.~\eqref{eq:EOMISGR} results the same form of equation $ G^{\mys{(1)}}_{\mu\nu}[h^L]=-1/2\Box h^L_{\mu\nu}$. The difference is, of course, the presence of the source term depending on the high-frequency perturbations. Making use of the freedom of simultaneously applying Eq.~\eqref{eq:FirstOrderGaugeFixing} on the coarse-grained operator on the right-hand side of Eq.~\eqref{eq:EOMISGR} results in (see e.g.~\cite{maggiore2008gravitational})
\begin{equation}
\big\langle G^{\mys{(2)}}_{\mu\nu}[h_H]\big\rangle=-\frac{1}{4}\Big\langle \partial_\mu h^H_{\alpha\beta}\partial_\nu h_H^{\alpha\beta}\Big\rangle\,.
\end{equation}
In summary, the leading order\footnote{Higher order contributions would involve various additional effects, such as a source term within the higher order propagation equation of primary waves~\cite{misner_gravitation_1973}. For the treatment of gravitational memory at low-frequency scales, a truncation at the first iteration is however generally largely sufficient.} Isaacson equations in an asymptotically flat approximation read
\begin{align}
   \Box h^H_{\mu\nu}&=0\,,\label{eq:EOMIISGR4}\\
   \Box h^L_{\mu\nu}&=-16\pi G\, t_{\mu\nu}[h^H]\,,\label{eq:EOMISGR4}
\end{align}
where the energy-momentum tensor identified in Eq.~\eqref{eq:DefPseudoEMTensorGR} adopts the well-known expression
\begin{equation}\label{eq:GWEMtensorGen}
t_{\mu\nu}[h^H]=\frac{1}{32\pi G}\Big\langle \partial_\mu h^H_{\alpha\beta}\partial_\nu h_H^{\alpha\beta}\Big\rangle\,.
\end{equation}
It is easily verified that this expression of an energy-momentum tensor of gravitational waves is indeed conserved and invariant under the gauge transformation in Eq.~\eqref{eq:CoordGaugeTransf} (see e.g. Ref.~\cite{maggiore2008gravitational}). In the following, we will solve this set of equations and find that the resulting low-frequency contribution to the outgoing radiation corresponds to gravitational memory.

\subsection{Solving the leading order Isaacson equations}\label{sApp:Solving IsaacsonEquation}

While Eq.~\eqref{eq:EOMIISGR4} results in a description of gravitational waves identified as the high-frequency perturbations in terms of a superposition plane-waves in a given radially outward direction, the emission of such GWs inevitably sources a corresponding low-frequency perturbation through Eq.~\eqref{eq:EOMISGR4} whose general solution is given by the Green's function of the wave-operator. Hence, we can easily write down the general solution as
\begin{align}
   h^L_{\mu\nu}(x) = -16\pi G  \int d^{4}x' \; G(x-x') \,t_{\mu\nu} (x')   \label{eqn:SolForm} \,,
\end{align}
where the retarded Green's function reads (see e.g.~\cite{Jackson:1998nia})
\begin{align}
    G(x-x') =-\frac{\delta(t_{r}-t')}{4\pi  \left | \vec{x}-\vec{x}' \right | }\, , \label{eq:Green's function}
\end{align}
with $t_{r}$ the retarded time of luminally propagating signals
\begin{equation}
    t_{r}\equiv t-|\vec{x}-\vec{x}'|\,.
\end{equation}

In order to extract the physical propagating degrees of freedom within the manifestly local but not manifestly gauge-invariant approach, an additional crucial step is required. Indeed, after imposing Eq.~\eqref{eq:FirstOrderGaugeFixing}, a residual gauge freedom remains, characterized by transformations satisfying $\Box\xi^{\mu}=\partial_\mu\xi^\mu=0$. In the radiation zone, this residual gauge freedom should be used to isolate the two true physical degrees of freedom of GR present in the spatial part of the asymptotic gravitational radiation $h_{ij}$ (see e.g.~\cite{maggiore2008gravitational}). For a superposition of plane waves, a projection onto these dynamical degrees of freedom is conveniently achieved via the transverse-traceless projector~\cite{maggiore2008gravitational,Heisenberg:2023prj}
\begin{equation}\label{eq:Projectors}
    \perp_{ijab}\,\equiv \,\perp_{ia}\perp_{jb}-\frac{1}{2}\perp_{ij}\perp_{ab}\,,
\end{equation}
where
\begin{equation}\label{eq:TransverseProjector}
\perp_{ij}\,\equiv \delta_{ij}-n_in_j=\theta_i\theta_j+\phi_i\phi_j\,,
\end{equation}
For convenience, we provide in Eq.~\eqref{eq:TransverseProjectorAExplicit} an explicit expression of this transverse projector. Moreover, the spatial orthonormal basis $\{n_i,\theta_i,\phi_i\}$ along a particular radial direction $n_i$ is defined in Eqs.~\eqref{eq:Def Direction of Propagation n} and \eqref{eq:Def Transverse Vectors u v}.
One can then write
\begin{equation}\label{eq:TT modes h and a SVH}
\perp_{ijab}h_{ab}=h^{TT}_{ij}\,.
\end{equation}
These two radiative degrees of freedom can be described in the standard $+$ and $\times$ basis, through
\begin{equation}\label{eq:gravitationalradiationGenA}
    h^{TT}_{ij}=h_+ e^+_{ij}+h_\times e^\times_{ij}\,,
\end{equation}
where
\begin{equation}\label{eq:DefsPolarizationTensorsA}
e^+_{ij}\equiv \theta_i\theta_j-\phi_i\phi_j\;,\qquad e^\times_{ij}\equiv \theta_i\phi_j+\phi_i\theta_j\,.
\end{equation}
Hence, due to the orthogonality conditions of this $TT$ basis, the two polarization modes are isolated via
\begin{equation}\label{eq:polmodes h SVH}
h_{+/\times}=\frac{1}{2}e^{ij}_{+/\times}h^{TT}_{ij}\,.
\end{equation}
These are the two true physical and dofs of GR refereed to at the beginning of the main text in Eq.~\eqref{eq:gravitationalradiationGen}.

First of all, since the energy-momentum tensor of gravitational wave in Eq.~\eqref{eq:GWEMtensorGen} is gauge-invariant, it can be expressed in terms of the two physical dofs only
\begin{align}\label{eq:StressEnergyGRThird}
   t_{\mu\nu}&=\frac{1}{32\pi G}\Big\langle \partial_\mu h^{TT}_{ij H}\partial_\nu h_{TT}^{ij H}\Big\rangle\nonumber\\
   &=\frac{1}{16\pi G}\,\Big\langle \dot{h}_{H+}^2+\dot{h}_{H\times}^2\Big\rangle\,l_\mu\,l_\nu\,.
\end{align}
where we have defined the null vector
\begin{equation}\label{eq:FirstDef l}
    l_\mu\equiv -\nabla_\mu t+\nabla_\mu r\,,
\end{equation}
with $\nabla_\mu r=\partial_\mu r=\delta_{\mu i} \,\partial_i r=\delta_{\mu i}\,n_i$ and $r=\sqrt{x^2+y^2+z^2}$ is the source-centered radial coordinate. The second equality in Eq.~\eqref{eq:StressEnergyGRThird} follows from Eq.~\eqref{eq:gravitationalradiationGenA}, such that
\begin{equation}
    \Big\langle \partial_\mu h^{TT}_{ij}\partial_\nu h^{TT}_{ij}\Big\rangle=2\, \Big\langle \partial_\mu h_+\partial_\nu h_++\partial_\mu h_\times\partial_\nu h_\times\Big\rangle\,,
\end{equation}
and the knowledge that to leading order, the asymptotic radiation following a massless wave equation satisfies 
\begin{equation}
    \partial_i h^{TT}_{ij} = -n_i \dot h^{TT}_{ij}\,.
\end{equation}

In solving Eq.~\eqref{eqn:SolForm} in the asymptotic limit, we should isolate these radiative degrees of freedom. We will do so by extracting the leading contribution to the limit of future null infinity and applying the appropriate projection in Eq.~\eqref{eq:Projectors}. It is most convenient to do so by performing the change of coordinates
\begin{align}\label{eq:ChangeofCoords1}
    (t,x,y,z) \rightarrow (u,r,\theta,\phi)\, ,
\end{align}
with $u$ the asymptotic regarded time 
\begin{equation}\label{eq:AsymptoticRetardedTimeA}
    u\equiv t-r\,.
\end{equation}
and $\{r,\theta,\phi\}$ representing standard source-centered spherical coordinates already introduced above. The limit to null infinity is then defined as the large distance limit $r\rightarrow\infty$ at fixed retarded time $u$. 

Furthermore, it is important to note that an emission of energy and momentum described by $t_{\mu\nu}$ depends temporally on the asymptotic source retarded time\footnote{While in this case of a null flux, the retarded time is characterized by the speed of light, in general, an emission of energy-momentum is characterized by the physical group velocity~\cite{PhysRev.105.1129}.}
\begin{align}
    u'=t'-r' \,.
\end{align}
More precisely, $t_{\mu\nu}$ in Eq.~\eqref{eq:StressEnergyGRThird} is entirely described by the energy flux [Eq.~\eqref{eq:EnergyFluxH}]
\begin{equation}\label{eq:EnergyFluxHA}
   \frac{dE}{du'd\Omega'}=\frac{r'^2}{16\pi G}\Big\langle \dot h^2_{H+} +\dot h^2_{H\times}\Big\rangle\,.
\end{equation}
It is this explicit dependence on the source retarded time, which will result in an explicit integration over retarded time and turn the memory contribution into a hereditary effect.
This also makes it explicit that the expression $r'^2\Big\langle \dot{h}_{H+}^2+\dot{h}_{H\times}^2\Big\rangle$ is independent of $r'$ due to the leading order fall-off behavior of $h_{H+}$ and $h_{H\times}$ as leading order perturbations.
It is thus useful to perform the same change to the relevant asymptotic coordinates for the source
\begin{align}\label{eq:ChangeofCoords2}
    (t',x',y',z') \rightarrow (u'_\psi,r',\theta',\phi') \,.
\end{align}

The asymptotic solution in Eq.~\eqref{eqn:SolForm} of the physical dofs then reads
\begin{widetext}
\begin{equation}\label{eq:memoryequationA}
h^{TT}_{ijL}(u,r,\Omega)=\frac{1}{4\pi r}\int_{-\infty}^u du'\int d\Omega' r'^2\Big\langle \dot{h}_{H+}^2+\dot{h}_{H\times}^2\Big\rangle \frac{\perp_{ijab}(\Omega)\,n'_an'_b}{1-\vec{n}' \cdot \vec{n}(\Omega)}=\frac{4G}{r}\int_{-\infty}^u du'\int d\Omega' \frac{dE}{du'd\Omega'}\left[\frac{n'_in'_j}{1-\vec{n}'\cdot  \vec n}\right]^{TT}\,.
\end{equation}
\end{widetext}
In appendix~\ref{App:DerivationIsaacsonLow} below we offer the explicit steps of this derivation. In the last equality within Eq.~\eqref{eq:memoryequationA}, we have used the frequently employed shorthand of denoting a projection onto the physical $TT$ modes by a $TT$ superscript. As for the high-frequency gravitational waves, this projection is necessary to isolate the physical gauge invariant degrees of freedom of memory.\footnote{The leading-order solution in Eq.~\eqref{eq:memoryequationA} identifies the low-frequency component of asymptotic radiation as an independent field with its own low-frequency gauge freedom, a point often obscured in earlier treatments of nonlinear memory~\cite{PhysRevD.44.R2945,Thorne:1992sdb,Favata:2008yd}.} 

This is our final result reported in Eq.~\eqref{eq:memoryequation} in the main text. As explained there, this low-frequency contribution to the radiation is identified as nonlinear gravitational memory.
Thus, while gravitational waves sourced by some localized system propagate as expected in the radiation zone, their presence inevitably also results in the existence of a propagating perturbation of the low-frequency background. Although this background component $h^L$ also satisfies a wave equation, and is part of the gravitational radiation reaching null infinity, in the terminology of Isaacson it is however \emph{not} a gravitational wave. Recall that the whole point of the Isaacson formalism was to identify a notion of well-defined gravitational waves as high-frequency perturbations $h^H$ on top of a slowly varying background. This naturally results in the statement that radiation, defined as the null energy-momentum lost by the localized system which formally reaches null infinity, is not only made out of gravitational waves, but is always accompanied by a low-frequency counterpart, which we identified here as gravitational memory. 


\section{Memory from a post-Newtonian perspective}\label{App:PNresults}

In this section, we review the description of gravitational memory during the inspiral phase. By explicitly revisiting this treatment, we aim to establish clear connections with the theoretical frameworks employed throughout this work, in particular the Isaacson approach and the interpretation of memory underlying the BMS balance laws. This comparison allows us to showcase how the same physical effect is captured from complementary perspectives and to clarify conceptual distinctions between these formulations.

When the compact objects are sufficiently separated, the GW signal can be accurately computed as an expansion in the PN parameter
   $ x = (M \omega)^{2/3}$
where $\omega$ is the orbital frequency.
Here, we focus on the leading-order PN contribution to the memory~\cite{Blanchet:1992br} and highlight its characteristic slow time growth, in contrast to the high-frequency oscillations of the dominant GW signal.
Higher-PN corrections to the memory have been computed in
Refs.~\cite{Favata:2008yd, Favata:2011qi, Ebersold:2019kdc, Henry:2023tka, Cunningham:2024dog}.

At leading order, we assume that most of the GW energy is emitted in the $(\ell,m)=(2,2)$ SWSH mode.
This approximation is well justified for circular, equal-mass binaries during the inspiral.
The memory contribution to the $(2,0)$ mode can, then, be written as [Eq.~\eqref{eq:memorymodes20Selection}]
\begin{equation}\label{eq:memPNfirst}
    h^L_{20}(u) = \frac{1}{r} \int_{-\infty}^{u} du' \,
    \frac{1}{7}\sqrt{\frac{5}{6\pi}} \,r'^2 \big| \dot{h}^H_{22} \big|^2 .
\end{equation}
Furthermore, the $(2,2)$ mode of a quasi-circular binary can be expressed in terms of the amplitude and phase of the GW strain,
\begin{equation}
    r' h^H_{22}(t) = A_{22}(t) e^{-i \phi_{22}(t)},
\end{equation}
such that the resulting energy-flux can be written as
\begin{align}\label{eq:enflux22}
    \big| \dot{h}^H_{22} \big|^2
    = \left( \omega_{22}A_{22} \right)^2 + \left( \dot{A}_{22} \right)^2 .
\end{align}
where the instantaneous GW frequency is defined as \begin{equation}
    \omega_{22} \equiv \dot{\phi}_{22}\,.
\end{equation}
As we show below, the ratio $\dot{A}_{22}/A_{22} \sim \dot{\omega}_{22}/\omega_{22}$ is parametrically smaller than $\omega_{22}$ during the inspiral, and therefore the second term in Eq.~\eqref{eq:enflux22} can be neglected at leading PN order.
Equation~\eqref{eq:memPNfirst} can then be rewritten as an integral over the GW frequency,
\begin{equation}\label{eq:memintomega}
    h^L_{20}(u) =
    \frac{1}{r} \int_{0}^{\omega(u)} d\omega' \,
    \frac{1}{7}\sqrt{\frac{5}{6\pi}}
    \frac{\big( \omega'_{22} A_{22}(\omega') \big)^2}{\dot{\omega}'_{22}} .
\end{equation}
This frequency-domain expression is valid as long as $\dot{\omega}_{22} \neq 0$, which holds throughout the inspiral but breaks down in the merger regime.

At leading PN order, the amplitude and frequency evolution are given by~\cite{maggiore2008gravitational}
\begin{align}
    A_{22} &= 2 \sqrt{\frac{\pi}{5}} \, 4 M_c \left( \frac{M_c \omega_{22}}{2} \right)^{2/3},\label{eq:A22} \\
    \dot{\omega}_{22} &= \frac{12}{5} \, 2^{1/3} \frac{(M_c \omega_{22})^{11/3}}{M_c^2},\label{eq:omega22}
\end{align}
where the chirp mass reads
\begin{equation}
    M_c \equiv \frac{(m_1 m_2)^{3/5}}{(m_1 + m_2)^{1/5}} \,,
\end{equation}
and $m_1+m_2=M$ are the masses of the individual BHs.
It is straightforward to verify that
\begin{equation}
    \dot{A}_{22}/A_{22} \sim \dot{\omega}_{22}/\omega_{22}\,,
\end{equation}
confirming that the amplitude grows on the same timescale as the instantaneous frequency. Moreover, the instantaneous evolution of the memory,
\begin{equation}
    \dot{h}^L \propto (M\omega)^{10/3}\,,
\end{equation}
is suppressed by a factor $(M\omega)^{5/3}$ relative to $\dot{h}^H$.
For this reason, the nonlinear memory is conventionally said to enter at 2.5PN order.
This classification refers to the local-in-time growth rate of the memory and does not account for its \emph{hereditary} character, which arises from the time integration in Eq.~\eqref{eq:memintomega}.

Substituting the expressions in Eqs.~\eqref{eq:A22} and \eqref{eq:omega22} into Eq.~\eqref{eq:memintomega}, we obtain
\begin{equation}
    h^L_{20} =
    \frac{1}{r} \frac{1}{7} 2^{5/6} \sqrt{\frac{5\pi}{3}}
    M_c (M_c \omega_{22})^{2/3}.
\end{equation}
Multiplying by the spin-weighted spherical harmonic
$_{\mys{-2}}Y_{20} = \frac{3}{4}\sqrt{\frac{5}{6\pi}} \sin^2\theta$,
the memory contribution to the plus polarization reads
\begin{equation}
    h^L_+ =
    \frac{1}{r} \frac{5}{14 \, 2^{2/3}}
    M_c (M_c \omega_{22})^{2/3} \sin^2\theta \,,
\end{equation}
while $h^L_\times=0$. This linear polarization of memory is a generic feature of quasi-circular binaries that posses a mirror symmetry across a plane as given by Eq.~\eqref{eq:SymmOrbitPlaneModes}.
For comparison, the dominant oscillatory plus polarization is
\begin{equation}
    h^H_+ =
    \frac{1}{r} \frac{4}{2^{2/3}}
    M_c (M_c \omega_{22})^{2/3}
    \left( \frac{1 + \cos^2\theta}{2} \right)
    \cos\!\big( \phi_{22}(t) \big),
\end{equation}
Therefore, we show the remarkable result that due to its hereditary nature, the memory modifies the GW waveform at leading order in PN expansion, and the total plus polarization waveform is 
\begin{align}
    h^{\mathrm{tot}}_+ &=
    \frac{1}{r} \frac{4}{2^{2/3}}
    M_c (M_c \omega_{22})^{2/3}\nonumber\\
    &
    \left[
        \frac{1 + \cos^2\theta}{2} \cos\!\big( \phi_{22}(t) \big)
        + \frac{5}{56} \sin^2\theta
    \right].
\end{align}

Within such a PN expansion, one estimates that for the maximal configuration of an edge-on binary ($\theta = \pi/2$), the inspiral memory amplitude is expected to reach at most $\simeq 17\%$ of the oscillatory signal.
However, a fully reliable estimate of the memory requires numerical relativity, since, as discussed in Sec.~\ref{ssSec:PropertiesMemoryModel}, the dominant contribution arises during the merger phase, where the PN expansion breaks down.
Numerical simulations yield a maximal memory contribution of approximately $27\%$ for equal-mass binaries, while the memory amplitude can completely dominate the radiation signal for high spin values of the merging black holes.

Furthermore, we note that the scaling of the memory with orbital frequency depends on the scalings of the waveform amplitude and frequency evolution.
If we parameterize them generically as
$A(t) \propto \omega(t)^\alpha$
and
$\dot{\omega}(t) \propto \omega(t)^\beta$,
a description that remains valid through merger,
the characteristic timescale for the growth of the memory is
\begin{equation}
    T_L
    \simeq
    \frac{h^L}{\partial_t h^L}
    =
    \frac{\omega}{\dot{\omega}} .
\end{equation}
The memory therefore grows on the radiation-reaction timescale, which is parametrically longer than the orbital timescale set by $\omega^{-1}$.
This separation of timescales makes explicit the distinction between the high-frequency and low-frequency components of the GW signal in the Isaacson picture. 

In summary, from a post-Newtonian perspective on quasi-circular inspirals, nonlinear gravitational memory appears as a linearly polarized hereditary contribution to the radiation that is parametrically distinct in its frequency content from the dominant oscillatory gravitational waves. It can be understood as arising from the quadrupole-quadrupole coupling that modifies the mass multipole moments of the system due to the emission of oscillatory GWs. Within the PN framework, however, gravitational memory only enters as one nonlinear effect among several others. In particular, the PN expansion also captures additional hereditary contributions that depend on the entire past history of the binary, such as \emph{tail} terms, which originate from the backscattering of GWs off the curved spacetime generated by the binary itself.

Gravitational memory is nevertheless unique in that it produces a genuine DC component in the waveform, associated with the cumulative emission of gravitational radiation to null infinity. In terms of amplitude, memory constitutes the leading-order correction to the radiation signal~\cite{Favata:2008yd} and uniquely characterizes its low-frequency spectrum, while its nonobservation in current GW data is entirely a consequence of the frequency-band limitations of existing detectors. Finally, the memory effect is distinguished by its close connection to asymptotic spacetime symmetries and infrared uniqueness theorems, rendering its potential future detection a significant observational milestone for fundamental physics.


\section{Derivations and Formulas}\label{App:Derivations}

\subsection{Spacial orthonormal radiation basis}\label{App:SpacialBasis}

Given a source-centered spherical coordinate system $\{t,r,\Omega=(\theta,\phi)\}$, the unit radial vector reads
\begin{equation}\label{eq:Def Direction of Propagation n}
    \mathbf{n}(\Omega)=(\sin\theta \cos\phi,\,\sin\theta \sin\phi,\,\cos\theta )\,.
\end{equation}
The associated transverse space of a given direction $\mathbf{n}(\Omega)$ can then be described through the basis vectors in the natural directions of the azimuthal and polar angles, respectively
\begin{subequations}\label{eq:Def Transverse Vectors u v}
\begin{align}
\bm{\theta}(\Omega)&=(\cos\theta \cos\phi,\,\cos\theta \sin\phi,\,-\sin\theta )\,,\\
\bm{\phi}(\Omega)&=(-\sin\phi,\,\cos\phi,\,0)\,.
\end{align}
\end{subequations}
This set of spacial vectors form an orthonormal spacial basis satisfying the completeness relation
\begin{equation}\label{eq:polcompleteness}
\delta_{ij}=n_in_j+\theta_i\theta_j+\phi_i\phi_j\,.
\end{equation} 


Given this spacial basis of radial outward radiation one can construct a basis for the two-dimensional spin-2 transverse traceless space through 
\begin{equation}\label{eq:DefsPolarizationTensors}
e^+_{ij}\equiv \theta_i\theta_j-\phi_i\phi_j\;,\qquad e^\times_{ij}\equiv \theta_i\phi_j+\phi_i\theta_j\,.
\end{equation}
These polarization tensors are associated with the characteristics of the two dynamical degrees of freedom of gravity $h_{ij}^{TT}$ described in GR.

In this context it is useful to define a transverse-traceless ($TT$) projector through [Eq.~\eqref{eq:Projectors}]
\begin{equation}\label{eq:ProjectorsA}
    \perp_{ijab}\,= \,\perp_{ia}\perp_{jb}-\frac{1}{2}\perp_{ij}\perp_{ab}=e^+_{ij} e^+_{ab} + e^\times_{ij} e^\times_{ab}\,.
\end{equation}
This form guarantees symmetry, tracelessness, and transversality by construction, and it directly projects any metric perturbation onto the two physical polarizations
\begin{equation}
    h_{ij}^{\rm TT} = \perp_{ijab} h_{ab} = h_+ e^+_{ij} + h_\times e^\times_{ij}\,.
\end{equation}
The associated individual transverse projectors are entirely constructed out of the original spacial basis vectors
\begin{equation}\label{eq:TransverseProjectorA}
\perp_{ij}\,\equiv \delta_{ij}-n_in_j=\theta_i\theta_j+\phi_i\phi_j\,.
\end{equation}
Explicitly, we have
\begin{widetext}
\begin{equation}\label{eq:TransverseProjectorAExplicit}
\perp_{ij} (\Omega) =
\begin{pmatrix}
1-\sin^2\theta\cos^2\phi & -\sin^2\theta\sin\phi\cos\phi & -\sin\theta\cos\theta\cos\phi \\
-\sin^2\theta\sin\phi\cos\phi & 1-\sin^2\theta\sin^2\phi & -\sin\theta\cos\theta\sin\phi \\
-\sin\theta\cos\theta\cos\phi & -\sin\theta\cos\theta\sin\phi & \sin^2\theta
\end{pmatrix}.
\end{equation}
\end{widetext}


\subsection{Spin-weighted spherical harmonics}
\label{App:SWSH}

Spin-weighted spherical harmonics (SWSHs) can be generated from ordinary spherical harmonics by repeated application of angular derivative operators. Explicitly, they are defined as
\begin{equation} 
\phantom{}_{\mys{s}}Y_{\ell m}(\theta,\phi)= \begin{cases} \sqrt{\frac{(\ell-s)!}{(\ell+s)!}}\eth^s Y_{\ell m}(\theta,\phi)\,, & \ell\geq s\geq 0\\ (-1)^s \sqrt{\frac{(\ell+s)!}{(\ell-s)!}}\bar{\eth}^{-s} Y_{\ell m}(\theta,\phi)\,, & 0> s\geq -\ell\\ 0\,, & |s|>\ell\\ \end{cases} 
\end{equation}
where the angular derivative operator \(\eth\) and its complex conjugate \(\bar{\eth}\) act on a function \(f_s\) of spin weight \(s\) as
\begin{align}\label{eth} 
\eth f_s(\theta,\phi) &\equiv -\sin^s\theta\left(\partial_\theta+i\,\csc\theta \partial_\phi\right)(f_s \sin^{-s}\theta)\,,\\ \bar\eth f_s(\theta,\phi) &\equiv -\sin^{-s}\theta(\partial_\theta-i\,\csc\theta\partial_\phi)\left(f_s \sin^{s}\theta\right)\,. 
\end{align}
The spin weight \(s\) of a scalar function \(f_s(\theta,\phi)\) on the sphere is defined through its transformation under the local \(U(1)\) gauge freedom parametrized by an angle \(\psi\) (see, e.g. Ref.~\cite{DAmbrosio:2022clk}),
\begin{equation}\label{eq:SpinWeightTransfos App}
    f_s(\theta,\phi)\;\longrightarrow\;
    f'_s(\theta,\phi)=f_s(\theta,\phi)\,e^{is\psi}\,.
\end{equation}
By construction, the operator \(\eth\) raises the spin weight by one unit, while \(\bar{\eth}\) lowers it by one. These definitions correspond to the standard Goldberg convention.
An explicit closed-form expression for the SWSHs is given by
\begin{align}
    &\phantom{}_{\mys{s}}Y_{\ell m}(\theta,\phi)
    =\, e^{i m\phi}
    \sqrt{\frac{(\ell+m)!(\ell-m)!(2\ell+1)}{4\pi(\ell+s)!(\ell-s)!}}
    \sin^{2\ell}\!\frac{\theta}{2}
    \notag\\
    &\times
    \sum_{q=0}^{\ell-s}
    \binom{\ell-s}{q}
    \binom{\ell+s}{q+s-m}
    (-1)^{\ell+m-s-q}
    \cot^{\,2q+s-m}\!\frac{\theta}{2}\,.
\end{align}

\vspace{1ex}
\noindent
\textbf{Orthogonality and completeness.}
The orthogonality and completeness properties of the SWSHs follow directly from those of the ordinary spherical harmonics,
\begin{align}
    \int_{S^2}d\Omega\;
    \phantom{}_{\mys s}Y_{\ell m}(\theta,\phi)\,&
    \phantom{}_{\mys s}Y^{*}_{\ell' m'}(\theta,\phi)
    = \delta_{\ell\ell'}\delta_{m m'}\,,
    \label{OrthogonalitySWSH}\\
    \sum_{\ell=0}^{\infty}\sum_{m=-\ell}^{\ell}
    \phantom{}_{\mys s}Y_{\ell m}(\theta,\phi)\,&
    \phantom{}_{\mys s}Y^{*}_{\ell m}(\theta',\phi')=\nonumber\\
    &= \delta(\phi'-\phi)\,
    \delta(\cos\theta'-\cos\theta)\,.
\end{align}

\vspace{1ex}
\noindent
\textbf{Angular derivatives.}
The action of the angular derivative operators on the SWSHs yields
\begin{align}
    \eth\,\phantom{}_{\mys s}Y_{\ell m}
    &= \sqrt{(\ell-s)(\ell+s+1)}\;
    \phantom{}_{\mys{s+1}}Y_{\ell m}\,, \label{eq:ASWSHid1}\\
    \bar{\eth}\,\phantom{}_{\mys s}Y_{\ell m}
    &= -\sqrt{(\ell+s)(\ell-s+1)}\;
    \phantom{}_{\mys{s-1}}Y_{\ell m}\,.
\end{align}
Therefore
\begin{equation}
    \bar\eth\eth \phantom{}_{\mys s}Y_{\ell m}=-(\ell-s)(\ell+s+1) \phantom{}_{\mys s}Y_{\ell m}\,,
\end{equation}
hence
\begin{equation}
    \bar\eth\eth Y_{\ell m}=-\ell(\ell+1) Y_{\ell m}\,.
\end{equation}
We thus identify the spin-weighted Laplacian on the sphere $\mathcal D^2\equiv\bar\eth\eth$.
In particular, note that
\begin{align}\label{eq:IdentityEth0}
    \bar\eth^2\eth^2 Y_{\ell m}&=\mathcal D^2(\mathcal D^2+2)Y_{\ell m}\nonumber\\
    &=(\ell-1)\ell(\ell+1)(\ell+2)Y_{\ell m}
\end{align}
as well as
\begin{align}\label{eq:IdentityEth2}
    \bar\eth^2\eth^2 \phantom{}_{\mys -2}Y_{\ell m}&=\mathcal D^2(\mathcal D^2-2)\phantom{}_{\mys -2}Y_{\ell m}\nonumber\\
    &=(\ell-1)\ell(\ell+1)(\ell+2)\phantom{}_{\mys -2}Y_{\ell m}
\end{align}
Finally, we state the commutation relations
\begin{equation}\label{eq:Ethcommutation}
    \left[\bar \eth,\eth\right]\phantom{}_{\mys s}Y_{\ell m}=2s\phantom{}_{\mys s}Y_{\ell m}\,.
\end{equation}

\vspace{1ex}
\noindent
\textbf{Complex conjugation and angular shifts.}
Complex conjugation of the SWSHs is given by
\begin{equation}
    (-1)^{s+m}\,
    \phantom{}_{\mys{-s}}Y^{*}_{\ell,-m}(\theta,\phi)
    =
    \phantom{}_{\mys s}Y_{\ell m}(\theta,\phi)\,,
    \label{CCSWSH}
\end{equation}
while shifts of the angular coordinates by \(\pi\) lead to
\begin{align}
    \phantom{}_{\mys s}Y_{\ell m}(\pi-\theta,\phi)
    &= (-1)^{\ell+m}\,
    \phantom{}_{\mys{-s}}Y_{\ell m}(\theta,\phi)\,, \label{thshift}\\
    \phantom{}_{\mys s}Y_{\ell m}(\theta,\phi+\pi)
    &= (-1)^{m}\,
    \phantom{}_{\mys s}Y_{\ell m}(\theta,\phi)\,. \label{phshift}
\end{align}
Combining Eqs.~\eqref{CCSWSH} and \eqref{thshift}, one obtains
\begin{equation}\label{CCSWSHandthshift}
    \phantom{}_{\mys s}Y_{\ell m}(\pi-\theta,\phi)
    =
    (-1)^{s+\ell}\,
    \phantom{}_{\mys s}Y^{*}_{\ell,-m}(\theta,\phi)\,.
\end{equation}

\vspace{1ex}
\noindent
\textbf{Triple product integral.}
The integral of three spin-weighted spherical harmonics satisfying
\(s_1+s_2+s_3=0\) can be expressed in terms of Wigner \(3j\) symbols as
\begin{widetext}
\begin{equation}\label{SWSHTrippleInt}
    \int_{S^2}\dd\Omega\;
    \phantom{}_{\mys{s_1}}Y_{\ell_1 m_1}\,
    \phantom{}_{\mys{s_2}}Y_{\ell_2 m_2}\,
    \phantom{}_{\mys{s_3}}Y_{\ell_3 m_3}
    =
    \sqrt{\frac{\prod_{i=1}^{3}(2\ell_i+1)}{4\pi}}
    \begin{pmatrix}
        \ell_1 & \ell_2 & \ell_3\\
        m_1 & m_2 & m_3
    \end{pmatrix}
    \begin{pmatrix}
        \ell_1 & \ell_2 & \ell_3\\
        -s_1 & -s_2 & -s_3
    \end{pmatrix}.
\end{equation}
\end{widetext}
This expression is non-vanishing only if \(m_1+m_2+m_3=0\) and
\(|\ell_1-\ell_2|\le\ell_3\le\ell_1+\ell_2\).
If the condition \(s_1+s_2+s_3=0\) is not satisfied, the above relation does not apply and one must instead use more general expressions involving gamma functions (see e.g. Appendix~A of~\cite{Favata:2008yd}).

\vspace{1ex}
\noindent
\textbf{Relation to Wigner-\(D\) matrices.}
Spin-weighted spherical harmonics are related to Wigner--\(D\) rotation matrices by\footnote{We adopt the conventions for the Wigner-\(D\) matrices
\(\mathfrak{D}^{\ell}_{m' m}(\psi,\theta,\phi)\) used in \textsc{Mathematica}.}
\begin{equation}
    \phantom{}_{\mys s}Y_{\ell m}(\theta,\phi)
    =
    (-1)^s
    \sqrt{\frac{2\ell+1}{4\pi}}\,
    \mathfrak{D}^{\ell}_{-s\,m}(0,\theta,\phi)\,.
\end{equation}
Using the group properties of the Wigner--\(D\) matrices, one finds
\begin{equation}
    \phantom{}_{\mys s}Y_{\ell m}(R_1 R_2)
    =
    \sum_{m'}
    \phantom{}_{\mys s}Y_{\ell m'}(R_2)\,
    \mathfrak{D}^{\ell}_{m' m}(R_1)\,,
\end{equation}
from which the transformation of the SWSHs under rotations follows as
\begin{equation}\label{rotSWSH}
    \phantom{}_{\mys s}Y_{\ell m}(\theta',\phi')
    =
    \sum_{m'}
    \phantom{}_{\mys s}Y_{\ell m'}(\theta,\phi)\,
    \mathfrak{D}^{\ell}_{m' m}(R)\,
    e^{is\psi}\,,
\end{equation}
where the additional phase factor \(e^{is\psi}\) accounts for the spin-weight transformation defined in Eq.~\eqref{eq:SpinWeightTransfos App}.

\subsection{Derivation of the reflection-symmetry identity}
\label{App:DerivationReflectionIdentity}

In this appendix, we provide a direct derivation of the symmetry relation stated in Eq.~\eqref{eq:SymmOrbitPlaneModes},
\begin{equation}\label{Appeq:SymmOrbitPlaneModes}
        h^H_{\ell m}=(-1)^{\ell}\,h^{H*}_{\ell,-m}\,.
\end{equation}
The result follows from the reflection symmetry of non-precessing binary systems when described in a frame in which the orbital plane coincides with the \(x\)--\(y\) plane. In such a frame, the high-frequency gravitational waveform is invariant under reflection across the orbital plane up to complex conjugation, namely
\begin{equation}
    h(t,\theta,\phi)=h^*(t,\pi-\theta,\phi)\,.
\end{equation}
Applying the above symmetry condition to a SWSH expansion yields
\begin{align}
 \sum_{\ell m} h_{\ell m}\,
 \phantom{}_{\scriptscriptstyle -2}Y_{\ell m}(\theta,\phi)
 &=
 \sum_{\ell m} h^*_{\ell m}\,
 \phantom{}_{\scriptscriptstyle -2}Y^{*}_{\ell m}(\pi-\theta,\phi)\,.
\end{align}
Using the identity in Eq.~\eqref{CCSWSHandthshift} for the combined action of complex conjugation and a reflection in \(\theta\), the right-hand side can be rewritten as
\begin{align}
 \sum_{\ell m} h^*_{\ell m}\,
 \phantom{}_{\scriptscriptstyle -2}Y^{*}_{\ell m}(\pi-\theta,\phi)
 &=
 \sum_{\ell m} (-1)^{\ell}\,h^*_{\ell,-m}\,
 \phantom{}_{\scriptscriptstyle -2}Y_{\ell m}(\theta,\phi)\,,
\end{align}
where, in the final step, we have relabeled the summation index \(m\rightarrow -m\).
Equating the mode coefficients on both sides immediately leads to
Eq.~\eqref{Appeq:SymmOrbitPlaneModes}.

\subsection{Solving the low-frequency Isaacson equation}\label{App:DerivationIsaacsonLow}
In this subsection, we explicitly extract the radiative contribution within the general solution to the low-frequency Isaacson equations in Eq.~\eqref{eqn:SolForm}, reproduced here for convenience
\begin{align}
   h^L_{\mu\nu}(x) = -16\pi G  \int d^{4}x' \; G(x-x') \,t_{\mu\nu} (x')   \label{eqn:SolFormA} \,.
\end{align}
Here, the retarded Green's function of the wave equation reads
\begin{align}
    G(x-x') =-\frac{\delta(t_{r}-t')}{4\pi  \left | \vec{x}-\vec{x}' \right | }\, , \label{eq:Green's functionA}
\end{align}
with $t_{r}$ the retarded time of luminally propagating signals
\begin{equation}
    t_{r}\equiv t-|\vec{x}-\vec{x}'|\,.
\end{equation}
This solution is adapted from~\cite{PhysRevD.44.R2945,Garfinkle:2022dnm,Heisenberg:2023prj,Zosso:2024xgy,Heisenberg:2024cjk}.

Under the change of coordinates in Eqs.~\eqref{eq:ChangeofCoords1} and \eqref{eq:ChangeofCoords1}
the volume form of the integral in Eq.~\eqref{eqn:SolFormA} becomes
\begin{align}
    d^{4}x' \rightarrow r'^{2}du'_\psi dr'd\Omega' \, ,
\end{align}
and the leading order Green's function in Eq.~\eqref{eq:Green's function} takes the form
\begin{align}
    G(x-x') = -\frac{1}{4\pi r}\mathcal{V}\,\delta (r'-(u-u') \mathcal{V})
    \, ,\label{eq:GreensFunctionFinal}
\end{align}
where
\begin{equation}\label{eq:DefRatio}
    \mathcal{V}\equiv 
    \frac{1}{1-\vec{n}' \cdot \vec{n}}\,.
\end{equation}
To obtain this result we have written
\begin{align}
    \left | \vec{x}-\vec{x}' \right | \simeq r-r'\vec{n}' \cdot \vec{n} \, ,
\end{align}
valid under the general assumption that
\begin{equation}\label{eq:assumptionSource}
    r'\ll r\,.
\end{equation}
In other words, the value of the memory component evaluated at a given spacetime point in the asymptotic limit only depends on the value of the source in a confined radius of spacetime, and can thus be evaluated ``outside of its own source" [See also Ref.~\cite{Garfinkle:2022dnm}]. This is true in general because, as explained in the main text, the low-frequency radiation signal $h^L$ is being sourced at the instant of emission of the unbound energy-momentum in the form of gravitational waves. More precisely, gravitational memory depends only on the time evolution and angular distribution of the emission of asymptotic energy flux and not its subsequent propagation. Assuming that the time scale for a given wave-pulse to accelerate to its final asymptotic velocity within the radiation zone is small compared to the time of propagation to the observer—an assumption encoded within the limit to null infinity introduced above—the emission can, without loss of generality, be assumed to be instantaneous.  
Moreover, we have used that $\delta(g(x))=  \sum_{i} \delta(x-x_{i})/ \left | g'(x_i) \right |$, 
where $x_i$'s are roots of $g(x)$. We can then use the delta function in Eq.~\eqref{eq:GreensFunctionFinal} to kill off the $r'$ integral, which effectively results in an upper bound for the $u'$ integral given by the value $u$ of asymptotic retarded time at which the low-frequency radiation is evaluated.

Putting everything together, and solving for the physical dofs in the radiation $h^{TT}_{ijL}$ by using a $TT$ projection defined in Eq.~\eqref{eq:Projectors}, Eq.~\eqref{eqn:SolFormA} in the limit to null infinity becomes
\begin{equation}
    h^{TT}_{ijL}(u,r,\Omega)=\frac{4 G}{r}\int_{-\infty}^u du'\int d^2\Omega' \frac{\perp_{ijab}\,r'^2 t_{ab}}{1-\vec{n}' \cdot \vec{n}}
\end{equation}
Plugging in the general form of the asymptotic energy-momentum tensor of emitted gravitational waves in Eq.~\eqref{eq:StressEnergyGRThird} we arrive at Eq.~\eqref{eq:memoryequationA}.

\subsection{SWSH expansion of memory formula}\label{App:SWSHExpansionMemory}

To complete this appendix, we explicitly derive the SWSH expanded form of the memory formula in Eq.~\eqref{eq:NonLinDispMemoryModesGR} starting from the general equation in terms of the TT metric perturbations [Eq.~\eqref{eq:memoryequation}]
\begin{equation}\label{eq:memoryequationApp}
    h_{ijL}^{\mathrm{TT}}
    =\frac{4G}{r}\int^{u}\!\mathrm d u'\int\!\mathrm d\Omega'\;
    \frac{dE}{du' d\Omega'}\,
    \Big[\frac{n'_i n'_j}{1-\vec n'\!\cdot\!\vec n}\Big]^{\mathrm{TT}},
\end{equation}
where $^{\mathrm{TT}}$ denotes projection onto the transverse-traceless modes with respect to the observation direction $\mathbf n(\Omega)$, and $\mathbf n'(\Omega')$ is the integration-direction unit vector. More precisely, we seek the harmonic modes of memory within the general SWSH expansion of the spin-weight $-2$ scalar
\begin{equation}\label{eq:MemoryPhasorA}
    h(u,r,\theta,\phi)=\sum_{\ell m} h_{\ell m}(u,r)\;{}_{-2}Y^{\ell m}(\theta,\phi).
\end{equation}

The crucial identity to arrive at such an expansion of Eq.~\eqref{eq:memoryequationApp} is the following identity of the TT angular factor in terms of symmetric tracefree (STF) tensors [A proof of this relation can be found in Appendix B of Ref.~\cite{Heisenberg:2023prj} or Appendix C2 of Ref.~\cite{Zosso:2024xgy}. See also Eq.~(2.34) in~\cite{BlanchetPaper}]
\begin{equation}\label{eq:BlanchetKernelAppendix}
   \Big[\frac{n'_i n'_j}{1-\vec n'\!\cdot\!\vec n}\Big]^{\mathrm{TT}}
   =\perp_{ijab}\sum_{\ell=2}^{\infty}\frac{2(2\ell+1)!!}{(\ell+2)!}\;n_{S-2}\;n'_{\langle abS-2\rangle}\,.
\end{equation}
For completeness, we briefly summarize the notion of STF tensors and the notation used throughout this appendix.
A multi-index of $\ell$ spatial indices is written as
\begin{equation}
    S \equiv i_1 i_2 \cdots i_\ell\,,
\end{equation}
while the STF projection of any tensor
$A_S$ is denoted by angle brackets,
\[
    A_{\langle S\rangle}
    \equiv \mathrm{STF}\!\left[A_{i_1 i_2 \cdots i_\ell}\right].
\]
This operation (i) symmetrizes all indices and (ii) removes all traces
with respect to the Euclidean metric $\delta_{ij}$.  For example,
\begin{align}
    A_{\langle ij\rangle}
    &= A_{(ij)} - \frac{1}{3}\delta_{ij}A_{kk},\\
    A_{\langle ijk\rangle}
    &= A_{(ijk)} - \frac{1}{5}
       \left(\delta_{ij}A_{llk} + \delta_{ik}A_{llj}
       + \delta_{jk}A_{lli}\right).
\end{align}
These STF combinations form an angular basis on the sphere that is
equivalent to the usual spherical harmonics.  The relation between them is
\begin{align}
    n_{\langle S\rangle}(\Omega)
    &=\frac{4\pi \,\ell!}{(2\ell+1)!!}\sum_{m=-\ell}^{\ell}\mathcal Y^{\ell m}_{\langle S\rangle}\,Y_{\ell m}(\Omega),\label{eq:nL_Ylm_app}\\
    Y_{\ell m}(\Omega)
    &=\mathcal Y^{\ell m}_S\,n_S(\Omega). \label{eq:nL_Ylm_app2}
\end{align}
where $\mathcal{Y}^{\ell m}_{S}$ are constant, angle-independent
STF tensors.  They satisfy the orthogonality condition
\[
    \bar{\mathcal{Y}}^{\ell m}_{S}\,
    \mathcal{Y}^{\ell m'}_{S}
    = \frac{(2\ell+1)!!}{4\pi\,\ell!}\,
      \delta_{mm'}.
\]

The usefulness of the STF basis is that a rank-2 TT field has a compact
multipole expansion.  Writing
\begin{align}\label{eq:general_STF_TT_app}
    h^{\mathrm{TT}}_{ij}
    = 4\perp_{ijab}
      \sum_{\ell=2}^{\infty}\frac{1}{\ell!}\!
      &\bigg[
        U_{abS-2}\,n_{S-2}\\
        &+\frac{2\ell}{\ell+1}\,
        \epsilon_{cd(a}V_{b)cS-2}\,n_{dS-2}
      \bigg],\nonumber
\end{align}
the STF multipoles $U_S$ and $V_S$ are the electric- and magnetic-parity
radiative moments of the gravitational field.  For the nonlinear memory,
only the electric-parity pieces $U_S$ contribute.

Substituting Eq.~\eqref{eq:BlanchetKernelAppendix} into Eq.~\eqref{eq:memoryequationApp} we obtain
\begin{align}\label{eq:memory_STF_sub_app}
     h_{ijL}^{\mathrm{TT}}
    =&\frac{4G}{r}\,\perp_{ijab}\sum_{\ell=2}^\infty\frac{2(2\ell+1)!!}{(\ell+2)!}\, n_{S-2}\nonumber\\
    &\;\times\int^{u}\!\mathrm d u'\int\!\mathrm d\Omega'\,
    \frac{dE}{du' d\Omega'}\,n'_{\langle abS-2\rangle}.
\end{align}
Comparing Eqs.~\eqref{eq:memory_STF_sub_app}) with \eqref{eq:general_STF_TT_app} shows that the memory contributes only to the electric-parity mass-type STF multipoles \(U_S\). Explicitly, one finds after matching combinatorial factors
\begin{equation}\label{eq:deltaU_STF_explicit_app}
     U^L_{S}(u)
    =\frac{G}{ r}\,\frac{2(2\ell+1)!!}{(\ell+1)(\ell+2)}
    \int\!\mathrm d\Omega'\,\mathcal F_{\mathrm{h}}(u,\Omega')\, n'_{\langle S\rangle},
\end{equation}
and
\begin{equation}
     V^L_{S}(u)=0 \,,
\end{equation}
where we have relabeled $ijL-2\rightarrow L$, and
\begin{equation}
    \mathcal F_{\mathrm{h}}(u,\Omega')\equiv \int^{u}\!\mathrm d u'\;
    \frac{dE}{du' d\Omega'}
\end{equation}
is the GW energy flux per unit solid angle.

Contracting Eq.~\eqref{eq:deltaU_STF_explicit_app} with \(\mathcal Y^{\ell m}_S\) and using orthogonality of the \(\mathcal Y\)-tensors yields the spherical-harmonic coefficients of the mass multipoles
\begin{equation}\label{eq:deltaU_lm_app}
    U^L_{\ell m}(u)
    =\frac{32\pi G}{r}\sqrt{\frac{(\ell-2)!}{2(\ell+2)!}}
    \int\!\mathrm d\Omega'\,\mathcal F_{\mathrm{h}}(u,\Omega')\,\bar{Y}_{\ell m}(\Omega').
\end{equation}
Using the usual relation between the radiative strain modes \(h_{\ell m}\) and the mass multipoles \(U_{\ell m}\)
\begin{equation}\label{eq:U_to_h_app}
    h_{\ell m}=\frac{1}{\sqrt{2}} \left(U_{\ell m}-iV_{\ell m}\right),
\end{equation}
and using Eq.~\eqref{eq:MemoryPhasor} to write
\begin{equation}
    \mathcal F_{\mathrm{h}}=\frac{1}{16\pi G}r'^2\langle\dot h_{+H}^2+\dot h_{\times H}^2\rangle=\frac{1}{16\pi G}r'^2\langle|\dot h_H|^2\rangle\,,
\end{equation}
we obtain the final expression for the memory contribution to each SWSH mode
\begin{equation}\label{eq:NonLinDispMemoryModesGRApp}
  h^L_{\ell m}(u)=\frac{1}{r}\sqrt{\frac{(\ell-2)!}{(\ell+2)!}}\int_{-\infty}^ud u'\int_{S^2}d\Omega'\,\bar Y_{\ell m}\, r'^2\big\langle |\dot{h}_{H}|^2\big\rangle\,,
\end{equation}
corresponding to Eq.~\eqref{eq:NonLinDispMemoryModesGR}.
\begin{figure*}
    \centering
    \includegraphics[width=1\linewidth]{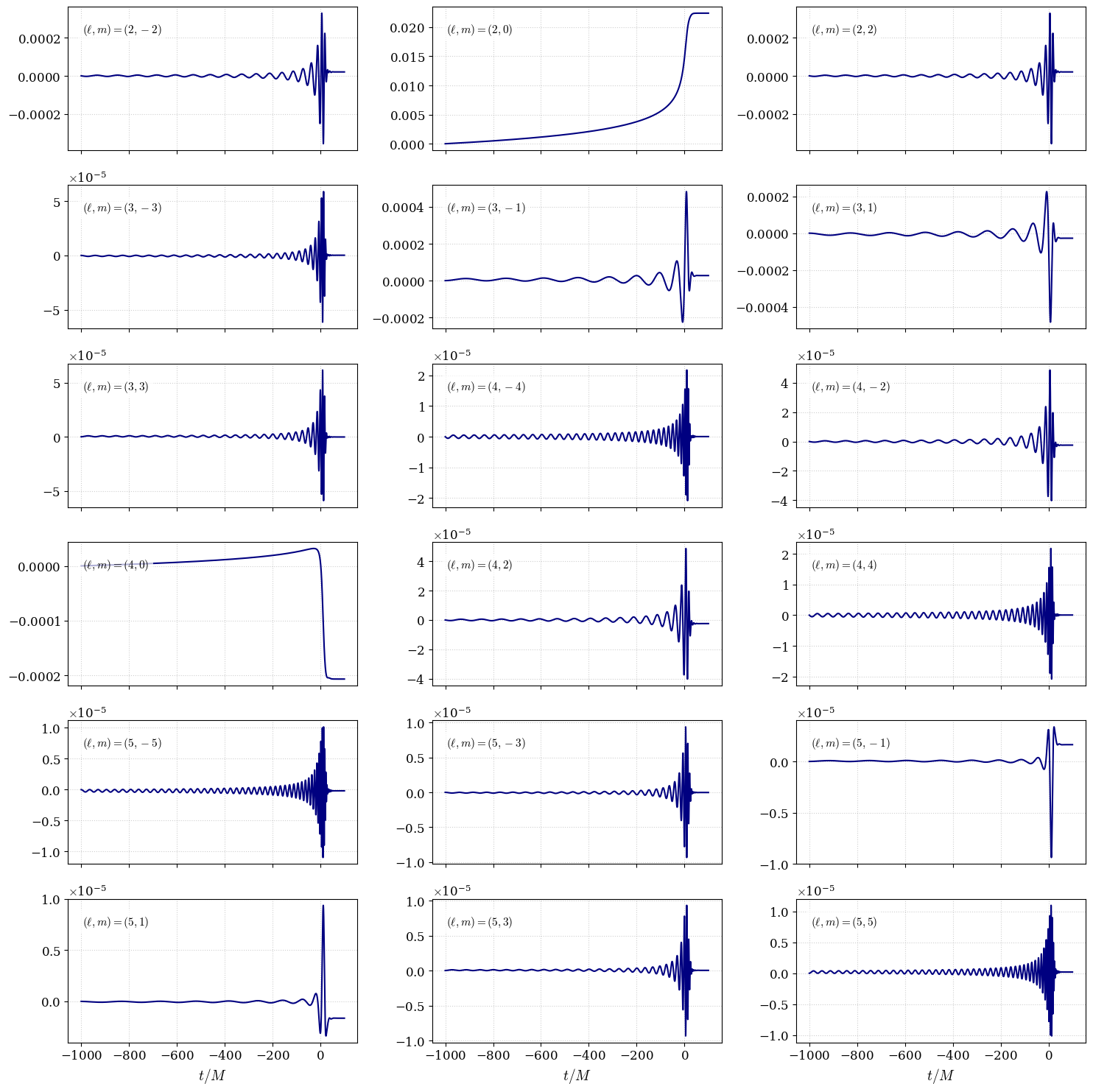}
    \caption{Output of SWSH modes $h_{\ell m}$ of the memory formula Eq.~\eqref{eq:memorymodes} (or equivalently Eq.~\eqref{eq:memory-pre-3j}) without averaging over high-frequency scales for a particular BBH merger. \textit{Parameters: $Q=6$, $\chi = 0.4$} }
    \label{fig:MultiModeMemory}
\end{figure*}

The angular integral above can be computed analytically by also expanding the
energy flux of high-frequency gravitational waves in spin-weighted spherical harmonics,
\begin{equation}
   \big\langle |\dot h_H|^{2}\big\rangle
   = \sum_{\ell_1 m_1}\sum_{\ell_2 m_2}
     \big\langle \dot h^H_{\ell_1 m_1}\,
                 \dot{\bar{h}}^H_{\ell_2 m_2} \big\rangle\,
     {}_{-2}Y_{\ell_1 m_1}\,
     {}_{-2}\bar Y_{\ell_2 m_2}.
    \label{eq:flux-SWSH-expansion}
\end{equation}
Using the standard complex–conjugation rule for spin-weighted harmonics,
\begin{equation}
   {}_{s}\bar{Y}_{\ell m}
   = (-1)^{s+m}\,{}_{-s}Y_{\ell,-m},
   \label{eq:SWSH-conjugation}
\end{equation}
we can rewrite the flux as a product of ${}_{-2}Y$ and ${}_{+2}Y$.
Inserting the expansion in Eq.~\eqref{eq:flux-SWSH-expansion} into the multipolar expression for the memory in Eq.~\eqref{eq:NonLinDispMemoryModesGRApp} we obtain
\begin{align}
  h^L_{\ell m}(u)
  &= \frac{1}{r}
     \sqrt{\frac{(\ell-2)!}{(\ell+2)!}}\;
     \sum_{\ell_1 m_1}\sum_{\ell_2 m_2}
      \,\nonumber\\
       &\times  (-1)^{m+m_2}\int_{-\infty}^{u}\! \mathrm{d}u'\,\;
       r'^{2}\,
       \big\langle
           \dot h_{\ell_1 m_1}\,
           \dot h^{*}_{\ell_2 m_2}
       \big\rangle
        \label{eq:memory-pre-3j}\\
  &\times
     \int_{S^{2}}\! \mathrm{d}\Omega'\;
       {}_{-2}Y_{\ell_1 m_1}\,
       {}_{+2}Y_{\ell_2,-m_2}\,
       Y_{\ell,-m}(\Omega') .\nonumber  
\end{align}
An example of the evaluation of this formula without implementation of the high-frequency spacetime average for a particular BBH merger event is shown in Fig.~\ref{fig:MultiModeMemory}. 

In principle, the Isaacson averaging in the time domain can be implemented through a coarse-graining procedure over several wavelengths of the high-frequency gravitational radiation~\cite{Isaacson_PhysRev.166.1263,Isaacson_PhysRev.166.1272,maggiore2008gravitational}. Since the dominant oscillatory scale is set by the instantaneous gravitational-wave frequency $f_H(u)$, which evolves during the inspiral and merger, the corresponding wavelength $\lambda_H(u)\sim 1/f_H(u)$ defines a time-dependent short scale. In this context, spacetime averaging is effectively equivalent to averaging over several local wavelengths of the radiation~\cite{maggiore2008gravitational}.

In practice, this is implemented through a general a kernel-based average of a quantity $O(u)$ as
\begin{equation}
    \langle O(u)\rangle
    = \int du'\,K(u,u')\,O(u')\,,
\end{equation}
where the kernel $K(u,u')$ is normalized and localized around $u$ with a width $\Delta u(u)$ satisfying
\begin{equation}
    \frac{1}{f_H(u)} \ll \Delta u(u) \ll T_L\,,
\end{equation}
with $T_L$ the characteristic timescale of the memory evolution. A simple example is a Gaussian window, or, alternatively, a top-hat window, both commonly used in analogous coarse-graining procedures. The key requirement is that the choice of kernel is compatible with the required properties (I-III) outlined in Sec.~\ref{App:IssacsonMemory}, and that it tracks the evolving frequency scale $f_H(u)$, ensuring that the averaging always extends over several oscillation periods while remaining local on the scale of the memory buildup.
When applied to the quadratic combinations entering Eq.~\eqref{eq:memory-pre-3j}, this procedure suppresses rapidly oscillating cross terms while retaining the secular, low-frequency contribution associated with the memory growth. 

In the case of quasi-circular binary black hole mergers, this separation of scales is effectively realized at the level of the multipolar expansion: the averaging suppresses oscillatory mode couplings and selects the $(2,0)$ mode as the dominant contribution to the memory signal. Consequently, restricting the memory formula to the $(2,0)$ harmonic mode provides a practical implementation of the Isaacson averaging in this regime, and no explicit averaging operation is required for the results presented in this work.

The remaining angular integral is exactly of the form to which the
three–spin–weighted–harmonic identity (Eq.~\eqref{SWSHTrippleInt})
applies.  Evaluating it expresses the memory mode
$h^L_{\ell m}$ as a finite sum over Wigner \(3j\) symbols and
products of the high-frequency radiative modes
$\dot h^H_{\ell m}$ as written in Eq.~\eqref{eq:memorymodes} in the main text.


\newpage
\bibliographystyle{utcaps}
\bibliography{references}

\clearpage

\end{document}